\documentclass[%
aip,
amsmath,amssymb,
superscriptaddress,
reprint,%
]{revtex4-1}
\usepackage{graphicx}
\usepackage{dcolumn}
\usepackage{bm}
\usepackage[utf8]{inputenc}
\usepackage[T1]{fontenc}
\usepackage{mathptmx}
\usepackage{etoolbox}
\usepackage{physics}
\usepackage[colorlinks,linkcolor=blue,citecolor=blue, urlcolor  = blue]{hyperref}
\usepackage{xcolor}
\usepackage{enumitem}
\usepackage{xfrac}
\usepackage{setspace}
\usepackage{natbib}
\usepackage{ifthen}
\usepackage{subfigure}
\graphicspath{{images/}}
\makeatletter
\def\@email#1#2{%
	\endgroup
	\patchcmd{\titleblock@produce}
	{\frontmatter@RRAPformat}
	{\frontmatter@RRAPformat{\produce@RRAP{*#1\href{mailto:#2}{#2}}}\frontmatter@RRAPformat}
	{}{}
}%
\makeatother
\begin{document}
	
	\preprint{AIP/123-QED}
	
	\title{Quantum Interference in Atomic Systems\vspace{0.3cm}}

	\author{Sajad Ahmadi}
	\affiliation{\mbox{Institute for Advanced Studies in Basic Sciences (IASBS), Departmant of Physics, 45137-66731, Zanjan, Iran}}
	\author{Mohsen Akbari}
	\altaffiliation{Corresponding author's email: \textcolor{blue}{\textit{mohsen.akbari@khu.ac.ir}}}
	\affiliation{\mbox{Quantum Optics Laboratory, Departmant of Physics, Kharazmi University, 16315-1355 Tehran, Iran}}
	\author{Shahpoor Saeidian}
	\affiliation{\mbox{Institute for Advanced Studies in Basic Sciences (IASBS), Departmant of Physics, 45137-66731, Zanjan, Iran}}
	\author{Ali Motazedifard\vspace{0.3cm}}
	\affiliation{\mbox{Quantum Sensing Lab, Quantum Metrology Group, Iranian Center for Quantum Technologies (ICQT), Tehran, 15998-14713,Iran}}
	\affiliation{\mbox{Quantum Optics Group, Department of Physics, University of Isfahan, Hezar-Jerib, Isfahan, 81746-73441, Iran}}
	
	\date{\today}
	
	\begin{abstract}
     Quantum interference takes center stage in the realm of quantum particles, playing a crucial role in revealing their wave-like nature and probabilistic behavior. It relies on the concept of superposition, where the probability amplitudes of different processes that contribute to the given phenomenon interfere with each other. When combined, their phases can interfere either constructively or destructively.      
     Quantum interference manifests in three distinct forms: optical interference, arising from the interaction of light waves and forming the basis for technologies such as lasers and optical filters. Interference via atoms involves manipulating atomic states to control light interaction, enabling techniques like Stimulated Raman Adiabatic Passage ($STIRAP$) and Electromagnetically Induced Transparency ($EIT$) in quantum information processing. Finally, self-interference of atoms occurs when matter waves associated with individual atoms interfere with themselves, enabling precise measurements in atom interferometry, a crucial tool for fields like quantum mechanics and navigation.
     These diverse forms of quantum interference have profound implications for numerous scientific disciplines, demonstrating its ability to encompass all quantum particles, not just light.
	\end{abstract}
	\maketitle
     \tableofcontents{}
   \section{Introduction}
   Interference, a characteristic of wave behavior, occurs when two or more waves overlap and combine to form a resultant wave, as dictated by the principle of superposition. This principle explains how waves combine and gives rise to interference, which can be constructive when peaks enhance each other and destructive when they cancel out\cite{Zubairy-Q}.\\
  In classical physics, interference typically manifests as variations in intensity (or intensity correlations), due to the superposition of physical waves, such as light or sound waves\cite{RevModPhys.84.777}. This phenomenon is rooted in wave theory based on Maxwell's equations, which leads to the principle of linear superposition for electric-field amplitudes\cite{Krzysztof}. A classic example is optical interference, where light waves propagating along separate paths recombine, and the phase difference between them determines the resulting bright and dark regions on a screen (i.e., fringes), provided that coherence time and length are maintained. Thomas Young's double-slit experiment serves as a cornerstone demonstration of this phenomenon in optics, showcasing the formation of an interference pattern with alternating bright and dark fringes due to overlapping light waves \cite{Young}. Similar interference effects are observed in acoustics and fluid dynamics, highlighting its significance in classical physics and providing evidence for the wave-like nature of various phenomena \cite{Rossing, lamb}.\\
  In contrast, the Hanbury Brown–Twiss(HBT) experiment demonstrates second-order interference by focusing on intensity correlations rather than amplitude interference. By splitting light from a source and measuring the intensity at two detectors, the experiment analyzes how the detected intensities are related over time. The key finding is that photons can exhibit statistical correlations, providing insights into the quantum nature of light. For instance, thermal light sources exhibit photon bunching, while single-photon sources exhibit photon antibunching. The HBT experiment has been fundamental in advancing our understanding of quantum optics\cite{fox,gerry}.
  
   Quantum mechanics also embraces the principle of superposition, playing a crucial role in quantum interference\cite{dirac}. However, the underlying mechanisms differ significantly from classical interference. In quantum mechanics, the superposition principle applies to probability amplitudes associated with the wavefunction of a particle, not the electric field amplitudes as in classical waves. When single photons are sent toward a double slit, their wavefunctions originating from each slit interfere, resulting in a superposition of the photon’s state and the formation of an interference pattern. This contrasts with the classical double-slit experiment, where the interference involves the amplitudes of the light waves themselves.\\
   The groundbreaking thought experiment proposed by Richard Feynman in 1965 envisioned that even single electrons, not just photons, would exhibit an interference pattern in a double-slit setup. This prediction, confirmed ten years later by the successful experiment with single electrons conducted by Steeds et al. (1974)\cite{Steeds}, paved the way for interference experiments with atoms, neutrons, and other quantum particles. This demonstrates how the phenomenon of interference, initially associated with light, extends to all quantum particles through the principle of wave-particle duality, as postulated by Louis de Broglie.
   
   Classical interference is the macroscopic expression of quantum interference; however, interference phenomena in the quantum domain are richer and more prominent.
   The fundamental distinction between quantum and classical interference lies in the nature of the entities involved and their interaction mechanisms. Classical interference arises from the direct interaction of physical waves, while quantum interference stems from the wave-like nature of particles, governed by the principle of particle-wave duality\cite{Krzysztof,Englert}. This duality leads to the superposition of probability amplitudes associated with the particle's wavefunction, essentially causing the particle to interfere with itself as it takes different paths simultaneously. This quantum superposition gives rise to the observed interference pattern. However, measuring the particle's path (which-path information) disrupts the superposition and eliminates the pattern. This crucial difference highlights that classical interference involves physical waves directly interacting and following classical wave equations, whereas quantum interference involves probabilities associated with the particle's wavefunction, demonstrating the fundamental distinction between the two phenomena.   
 
  The article explores quantum interference in four parts: optical interference, interference via atoms, interference of atoms with themselves, and interference in quantum technologies\\
  We will first explore optical interference in Sec.\ref{Op_in}, the most fundamental form. Understanding its underlying principles will prove informative for the more complex phenomena discussed later.
  Optical interference occurs when light waves interact, superimposing their electric fields. This interaction can lead to constructive or destructive interference, resulting in either brighter or dimmer intensity, depending on the spatial overlap and phase relationship between the waves.
  
  In Sec.\ref{IvA}, we explore interference via atoms and its applications. This type of quantum interference involves the manipulation of atomic states to induce interference effects, leading to phenomena such as STIRAP (Stimulated Raman Adiabatic Passage), EIT (Electromagnetically Induced Transparency), and CPT (Coherent Population Trapping). These techniques exploit the coherent superposition of atomic states to control how light interacts with atomic systems. Interference via atoms allows for precise control and manipulation of atomic systems and is essential for various applications in quantum technology, including quantum computing, quantum communication, and high-precision metrology. These techniques demonstrate the rich behavior of atoms when interacting with light and pave the way for advanced quantum technologies. In the subsequent subsections of this section, we will examine the Raman process in three-level atoms to delve deeper into interference via atoms.\\
  This section concludes by discussing a novel type of interference observed with artificial atoms, distinct from the interference phenomena observed in quantum optics using natural, smaller atoms. These artificial atoms, significantly larger than their natural counterparts, are often referred to as 'giant atoms.' Their ability to couple to waves at multiple points spaced at wavelength distances apart leads to unique interference effects.

  In Sec.\ref{IAWT}, we investigate a type of quantum interference with a particular focus on the phenomenon of self-interference within atoms. At the heart of this exploration lies the remarkable behavior of atoms, which exhibit wave-like properties and interfere with each other, even when treated as individual particles. This phenomenon is commonly observed in experiments such as atom interferometry, where the interference pattern arises from the superposition of different quantum states of the atoms.\\
  Experiments involving this interference typically split a beam of atoms into separate paths before recombining them. As the matter waves associated with these atoms travel along their paths, they interfere with each other, creating observable patterns and leading to regions of constructive and destructive interference. These patterns enable precise measurements of physical quantities such as gravitational acceleration, rotation, and magnetic fields, providing valuable insights into the wave nature of matter. They have various applications in fields like quantum mechanics, atom optics, and quantum information processing \cite{RevModPhys.81.1051}.
  In this section, we will explore some atomic interferometers such as the Mach-Zehnder and Ramsey interferometers, and observe how atoms interfere together in these devices.
  
   Finally, in Sec.\ref{Iqt}, we delve into the role of quantum interference in some applications of quantum technologies. Quantum technology holds immense potential across various fields, including Quantum Computing, Quantum Cryptography, and Quantum Sensing (including metrology, imaging, and more). In each of these applications, quantum interference plays a distinct role.\\
  \textbf{Quantum Computing}: Quantum interference plays a crucial role in quantum computing. By manipulating qubits that exploit interference phenomena, quantum algorithms like Shor's and Grover's achieve significant computational advantages over classical algorithms \cite{Nielsen_2010, Bravyi, Aaronson2011, 365700}.\\
  \textbf{Quantum Cryptography}:Quantum Key Distribution protocols can sometimes harness the power of quantum interference to create unbreakable communication channels. This unique phenomenon of quantum mechanics ensures that eavesdroppers cannot intercept messages without detection. \cite{Gisin2002, Scarani2009, Lo1999, Maurer}.\\
  \textbf{Quantum Metrology and Imaging}: Interference-based techniques are employed in quantum metrology for high-precision measurements and in quantum imaging for high-resolution imaging techniques beyond classical limits. Quantum interferometers can surpass the sensitivity limits of classical devices, enabling applications in gravitational wave detection, magnetic field sensing, quantum lithography, quantum optical coherence tomography, and ghost imaging, which utilize quantum entanglement and interference for enhanced resolution and sensitivity \cite{Giovannetti, Demkowicz, Xiaoying, Hosten, RevModPhys.89.035002, Rui, PhysRevA.73.062305, Pittman, Boto, Kawabe, Dbrowska, Kok}.
\section{\label{Op_in}Optical Interference}
 Optical interference is a fundamental phenomenon that reveals the wave nature and the correlation between radiation fields when referring to classical events. Concepts of optical interference, demonstrated in experiments by Michelson and Young, are well-known. In these experiments, observed interference patterns and fringes indicate either temporal coherence (in the Michelson interferometer) or spatial coherence (in the Young interferometer) of the light beams entering the interferometer, exemplifying the wave-like behavior of light \cite{wolf}.
  However, the analysis of these experiments became challenging with the advent of quantum mechanics, as light exhibited particle-like properties in many cases. Mandel and Pfleegor reported an observation of interference patterns from non-independent photon beams \cite{Mandel, Swain}.\\ 
 Based on theoretical analyses arising from quantum and classical optics, the interference pattern resulting from optical interference is a consequence of first-order field coherence \cite{Beginners}. Additionally, there are interference effects that distinguish the quantum nature of light from its wave-like behavior, arising due to higher-order (second-order) correlations between the driving fields \cite{Swain}.\\ 
 From a classical perspective, we would state that an interference pattern is created as a result of the superposition of classical wave amplitudes. However, from a quantum standpoint, based on Young's double-slit experiment, an interference pattern is observed, indicating the result of the superposition of photon probability amplitudes \cite{Swain}. This principle of superposition is at the heart of quantum mechanics and serves as a tool to distinguish the quantum nature of light from its wave nature \cite{Beginners}.\\ 
 In this article, we will explore quantum interference by atoms. However, a preliminary understanding of optical interference is imperative for comprehending atomic interference.
 \subsection{ Classical Interference}
 In optical interference, light waves propagate along separate paths and recombine on a screen or detectors. Depending on the phase difference between the waves along the two paths, we may observe constructive or destructive interference on the screen. By placing these two states side by side, we will observe an interference pattern.
\begin{figure}[h!]
	\centering
	\subfigure[Mach-Zehnder interferometer]
	{
		\includegraphics[width=.3\textwidth]{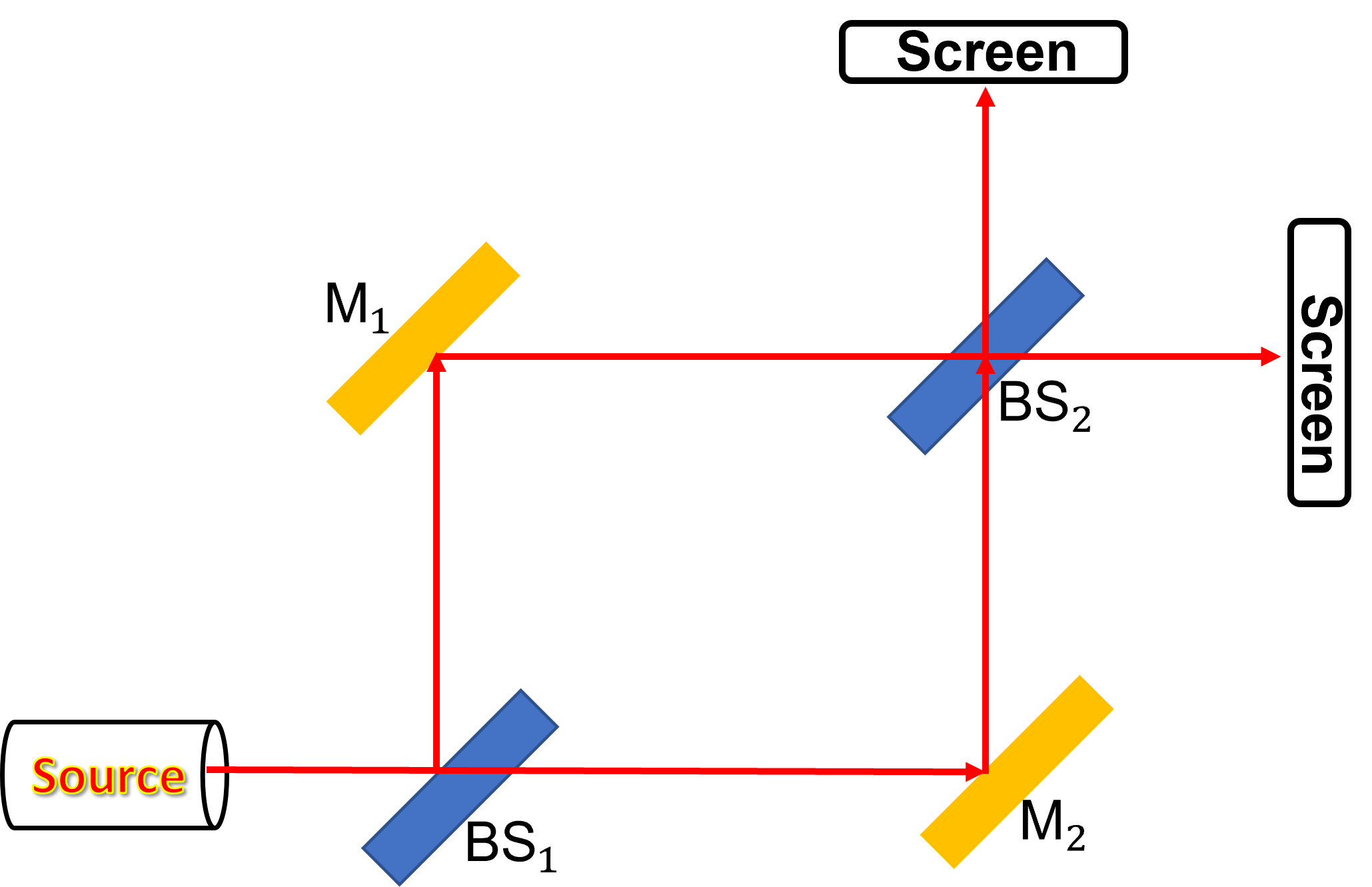}
		\label{zen}
	}
	\subfigure[Michelson Interferometer]
	{
		\includegraphics[width=.28\textwidth]{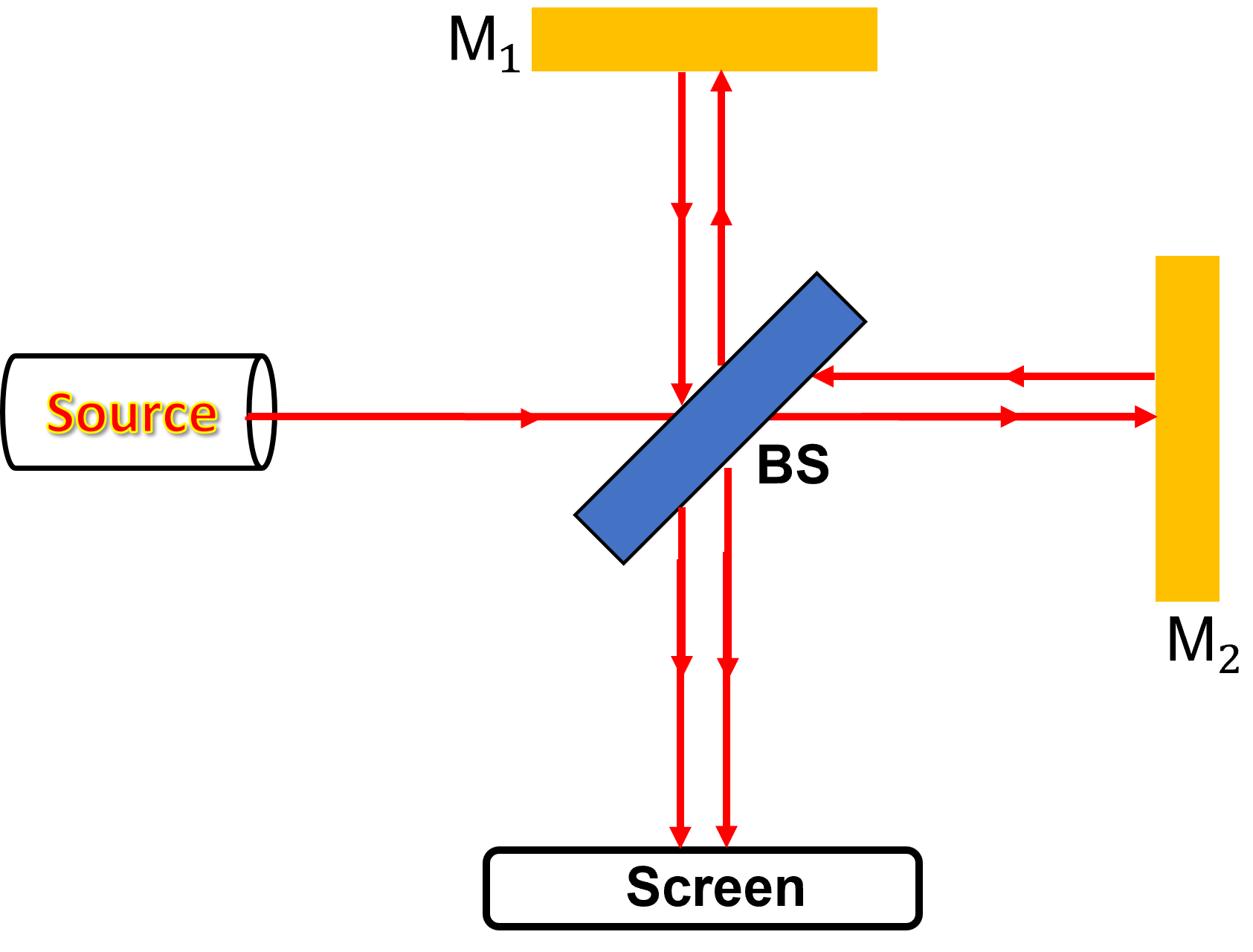}
		\label{mic}
	}
	\subfigure[Young's double-slit]
	{
		\includegraphics[width=.32\textwidth]{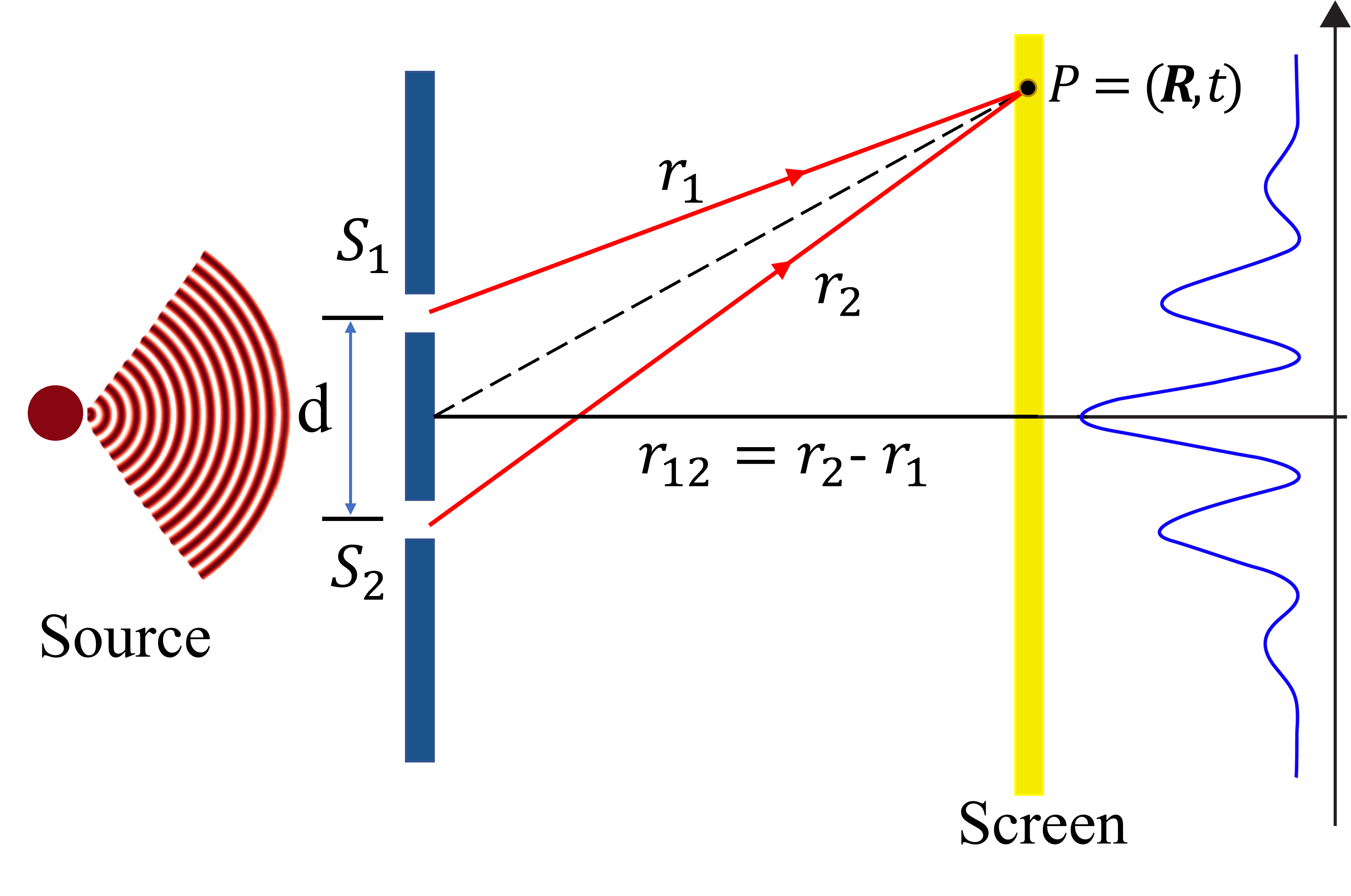}
		\label{p1}
	}
	\caption{(a) Mach-Zehnder interferometer: This interferometer works by splitting the beam into two paths using the beam splitter \(BS_1\). Each path then interacts with mirrors \(M_1\) and \(M_2\) before recombining at the beam splitter \(BS_2\). Finally, this setup produces an interference pattern visible on a screen. (b) Michelson Interferometer: A beam of light from a source passes through a beam splitter, then travels along two different paths before recombining at the detector, resulting in interference between the two beams on the screen. (c) Young's double-slit experiment: Two monochromatic light beams exit from slits \(s_1\) and \(s_2\), resulting in the formation of an interference pattern on the screen.}
	\label{cla_Inter}
 \end{figure}
 Several well-known interferometers, like the Mach-Zehnder, Michelson, and Young's double-slit experiment (shown in Fig.\ref{zen}, Fig.\ref{mic}, and Fig.\ref{p1}, respectively), demonstrate the wave nature of light. In these experiments, the light wave is initially split into two separate beams. Each beam travels along its own distinct path before recombining on a screen or detectors.\\
 In the analysis of first-order coherence, the Young's double-slit experiment serves as the elementary model, playing an important role in understanding the essential classical and quantum aspects of light \cite{Beginners}. The interference pattern in this experiment can be described by the normalized first-order coherence function,  $g^{(1)}(\textbf{R})$, which indicates the correlation between the amplitudes of the fields from the two slits.
 \begin{align} 
 	I(\textbf{R}) &= I_1 + I_2 + 2\sqrt{I_1I_2} \ g^{(1)}(\textbf{R})\cos(k_0 \textbf{R}.\textbf{r}_{12})
 	\label{z4}
 \end{align}
 The most important quantity in this equation is \( g^{(1)} \), called the normalized first-order coherence function, which measures coherence between two beams based on their intensity \cite{Beginners}. 
 \begin{equation}
 	g^{(1)}(\textbf{R}) = \dfrac{\expval{E_1^* E_2}}{\sqrt{I_1I_2}}
 	\label{g-1}
 \end{equation}
In other words, \( g^{(1)}(\textbf{R}) \) indicates the correlation between the electric field amplitudes \( \textbf{E}_1 \) and \( \textbf{E}_2 \) originating from the slits \cite{Beginners,Swain}. First-order coherence occurs when \( \vert g^{(1)}(\textbf{R})\vert=1 \), indicating complete correlation between the fields and resulting in an interference pattern \cite{Beginners,orszag}. However, when there is no correlation between the fields, i.e., \( \vert g^{(1)}(\textbf{R})\vert=0 \), the final intensity will be the sum of the intensities of the two fields, and therefore will not depend on the position \( P(\textbf{R},t) \).\\
 It is appropriate to use Rayleigh's definition to illustrate the visibility of the fringes\cite{gerry}.
\begin{align} 
	\mathcal{V}=\dfrac{I_{max}-I_{min}}{I_{max}+I_{min}} 
	\label{z7}
\end{align}
In this case, based on Eq.\eqref{z4}, we can write
\begin{align} 
	I_{\frac{max}{min}}=I_1+I_2\pm 2\sqrt{I_1I_2} \ \ \vert g^{(1)}(R)\vert
	\label{z8}
\end{align}
This gives us the new equation in terms of visibility and first-order coherence\cite{gerry}.
\begin{align} 
	\mathcal{V}=g^{(1)}(\textbf{R}) \dfrac{2\sqrt{I_1I_2}}{I_1 + I_2} 
	\label{zzz7}
\end{align}
It is clear that if \( I_1 = I_2 \), then \(\mathcal{V} = g^{(1)}(\mathbf{R})\). Therefore, the first-order correlation function determines the visibility of the interference fringes\cite{fox}.\\
First-order coherence experiments cannot distinguish between light states that have identical spectral distributions but completely different photon distributions\cite{gerry}. In the 1950s, R. Hanbury Brown and R. Q. Twiss introduced a novel correlation experiment in Manchester, known as the Hanbury Brown–Twiss (HBT) experiment. This experiment focuses on intensity correlations rather than field correlations and plays a crucial role in understanding the statistical properties of photons\cite{BROWN}.
 \begin{figure}[h!]
 	\includegraphics[width=6cm]{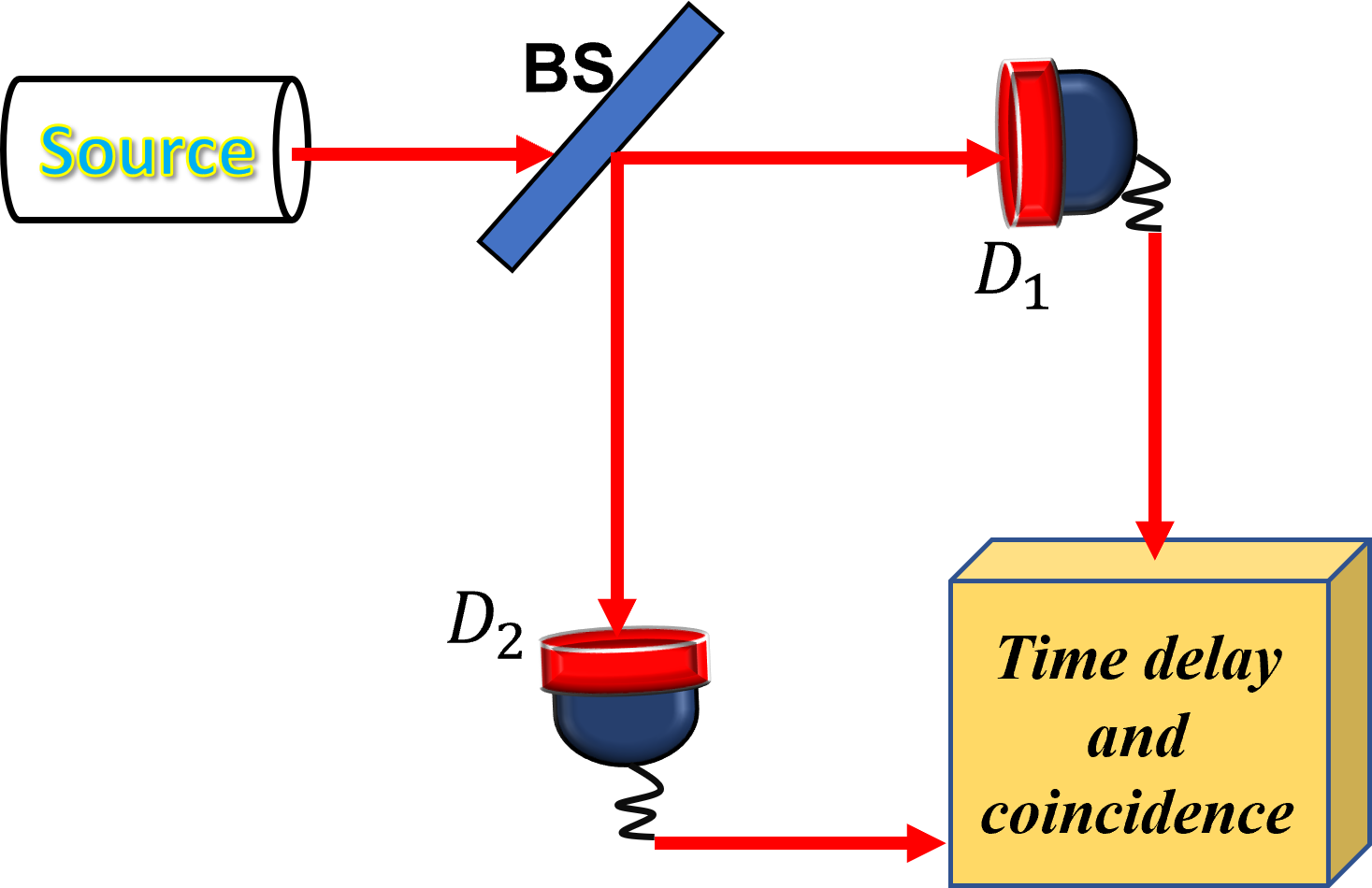}
 	\centering 
 	\caption{
 		A simple illustration of the HBT experiment: In this setup, detectors \( D_1 \) and \( D_2 \) are positioned equidistantly from the beam splitter \( BS \). This configuration measures the coincidence rate of time delays: one detector registers a count at time \( t \), while the other records a count at \( t + \tau \). If the time delay \( \tau \) is smaller than the coherence time \( \tau_0 \), information about the statistical properties of the light beam interacting with the beam splitter can be determined.
 	}
 	\label{p2}
 \end{figure} 
 The second-order coherence function \( g^{(2)}(\tau) \) associated with a classical field having a complex amplitude is determined in the HBT experiment, which exhibits different coherence properties compared to the first-order coherence function. Even if the driving fields are independently generated by two sources with a random phase difference, we will still observe an interference pattern \cite{Beginners,gerry}. Moreover, in second-order coherence experiments, \( 1 \leq g^{(2)}(\tau) < \infty \) holds true \cite{gerry}.
\subsection{\label{sec:QI-subsection}Quantum interference}
Research on the quantum properties of light began around half a century ago. The advancements in this domain allow individuals to control the coherence of quantum optical systems and enable practical quantum engineering. As a result, quantum optics methods have provided the means to conduct thought experiments concerning the foundational principles of quantum theory. Controlling quantum phenomena allows for the exploration of new protocols for information processing, signaling the promise of new technologies based on quantum information science\cite{RevModPhys.84.777}.\\
Quantum interference is a phenomenon that emphasizes the concept of superposition of probability amplitudes, rather than the superposition of electric field amplitudes of classical light.\\

 Let's go back to the double-slit experiment. We can understand the fundamental concepts of quantum mechanics from this experiment because it is where the concepts of superposition, uncertainty, measurement, and quantization are well-linked. We know that light is made up of particles called photons. Now, if we have a single-photon source that sends a beam with a single-photon state towards the double slit, then farther from the slits and on the screen, we’ll observe an interference pattern forming slowly—one photon at a time.\\

When the wavefunction of a photon originating from the left slit interferes with the wavefunction originating from the right slit, the result is a \textbf{superposition} of photon states. Unlike in the classical double-slit experiment, it is not the wave amplitudes that interfere here, but rather the probability amplitudes of the wavefunction. 
However, if we \textbf{measure} which slit the photon passes through, we will observe it in one slit or the other, not in both simultaneously. This measurement disrupts the superposition and eliminates the interference.
In fact, it's impossible to precisely know which slit a photon has passed through while simultaneously observing the interference pattern. This exemplifies the \textbf{uncertainty principle}. Let's illustrate this with equations.\\

\begin{figure}[h!]
	\includegraphics[width=8.4cm]{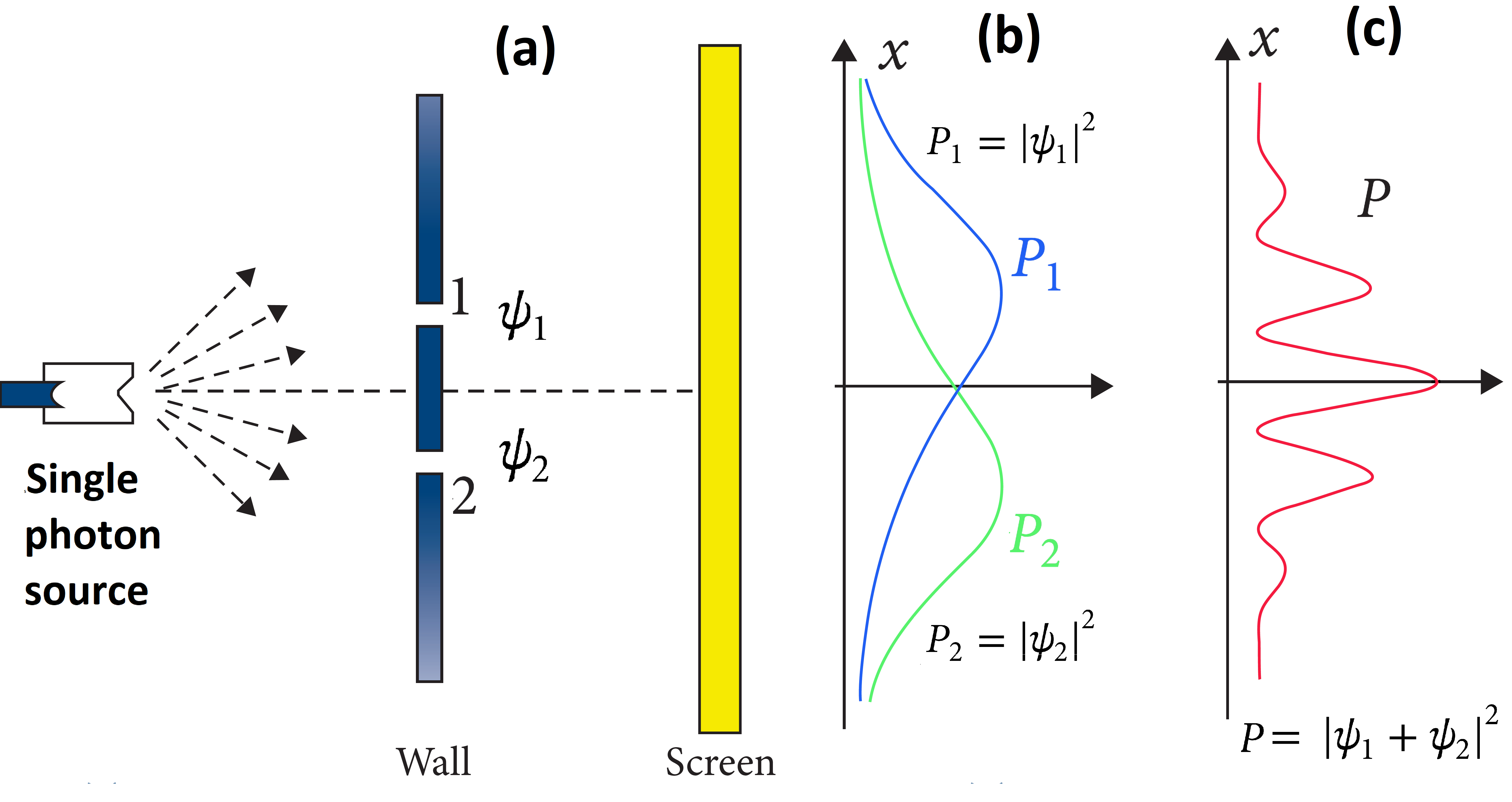}
	\centering 
	\caption
	{
		(a)The double-slit experiment using a single-photon source. (b)If a detector is placed in front of the first or second slit for measurement, the probability of observation for the corresponding wavefunction increases and the interference pattern disappears. (c)The image shows the moment when a photon passes through both slits without a detector and reaches the observation screen. In this situation, the interference pattern will be visible.  Figure adapted from [\onlinecite{Zubairy-Q}].
	}
	\label{pp2}
\end{figure}

In the double-slit experiment with a single-photon source (Fig.\ref{pp2}), we denote the wavefunction arising from the first slit as $\ket{\psi_1}=\sum_{n} c^{(1)}_n\ket{n}$ and the wavefunction of the photon passing through the second slit as $\ket{\psi_2}=\sum_{n} c^{(2)}_n\ket{n}$, where, $\{\ket{n}\}$ are the basis vectors on which the states $\psi$ are expanded. In simpler terms, $\{\ket{n}\}$ represents the possible states of the system. Additionally, $c_n=\braket{n}{\psi}$ are complex expansion coefficients that represent the components of the state vector $\psi$ in this basis. \\
Using these coefficients, we can determine the probability of finding the system in a specific basis.\\
In other words, by performing a measurement, the system is found in only one of the states $\ket{n}$ with a probability determined by $\abs{c_n}^2$. This allows us to determine the probability of the system being in any given state\cite{sakur,Lukin,Zettili,Zubairy-Q}.\\
If we close the \( i \)-th slit (\( i=1,2 \)), the final wavefunction at  position \( R \) on the screen will be \( \psi_j(R) \) with $i \neq j$, and the probability of finding a photon there will be \( P_j= \left| \psi_j(R) \right|^2 \). But when both slits are open, the overall photon wavefunction at position \( R \), according to quantum mechanics, must be a superposition of the two state functions, namely \( \psi(R) =({{\psi_1(R) + \psi_2(R)}})/{{\sqrt{2}}} \). Clearly, in this situation, the probability of observing the photon(s) on the observation screen will be \( P = \left| \psi(R) \right|^2  \).\\
The superposition of these two states has a unique property: the probability of being in state ($\ket{n}$) is given by the absolute value of the sum of the expansion coefficients, not the sum of probabilities\cite{Lukin,Zubairy-Q}.
\begin{equation}
	P_n=\dfrac{1}{2}\abs{c^{(1)}_n + c^{(2)}_n}^2\neq \dfrac{1}{2}(P^{(1)}_n + P^{(2)}_n)
\end{equation}
The origin of this inequality, indicated here by \(c^{(1)*}_{n} c^{(2)}_n\), arises from the interference term between the two state vectors\cite{Lukin}. 
It should be noted that these relations and their results are only valid if we cannot initially determine which slit the photon emitted from the source passes through. In fact, if we can perform an experiment that determines whether the photons passed through slit 1 or slit 2, then the probability of finding the photon at a point like R on the screen will be equal to the sum of the probabilities for each state ($P=P_1 + P_2$). In this situation, we will no longer have an interference pattern\cite{Feynman,Zubairy-Q}.
We saw how we could review fundamental quantum concepts in a very simple way using interference.

In general, quantum interference is a central concept in quantum mechanics, highlighting unique behaviors at the quantum level where classical physics no longer applies. \\
For optical quantum interference, we can explore first and second-order coherence functions. The calculation process here is analogous to that of the classical case, with the only distinction being that in the quantum description of the interference phenomenon, we utilize field operators instead of field amplitudes. For the first-order coherence functions, we have
 \begin{align} 
 	I(R,t) = I_1 + I_2 + 2\sqrt{I_1I_2} g^{(1)}(x_1 , x_2)
 	\label{z14}
 \end{align}
 Where \(x_i=(\mathbf{R}_i , t)\), \(I_i=\text{Tr} [ \hat{\rho}\hat{E}^- (x_i) \hat{E}^+ (x_i)]\) denotes the photon intensity arising from each slit ($i=1,2$), and \(\hat{E}^{\pm}(x_i)\), represents the positive and negative frequency components of the field operators. Therefore, the first-order quantum coherence function can be written as\cite{loudon,gerry,orszag}\\
 \begin{align} 
 	g^{(1)}(x_1 , x_2) = \dfrac{\langle E^{(-)}(x_1)E^{(+)}(x_2)\rangle}{\sqrt{I_1 I_2}}
 	\label{z16}
 \end{align}
 We see that the first-order correlation function, given in Eq.\eqref{z16}, obtained by field operators, is similar to the classical quantity described in Eq.\eqref{g-1}. \\
 This similarity arises from the fact that Young’s double-slit experiment cannot distinguish between the quantum and classical effects described by the first-order correlation functions\cite{Beginners}. Because in both cases, \(0 \leq \vert g^{(1)}(x_1 , x_2)\vert \leq 1\).\\
 
 What distinguishes quantum interference from classical is the second-order correlation function\cite{Swain}. Based on the definition, the normalized second-order quantum correlation function is written as follows\cite{gerry}
 \begin{align} 
 	g^{(2)}(x_1,x_2 ; x_2 , x_1)=\dfrac{G^{(2)}(x_1,x_2 ; x_2 , x_1)}{G^{(1)}(x_1,x_1) G^{(1)}(x_2,x_2)}
 	\label{z17}
 \end{align}
 In which\cite{gerry}
 \begin{align} 
 	G^{(2)}(x_1,x_2 ; x_2 , x_1)=Tr[\hat\rho \hat{E}^- (x_1) \hat{E}^- (x_2)\hat{E}^+ (x_2)\hat{E}^+ (x_1)]
 	\label{z18}
 \end{align}
 is the second-order quantum correlation function and
 \begin{align} 
 	G^{(1)}(x_i,x_i)=Tr\lbrace \rho \hat{E}^- (x_i) \hat{E}^+ (x_i)\rbrace
 	\label{z19}
 \end{align}
 is the intensity is due to each slit. \\
 The function \( g^{(2)}(x_1,x_2 ; x_2 , x_1) \) represents the joint probability of detecting the first photon at position \( \textbf{r}_1 \) and time \( t_1 \), as well as the second photon at position \( \textbf{r}_2 \) and time \( t_2 \)\cite{gerry}.
 If our quantum field satisfies the following two conditions, then it will exhibit second-order coherence.
 \begin{subequations}
 	\begin{align} 
 		& \vert g^{(1)}(x_1 , x_2)\vert=1 \ \ , \ \ g^{(2)}(x_1,x_2; x_2,x_1)=1 \\
 		&  G^{(2)}(x_1,x_2; x_2,x_1)=G^{(1)}(x_1 , x_1)G^{(1)}(x_2 , x_2)
 		\label{z20}
 	\end{align}
 \end{subequations}
Quantum interference plays an important role in the development of new trends in quantum optics. For example, it is utilized in quantum phase estimation, a technique used in various quantum algorithms to estimate the phase of a quantum state. Quantum optical experiments can then be used to test fundamental aspects of quantum physics, such as the EPR paradox, entanglement, and Bell's inequality\cite{Beginners}.
\section{\label{IvA} Interference via atoms}
In this section, we delve into atomic interference and its exciting possibilities. We discuss how light interacts with atoms, producing effects such as coherent population trapping (CPT), stimulated Raman adiabatic passage (STIRAP), and electromagnetically induced transparency (EIT). These effects showcase how we can manipulate light with atoms, rendering materials transparent or trapping light within specific atomic states.\\
Since some light manipulation effects require three-level atomic systems, we will focus on these systems in this paper. Sec.\ref{thr} will introduce three-level systems to establish a strong foundation. Understanding their interaction with light is crucial for harnessing atomic interference in future technologies. Additionally, to gain a more precise understanding of atomic interference, we will review Raman processes in Sec.\ref{ram}.
\subsection{\label{thr}Three-level systems}
\begin{figure}[h!]
	\centering
	\subfigure[${\Xi}$-type ]
	{
		\includegraphics[width=0.1\textwidth]{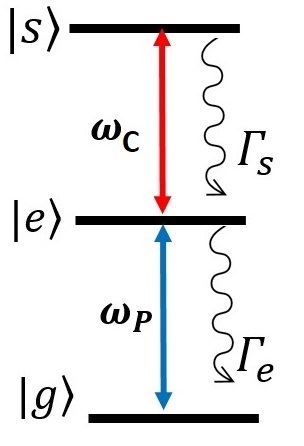}
		\label{ladd}
	}
	\subfigure[V- type]
	{
		\includegraphics[width=0.17\textwidth]{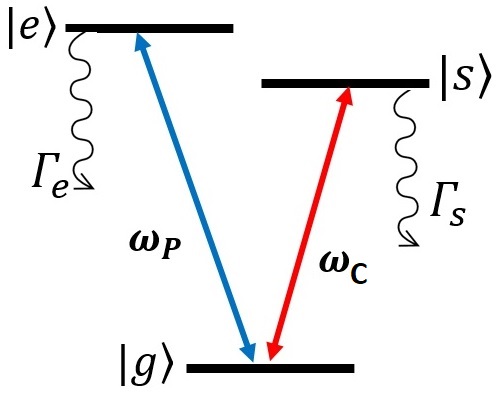}
		\label{tls2}
	}
	\subfigure[$\Lambda$- type]
	{
		\includegraphics[width=0.16\textwidth]{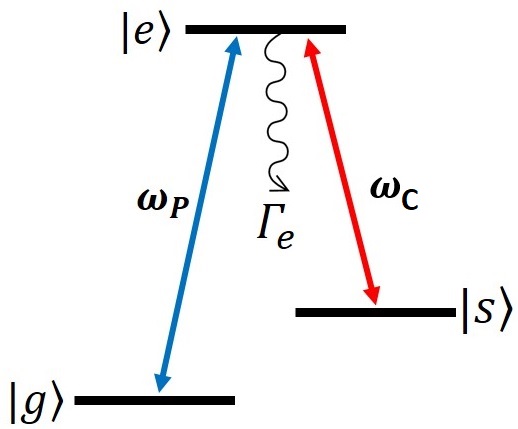}
		\label{tls1}
	}
	\caption{The three possible configurations of the energy levels of a three-level atom occur when the driven fields are in the resonance state. In these configurations, the spontaneous emission of each level is denoted by the symbol $\Gamma$. 
	}
	\label{TLS}
\end{figure}

\begin{figure}[h!]
	\centering
	\subfigure[Equal single-photon detuning for the driven fields]
	{
		\includegraphics[width=0.29\textwidth]{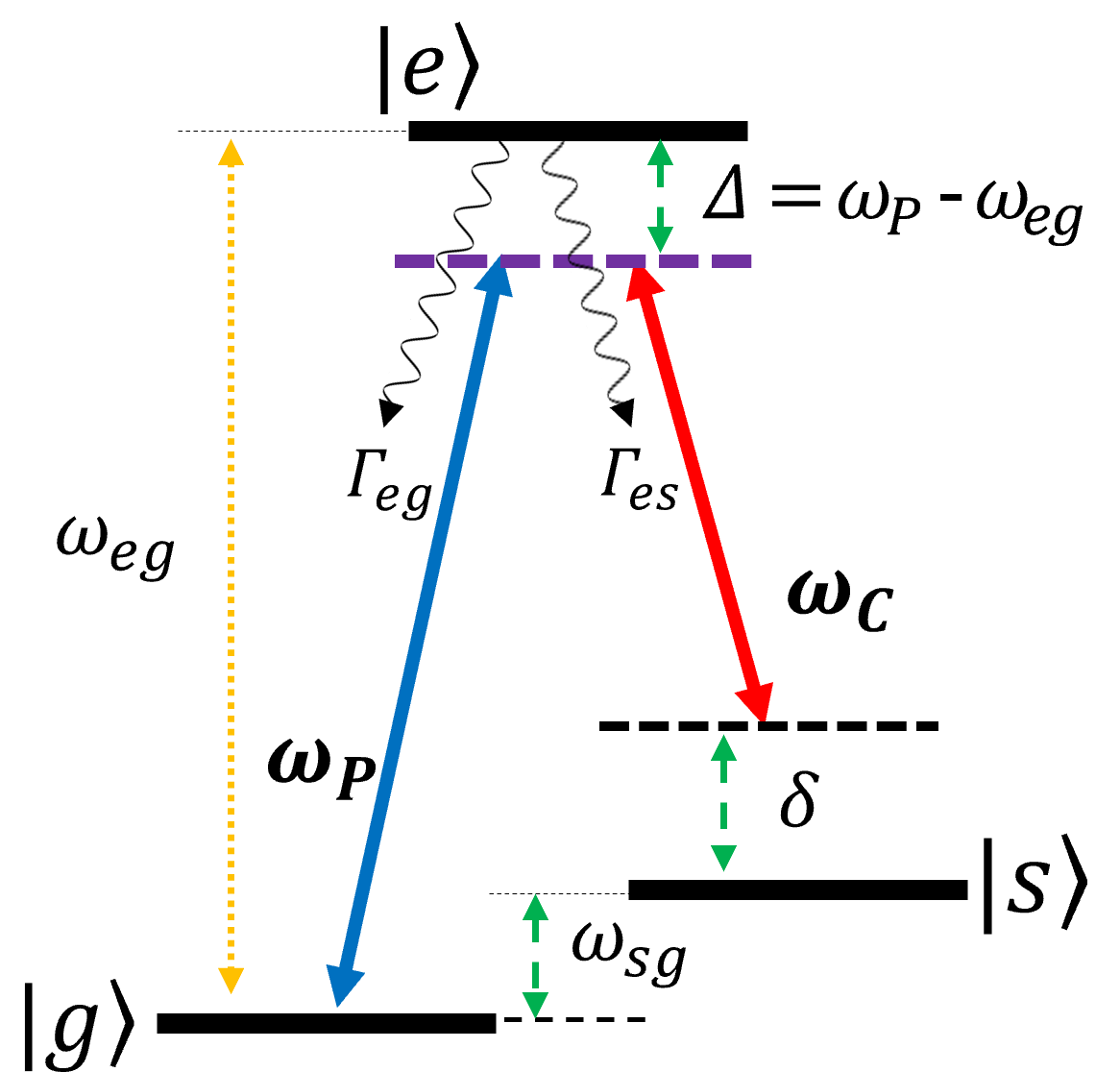}
		\label{nores}
	}
	\subfigure[Different single-photon detunings for the driven fields.]
	{
		\includegraphics[width=0.37\textwidth]{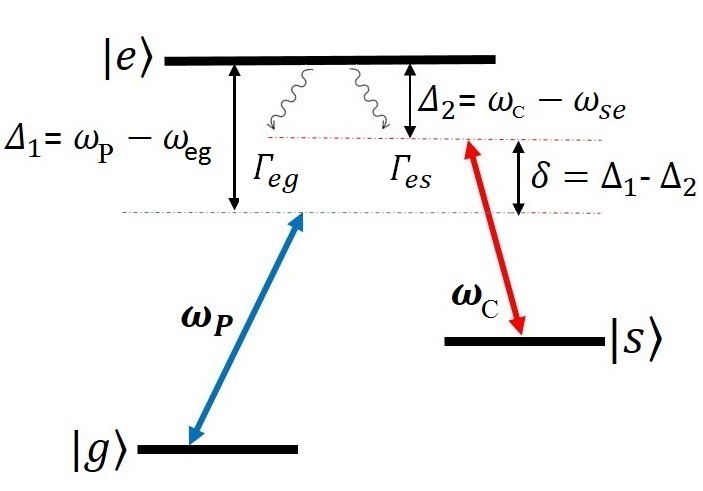}
		\label{nores2}
	}
	\caption{The $\Lambda$-type atom in two configurations with different photon detunings. In this figure, $ \Gamma_{ij} $ represents the decay rate from state $ \ket{i} $ to state $ \ket{j} $.}
\end{figure}

Understanding how external fields induce quantum interference in multi-level atomic systems is crucial. This phenomenon provides a powerful tool for manipulating the optical properties of these systems, leading to fascinating applications like EIT and STIRAP. Three common configurations exist for three-level systems: ladder, V-type, and $\Lambda$-type.\\
Three-level systems share some similarities and differences with their two-level counterparts. When exposed to constant radiation, both systems can exhibit Rabi oscillations in their energy level populations. However, the additional energy level in three-level systems introduces a greater degree of freedom, enabling a wider variety of controllable excitations. Refs.[\onlinecite{shore77,rad82,yoo85}] provide in-depth discussions on these comparisons.\\
Exploring the simplest form of system excitation, we'll investigate how two laser fields (denoted "P" for the pump field and "C" for the coupling/Stokes field) can induce a Raman transition. The configuration of the three-level system plays a crucial role in this process. As illustrated in Fig.\ref{TLS}, these configurations can be categorized into three distinct cases, as discussed in detail by [\onlinecite{shore}].
These three common configurations are defined as follows\\
\textbf{Laddar($\Xi$)-type atomes:}
According to Fig.\ref{ladd}, we observe that in this configuration, the dipole transitions $ \ket{g} \leftrightarrow \ket{e}$ and $\ket{e} \leftrightarrow \ket{s}$ are allowed, while the directly transition $\ket{g} \leftrightarrow \ket{s}$ is forbidden. This means that the states $\ket{g}$ and $\ket{s}$ have the same parity. In other words, the quantum angular momentum  number $l$ is either even or odd for both states. Therefore, according to the selection rule, these transitions will not be possible\cite{peter}.
Clearly, when an atom with such an energy level is excited to the $\ket{s}$ level, it will spontaneously emit and decay to the ground state through a cascade process as follows
\begin{enumerate}
	\item $\ket{s} \rightarrow \ket{e}$
	\item $\ket{e} \rightarrow \ket{g}$
\end{enumerate}
The frequency of the photon produced in each of the above steps is very close to the transition frequency $\ket{s} \rightarrow \ket{g}$. An important example of this situation is the cascade radiation $4p^2\ ^1S_0 \rightarrow 4s\ 4p\ ^1P_1 \rightarrow 4s^2\ ^1S_0$ in calcium atoms, which was used in the experiments of Alain Aspect to generate entangled photon pairs (polarization entangled) for experimental tests of Bell's inequalities\cite{peter}.\\
\textbf{V-type atomes:} 
 The V-type configuration demonstrates notable quantum-beat interference patterns resulting from the interaction between two excitation pathways connected to a single quantum state\cite{VITANOV200155}. In this configuration, as shown in Fig.\ref{tls2}, the atom has a stable ground state denoted by ${\ket{g}}$ and two distinct excited states ${\ket{e}}$ and ${\ket{s}}$. Both excited states are connected to the ground state via allowed dipole transitions, but there is no direct connection between them. Therefore, if the atom is excited to either state ${\ket{s}}$ or ${\ket{e}}$, it will decay to the ground state ${\ket{g}}$. During this process, a single photon is spontaneously emitted with a frequency close to the resonant frequency of the corresponding transition. One interesting application of this model is related to experiments involving quantum jumps with single trapped ions. These ions are suitable candidates for the implementation of qubits in the framework of trapped-ion quantum computers\cite{peter}.\\
\textbf{Lambda($\Lambda$)-type atomes:}
In Fig.\ref{tls1}, a $\Lambda$-type atom is depicted with two ground states, $\ket{g}$ and $\ket{s}$, and one excited state, $\ket{e}$. The electron can decay from the excited state $\ket{e}$ to either of the ground states. The ground state $\ket{g}$ is stable, while $\ket{s}$ is metastable, meaning it has a shorter lifetime than $\ket{g}$ but is still longer-lived than excited states.
This configuration is often referred to as the Raman configuration because by driving fields with frequencies $\omega_{C}$ and $\omega_{P}$ to the atomic transitions $\ket{g} \leftrightarrow \ket{e}$ and $\ket{e} \leftrightarrow \ket{s}$, respectively, we can induce a two-photon transition directly between $\ket{g}$ and $\ket{s}$, known as the Raman transition.\\
Unlike Fig.\ref{TLS}, where the fields are in resonance with the atom's transitions, we can also explore situations where the fields have off-resonant frequencies. We adjust the laser frequency $\omega_P$ so that the energy difference between the energy levels $(E_{e} - E_{g})/{\hbar}$ is separated by an amount $\Delta=\omega_P - \omega_{eg}$, which is called the \textbf{single-photon detuning}, as shown in Fig.\ref{nores}. It is worth noting that $\omega_{eg} = \omega_e -\omega_g$ represents the transition frequency for the levels $\ket{g}$ and $\ket{e}$.
In fact, the transitions $\ket{e} \leftrightarrow \ket{s}$ and $\ket{g} \rightarrow \ket{e}$ for frequencies $\omega_C$ and $\omega_P$ are separated by a $\Delta$. 
Therefore, an atom in the level $\ket{g}$ can absorb a photon with frequency $\omega_P $ and, due to stimulated emission caused by the other laser with frequency $\omega_C$ , it can make a transition to state $\ket{s}$ through a process.
The quantity $\delta = (\omega_P -\omega_C) - (\omega_{es} - \omega_{eg})$, called the \textbf{two-photon detuning}, allows us to determine the frequency difference between the two laser beams and the energy difference of the ground states ($(E_s - E_g)/{\hbar}$).\\
To study this in more detail, we consider a semiclassical interaction, using the Hamiltonian of a three-level \(\Lambda\)-type atom , as shown in Fig.\ref{nores} and a classical field. While we perform calculations for this specific configuration, the Hamiltonian for other states can be derived similarly. We further generalize by assuming each field has its own single-photon detuning, as illustrated in Fig.\ref{nores2}. Applying a classical field to the system yields the following Hamiltonian\\
\begin{align}
	\hat{H}^{\Lambda}=\hat{H}^A + \hat{H}^{AF}(t) 
	\label{e14}
\end{align}
In this expression, $\hat{H}^A$ represents the Hamiltonian of the atom, and $\hat{H}^{AF}(t)$ is the Hamiltonian of the atom-field interaction.  The value of each of these Hamiltonians for the atom is
\begin{equation}
	\hat{H}^A =\hbar\Big(\omega_g\hat{ \sigma}_{gg} + \omega_e\hat{ \sigma}_{ee} + \omega_s\hat{ \sigma}_{ss} \Big)
	\label{e15}
\end{equation}
Where $\sigma_{ij}=\dyad{i}{j}$ and $(i,j=g,e,s)$ represent the atomic transition operators. If the total classical field applied to the system is $\mathbf{E}=\mathbf{E}_P + \mathbf{E}_C$, for the interaction part of the Hamiltonian, we can write
\begin{align}
	\hat{H}^{AF}(t)&=-\hspace*{-.35cm}\sum_{i,j=g,e,s}\dyad{i} \boldsymbol{\mu} \boldsymbol{.} \mathbf{E}\dyad{j}
	\label{e16}
\end{align}
Here, $\boldsymbol{\mu}$ represents the dipole moment of the atom.\\

To simplify Eq.\eqref{e16}, we need to consider two points. First, since the integral $\ev{\boldsymbol{\mu}}{i}$ is zero, all diagonal terms in the above equation become zero. Second, only allowed dipole transitions should be considered in the off-diagonal terms. For the lambda configuration, the dipole transition $\ket{g} \leftrightarrow \ket{s}$ is not allowed. 
As a result, we obtain for Eq.\eqref{e16}
\begin{align}
	\hat{H}^{AF}(t)=&-\Big(\mel{g}{ \boldsymbol{\mu.E} }{e}\hat{ \sigma}_{ge} +\mel{e}{ \boldsymbol{\mu.E} }{g}\hat{ \sigma}_{eg}\nonumber \\ 
	& +\mel{e}{ \boldsymbol{\mu.E} }{s}\hat{ \sigma}_{es}  + \mel{s}{ \boldsymbol{\mu.E} }{e}\hat{ \sigma}_{se}\Big)
	\label{e19}
\end{align}
We note that each driving field targets a specific atomic transition. For example, \(E_P\) drives the transition between the ground state \( \ket{g} \) and the first excited state \(\ket{e}\).
Next, we assume the driven fields take the form \(E_l = \varepsilon_l (e^{i\omega_l t} + e^{-i\omega_l t})\) for \(l = P, C\), where \(\varepsilon_l\) represents the amplitude of the \(l\)-th field and \(\omega_l\) is its frequency.

We introduce the Rabi frequencies \(\Omega_P\) and \(\Omega_C\). Mathematically, they are expressed as \( \Omega_P={\mel{g}{\boldsymbol{\mu.\varepsilon_p}}{e}}/{\hbar} \) or \(\Omega_P =  {\mu_{ge} \varepsilon_P}/{\hbar}\) and \(\Omega_C = {\langle s \vert \boldsymbol{\mu} \cdot \boldsymbol{\varepsilon}_C \vert e \rangle}/{\hbar} = {\mu_{se} \varepsilon_C}/{\hbar}\), where \(\mu_{ge}\) and \(\mu_{se}\) are the dipole moments for the ground-to-excited state and excited-to-\(\vert s \rangle\) state transitions, respectively.\\

Assuming the conditions of the Rotating Wave Approximation (RWA) are met, we can express the system's Hamiltonian, as given in Eq.\eqref{e14}, within the Schrödinger picture. However, for a simpler description of the dynamics and to eliminate time dependence from the Hamiltonian, we transform the system into the interaction picture
\begin{align}
	\hat{H}^{\Lambda}_{I}=-\hbar[\Delta_1\hat{ \sigma}_{ee} + \delta\sigma_{ss}]-\hbar[\Omega_{P}\sigma_{eg} + \Omega_{C}\sigma_{es} + h.c.]
	\label{e28}
\end{align}
In which the subscript "$I$" refers to the interaction picture. Also
\begin{subequations}
	\begin{align}
		&\delta=(\omega_P - \omega_C)+\omega_{es}-\omega_{eg}=\Delta_1-\Delta_2\\[5pt]
		&\Delta_1=\omega_P - \omega_{eg} \  \  \  , \  \  \  \Delta_2=\omega_C - \omega_{se}
	\end{align}
	\label{e27}
\end{subequations}
It is important to note that the Hamiltonians expressed in Eq.\eqref{e28} were derived under the assumption that quantities such as amplitude, field strength, dipole moment, etc., remain constant with respect to time. However, if these parameters change—even while remaining within a rotating frame—the Hamiltonian will exhibit time dependence. \\
An example of this situation can be observed in phenomena like STIRAP, in which the driving fields have time-dependent amplitudes.\\
If we repeat the above process for ladder and V-type three-level atoms, we arrive at the following results\cite{peter}
\begin{subequations}
\begin{align}
	\hat{H}^{\Xi}_{I}&=-\hbar[\left(\Delta_1\hat{ \sigma}_{ee} +( \Delta_1 + \Delta_2)\sigma_{ss} \right) + \left(\Omega_P\hat{ \sigma}_{eg} + \Omega_{C}\sigma_{se} + h.c.\right)]\\
	\hat{H}^{V}_{I}&=-\hbar[\left(\Delta_1\hat{ \sigma}_{ee} + \Delta_2\hat{ \sigma}_{ss} \right) + \left(\Omega_P\hat{ \sigma}_{eg} + \Omega_{C}\sigma_{sg} + h.c.\right)]
\end{align}
\label{ee29}
\end{subequations}
Understanding the system's Hamiltonian simplifies the examination of its associated wavefunction. When we are in the dispersive regime, and the conditions \( \Delta \gg \Gamma_e, \Omega_{P,C} \) (where \( \Gamma_e \) denotes the spontaneous emission rate from the intermediate level) are met, the criterion for the adiabatic elimination of this intermediate level is fulfilled.\\
The system's wavefunction can be written as follows
\begin{equation}
	\ket{\psi}=c_g \ket{g}+c_e\ket{e}+c_s \ket{s}
	\label{sdd22}
\end{equation}
The \( c_i \) coefficients (where \( i = g, e, s \)) represent the probability amplitudes of each state. If we use the Schrödinger equation to solve for the probability amplitudes in Eq.\eqref{sdd22} for a $\Lambda$-type atom, taking into account necessary approximations (such as the RWA, etc.) in the dispersive regime, we obtain the expression for \( c_e \) as follows\cite{peter}.
\begin{equation}
	c_e=\dfrac{\Omega_{P}c_g+\Omega_C c_s}{\Delta+ i \Gamma_e}
\end{equation}
Since \( \Delta \) is much larger than the Rabi frequencies and the spontaneous emission rate of the intermediate level, we can quickly conclude that \( c_e \approx 0 \) and the \( \ket{e} \) level can be effectively eliminated. Under these conditions, the system will behave like a two-level system \cite{peter, qea}.\\
 Similar calculations can be used for ladder and V-type atoms.
Therefore, by carefully selecting specific conditions, it is often possible to simplify the analysis of a three-level system and treat it as if it were a two-level system. This simplification enables us to focus our attention on the most essential aspects of the system and facilitates a deeper understanding of its behavior and properties.
\subsection{\label{ram}Raman processes}
Interaction between external fields and atoms can induce various processes, including transitions between atomic energy levels. This phenomenon, known as the\textbf{ Raman transition}, is fundamental to understanding quantum interference in atomic systems.\\
Understanding the role of Raman transitions in quantum interference requires exploring Bragg scattering, as stimulated Raman transition represents the simplest form ("first order") of Bragg scattering \cite{steck}. Bragg scattering explains the effects of electromagnetic wave reflection from periodic structures with spacings in the wavelength range. It describes how the superposition of wavefronts scattered by the lattice planes establishes a precise relationship between wavelength and scattering angle. Such periodic structures can manifest as crystals or alternating optical lattices, where the wave is scattered after interacting with the atoms in the lattice.\\

\begin{figure}[h!]
	\centering
	\includegraphics[scale=0.11]{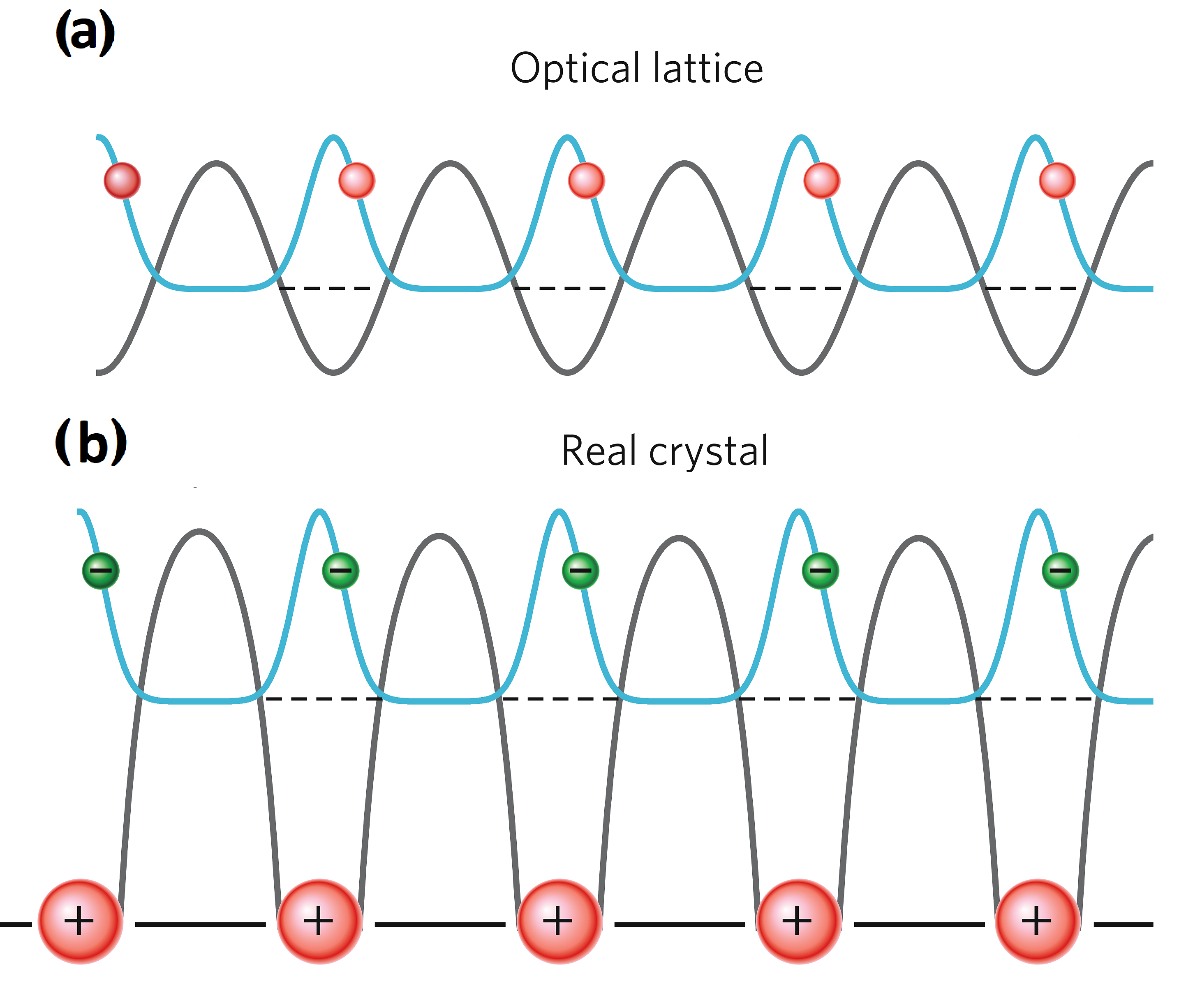}
	\caption
	{\textbf{Comparison of crystal and optical lattice:}
		\textbf{(a)} In an optical lattice, atoms are trapped in a sinusoidal potential well (gray) created by a standing wave laser beam. \textbf{(b)} The wave functions of the atoms (blue) correspond to those of the valence electrons in a real crystal. Here, the periodic potential is created by the attractive electrostatic force between the electrons (-) and the ions (+) comprising the crystal. Figure adapted from [\onlinecite{Greiner2008}].
	}
	\label{fff03}
\end{figure}

An optical lattice, depicted in Fig.\ref{fff03}, is a periodic optical structure created by the superposition of laser beams, resulting in the production of a standing wave. This ultimately produces a periodic potential that affects atoms. In these standing laser fields, atoms are cooled by laser cooling and then trapped in a minimum potential with a spatial extent smaller than the optical lattice wavelength\cite{Lattices,Yehuda}. However, it should be noted that we are in a regime where the energy of the atoms is sufficiently high that they are not confined to the potential wells arising from the optical lattice.
The analogy of Bragg scattering extends to atomic beams as well. By replacing the electromagnetic wave with an atomic beam and the crystal with an optical lattice, we can still observe this phenomenon.\\

 Bragg scattering from a standing wave offers a valuable technique for coherently splitting an atomic beam into two distinct beams. This technique finds applications in creating atomic beam splitters and mirrors. \\
 
 When an atomic beam encounters an optical lattice at an angle $\theta$, Bragg scattering occurs only when $\theta$ is equal to one of the $n$-th order Bragg scattering angles $\theta_n$ \cite{Bragg}.
\begin{equation}
	\lambda_L \sin\theta_n = n \lambda_{dB}
\end{equation}
Where $\lambda_{dB}$ is the de Broglie wavelength of the atom, and $\lambda_{L}$ is the laser wavelength. In general, nth-order Bragg scattering is a $2n-$photon transition. Then, as mentioned earlier, first-order Bragg scattering ($n=1$) corresponds to a two-photon stimulated Raman transition.\\
But how does a matter wave interact with a standing light wave? This is only possible if the atom has an internal transition that allows it to scatter photons from the light beam. Since each absorption and emission process transfers recoil momentum to the atom, we can understand Bragg scattering as a Raman scattering process. A photon from the laser beams of the optical lattice is created, absorbed by the atom, and then re-emitted \cite{Courteille}. Therefore, to understand the stimulated Raman transition, we need to investigate the Raman process.\\
When an electromagnetic field with frequency $ \omega_0 $ interacts with an atom or molecule, after the interaction, the outgoing photon can be scattered elastically or inelastically. \\

\begin{figure}[h!]
	\centering
	\includegraphics[scale=0.37]{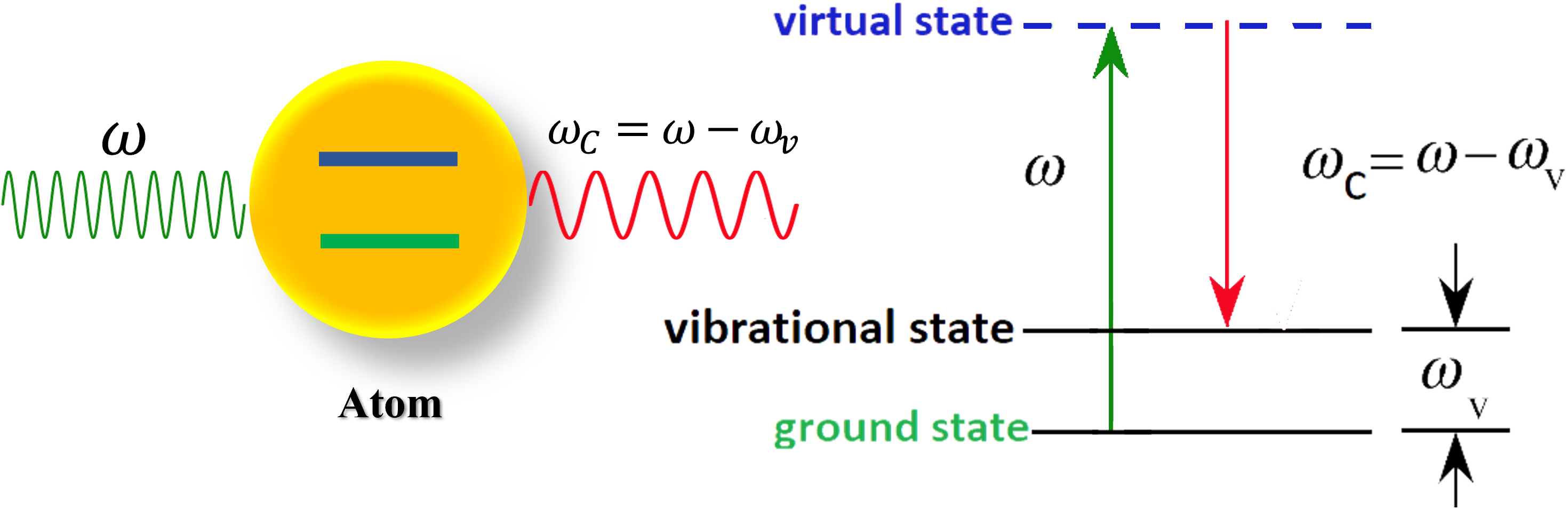}
	\caption{Stimulated Raman Scattering occurs when a photon with frequency $\omega$ interacts with a molecule or atom. In this process, the photon is absorbed, transfers its energy to the molecule. Consequently, the molecule becomes excited, then transfers to a higher vibrational state and  its internal energy increases by $\hbar \omega_v$. Simultaneously, a new photon with a lower frequency, $\omega_C = \omega - \omega_v$, is emitted, known as the Stokes-shifted frequency. The difference in frequency, $\omega_v$, represents the specific amount of energy retained by the molecule for its vibrational excitation.}
	\label{f1}
\end{figure}

\begin{figure}[h!]
	\centering
	\subfigure[Stokes and anti-Stokes scattering.]
	{
		\includegraphics[width=0.32\textwidth]{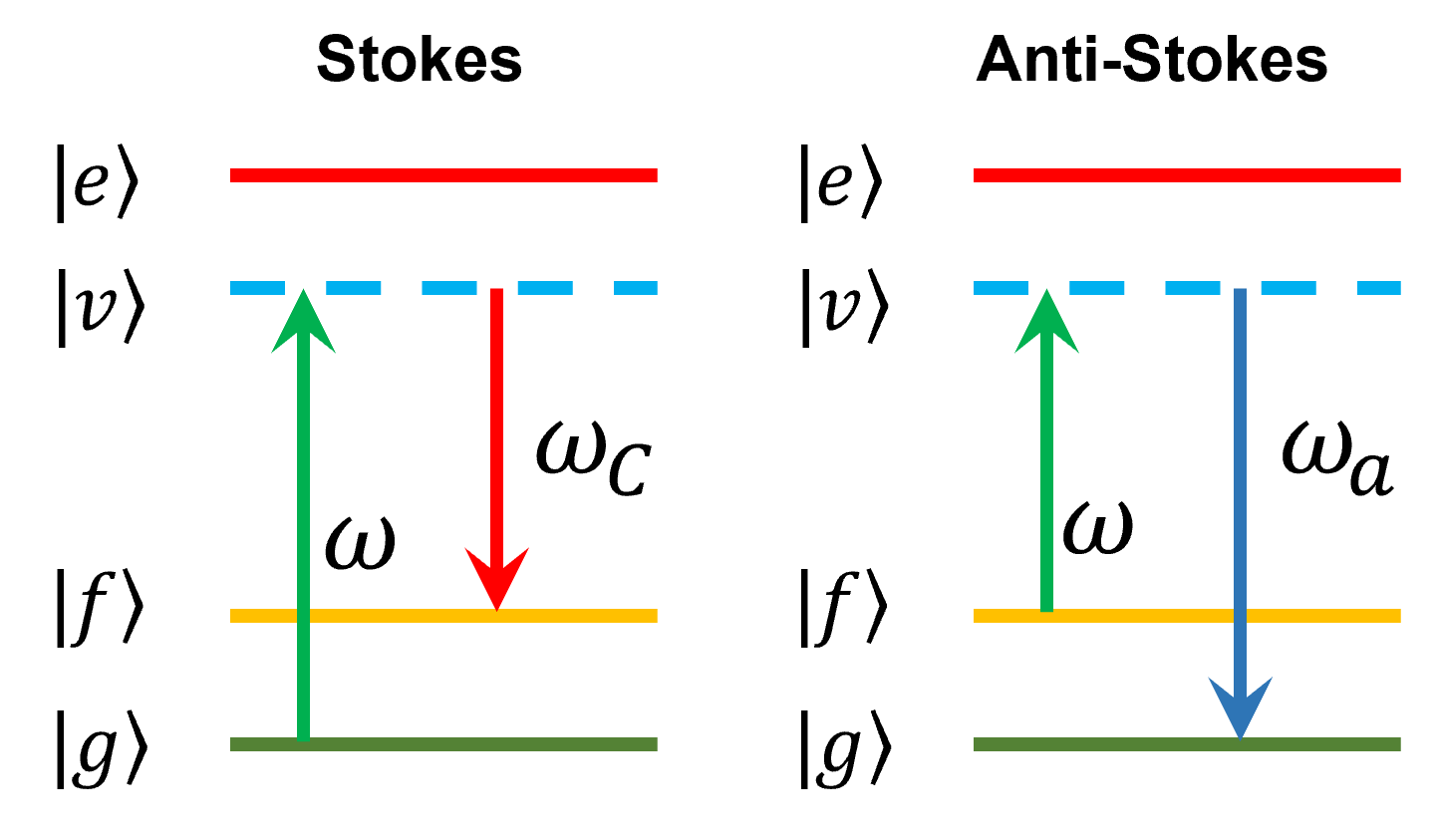}
		\label{ff1}
	}
	\vspace*{0.6 cm}
	\subfigure[Different diagrams of Raman scattering.]
	{
		\includegraphics[width=0.48\textwidth]{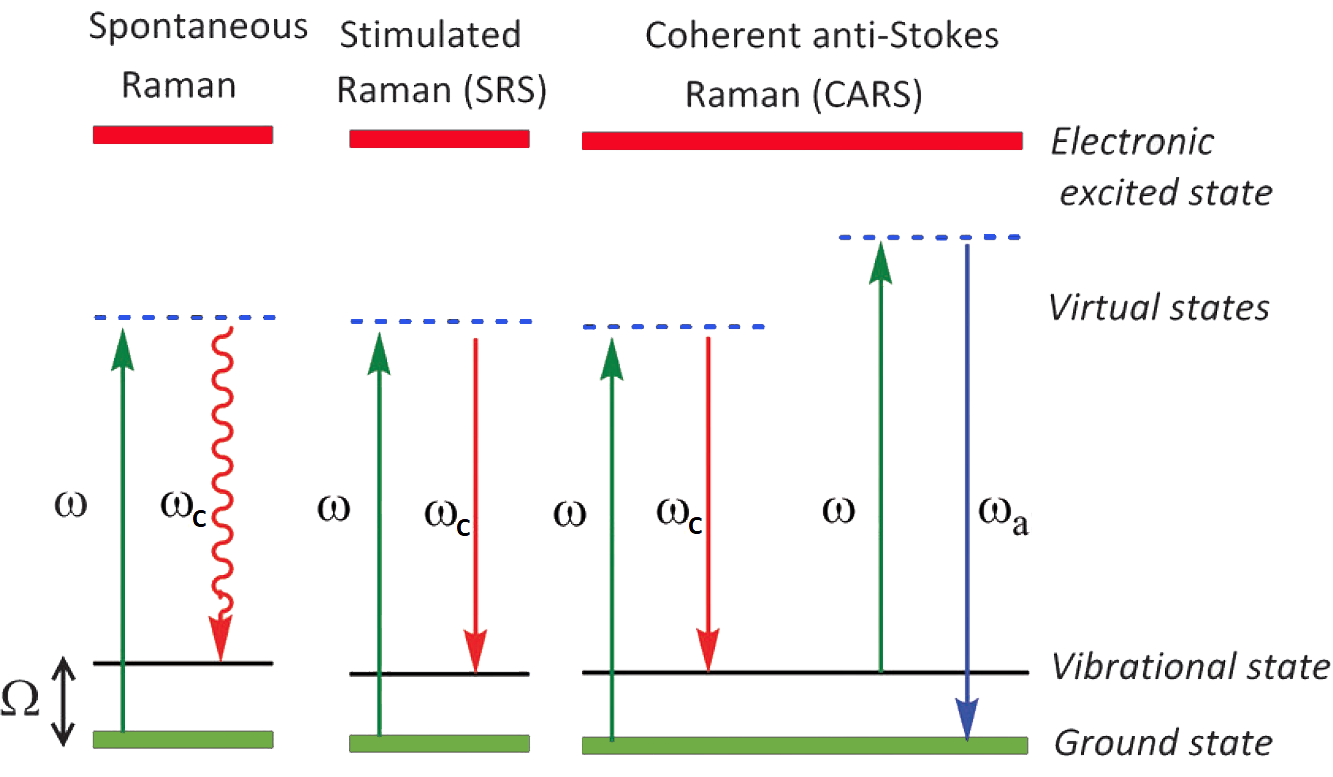}
		\label{fig}
	}
	\caption{We illustrate Stokes and anti-Stokes scattering in a simplified form in (a), while in (b), we explore different cases of this scattering phenomenon. Raman scattering    describes the interaction between light and a sample, resulting in the emission of light with a different frequency. 
		In \textbf{Spontaneous Raman scattering}, a coherent pump field (with frequency $\omega$) is applied to the sample. As a result of this radiation, a red-shifted frequency $\omega_{\text{C}}$, arising from inelastic scattering, is spontaneously emitted. 
		However, in \textbf{Stokes Raman scattering}, two coherent beams with frequencies $\omega$ and $\omega_{\text{C}}$ are required to be applied to the sample. If the difference between these two frequencies ($\Delta\omega = \omega - \omega_{\text{C}}$) equals the system's transition frequency ($\Omega$), stimulated emission occurs.	The most complex type is \textbf{Anti-Stokes Raman scattering}, which involves four light beams and a complex process to induce emission at a higher frequency (anti-Stokes) by manipulating the molecule's energy states. Figure (b) is adapted from[ [\onlinecite{che}].}
\end{figure}

Elastic scattering occurs when the frequency of the output field matches that of the driving field. This is known as Rayleigh scattering. However, if the scattering is inelastic, the frequency of the output field is different from the frequency of the applied field. This is known as Raman scattering. In Raman scattering, the frequency of the scattered photon will change by the system's transition frequency depending on the type of scattering\cite{che,nan}.
\begin{figure}[h!]
	\centering
	\includegraphics[scale=0.42]{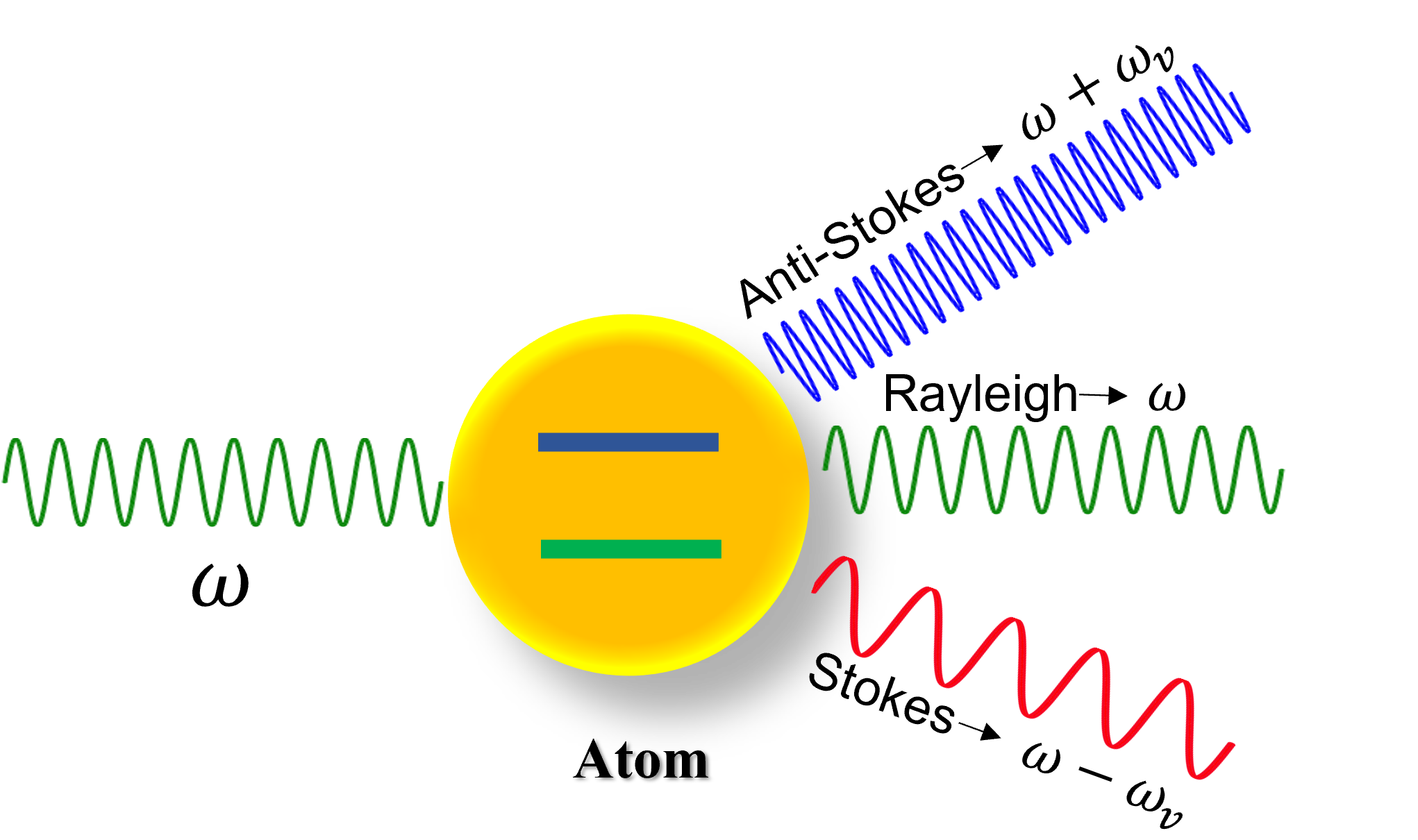}
	\caption{The interaction of a field with frequency $\omega$ and a vibrating atom with transition frequency $\omega_v$ can result in Rayleigh, Stokes, or Anti-Stokes scattering in the output.}
	\label{f2}
\end{figure}
In this interaction, the incident photon excites one of the electrons to a virtual state (in quantum mechanics, a virtual state is a short-lived state that cannot be observed) because it does not have enough energy to reach a real state. After the electron is transferred to the virtual excited state, a photon is immediately emitted and the electron will be in a lower state. \\
During this process, energy is transferred to the atom, causing the electron to transition to a higher vibrational state. According to Fig.\ref{f1}, we see that the emitted photon will have less energy than the incident photon (i.e., there is a redshift). This phenomenon is called \textbf{stimulated Raman scattering} or Stokes scattering, which is a very useful tool for manipulating cold atoms and ions \cite{James}.
According to Fig.\ref{ff1}, Stokes Raman scattering involves a transition from the ground state $\ket{g}$ to the final state $\ket{f}$ via a virtual intermediate state ($\ket{v}$) associated with the excited state $\ket{e}$. Anti-Stokes Raman scattering, on the other hand, requires a transition from state $\ket{f}$ to level $\ket{g}$ with $\ket{e}$  as the intermediate state\cite{boyd}.
In other words, if the interaction between the incident photon and the system causes the output to have a higher energy photon than the incident photon (i.e., a blue shift), then we are dealing with \textbf{anti-Stokes Raman scattering}. \\
In Raman scattering, if only one field is used to excite the system, then the electron, after being excited, spontaneously emits a photon with a specific frequency and decays to a lower level. This type of Raman scattering is called\textbf{ Spontaneous Raman scattering}. However, if, as shown in Fig.\ref{fig}, two electromagnetic fields are driven  to the system, then we will observe the phenomenon of \textbf{Stimulated Raman scattering}\cite{che,nan}.\\

So far, we have discussed Raman scattering. However, our attention is now directed towards stimulated Raman transition, which is a subset of Raman scattering. The overall process in stimulated Raman transition is similar to that of stimulated Raman scattering.\\

Stimulated Raman transition occurs when two fields are used to induce a transition, during which one photon is virtually absorbed and another photon is virtually emitted via an excited state.\\
It is through this virtual absorption and emission that the frequencies of the real lasers are adjusted so that their frequency and the transition frequency have a significant difference. Therefore, the probability of a real photon being absorbed and spontaneously emitted is very low. Because both absorption and emission occur through induced processes by the fields, this phenomenon is sometimes also referred to as stimulated Raman scattering. \\
As we have observed, in the process of stimulated Raman scattering, two photons are involved, making this transition an example of a two-photon transition process as well\cite{qea}. \\

To better understand the concept, let’s consider a standing wave formed by two waves with equal amplitudes but opposite propagation directions.
\begin{align}
	\textbf{E}(x,t)=\hat{z} 2 E_0\cos(kx) \cos(\omega t)
\end{align}
Considering the waveform of the final wave and utilizing the definition of the Rabi frequency, we can readily express $\Omega(x)=\Omega_0\cos(kx)$. Additionally, we know that the effective optical potential is equal to\cite{steck}
\begin{equation}
	V_{eff}(x)=\dfrac{\hbar \abs{\Omega(x)}^2}{4\Delta}
\end{equation}
Substituting the Rabi frequency and ignoring constant terms, we can write
\begin{equation}
	V_{eff}(x)=V_0\cos(2kx) =\dfrac{V_0}{2}(e^{-2ik x}+ e^{2ik x}) 
\end{equation}
Where $ V_0={\hbar \abs{\Omega_0}^2}/{8\Delta}$. With knowledge of the potential function, writing the Schrödinger equation for an atom of mass \( m \) in position representation is not difficult.
\begin{align}
	i\hbar \pdv{\ket{\psi}}{t}=\bigg[\dfrac{p^2}{2m}+\dfrac{V_0}{2}(e^{-2ik x}+ e^{2ik x})\bigg]\ket{\psi}
\end{align}
For a more detailed analysis, we express this equation in momentum representation.
\begin{equation}
	i\hbar \pdv{\phi(p)}{t}=\dfrac{p^2}{2m}\phi(p)+\dfrac{V_0}{2}\biggl[\phi(p-2\hbar k) + \phi(p+2\hbar k)  \bigg]
\end{equation}
In this context, $\phi(p)=\braket{p}{\psi}$. The evolution of the standing wave imposes a "ladder" structure on the momentum, so that if the initial momentum of the atom is $\ket{p}$, then at the end of the process the atom can only occupy states such as $\ket{p+2n\hbar k}$. In this relation, $n$ is an integer and its value for the stimulated Raman transition using a pair of photons is $n=1$. \\
It should be noted that the quantization of momentum has a clear interpretation in terms of the stimulated scattering of lattice photons. Specifically, if an atom absorbs a photon moving in one direction and then re-emits it in the opposite direction, then the atom will recoil, which results in a doubling of the change in the atom's momentum or $2\hbar k $ (assuming $\abs{k_1}=\abs{k_2}=k$). However, momentum transfer to atoms can be observed as a Raman transition between different states (for example, from $\ket{g, p}$ to $\ket{s, p+2\hbar k}$ see Fig.\ref{pp3})\cite{steck}. \\
In general, after the first laser pulse is applied, the atom absorbs a photon with frequency $ \omega_1 $ (or momentum $ \hbar \textbf{k}_1 $) and then re-emits it in a stimulated manner with frequency $ \omega_2 $ (or momentum $ \hbar \textbf{k}_2 $). Because the laser beams are applied in opposite directions (counter-propagate), the momentum vectors of the first ($\hbar \textbf{k}_1$) and the second ($\hbar \textbf{k}_2$) photons have opposite directions. Therefore, the momentum of the atom in the state $ \ket{s} $ must be equal to $ \textbf{p}+\hbar (\textbf{k}_1 -\textbf{k}_2) $, and the final state will be of the form $ \ket{s,\hbar \textbf{k}_{eff}} $, where $ \textbf{k}_{eff}=\textbf{k}_1 -\textbf{k}_2 $\cite{Garrido}. The utilization of these transitions becomes particularly significant when the driven electromagnetic fields induce coherent superposition in atoms, ultimately leading to phenomena such as atomic interference.
\subsection{Autler-Townes or  Stark AC effect}
We know that by driving an atom with a field (electric or electromagnetic), we will observe the splitting of its energy levels. If the applied field is of the $DC$ (Direct Current) type,  this splitting is called the $DC$ Stark effect. However, if the applied field is oscillating or of the $AC$ (Alternating Current) type (such as electromagnetic waves), it is known as the Autler-Townes effect or the Stark $AC$ effect.  This effect was discovered in 1955 by Stanley Autler and Charles Townes using microwave radiation, prior to the invention of the laser\cite{Cohen-Tannoudji1996}.\\

In a three-level system, the Autler-Townes effect can be observed when the system is driven by a strong electromagnetic field. For example, in a lambda-type three-level atom, this phenomenon can be investigated using various methods. One approach is to adiabatically eliminate the excited  $\ket{e}$ level and analyze the splitting of the two remaining levels.\\ Alternatively, one can simplify the problem by considering the non-adiabatic case and focusing on the splitting of the system with only the ground ( $\ket{g}$) and excited ( $\ket{e}$) levels (i.e., one side of the atom). In the Schrödinger picture, this is described by writing the wavefunction for the remaining part of the system.
\begin{align}
	\ket{\psi(t)}=A(t)\ket{g} + B(t)\ket{e} 
	\label{ee30}
\end{align}
If we apply the slowly varying amplitude approximation to the probability amplitudes, $A(t)=c_{1}(t)e^{-i\omega_g t}$ and $B(t)=c_{2}(t)e^{-i\omega_{e}t}$, these expressions are taken into account.
\begin{equation}
	\ket{\psi(t)}=c_{1}(t)e^{-i\omega_{g}t }\ket{g} +c_{2}(t)e^{-i\omega_{e}t }\ket{e}
	\label{e30}
\end{equation}
If \( \omega_{eg} = \omega_e - \omega_g \) represents the transition frequency between the  $\ket{g}$ and  $\ket{e}$ states, then by applying a field \( \omega_P \) to the system and solving the Schrödinger equation for the above state function  in the semiclassical manner, we obtain the following solutions for the coefficients \( c_{1}(t) \) and \( c_{2}(t) \)\cite{zub}
\begin{subequations}
	\begin{align}
		&c_2(t)=\Bigg[c_{2}(0) \left(\cos\dfrac{\Omega_{eff}t}{2} -\dfrac{i\Delta}{\Omega_{eff}}\sin\dfrac{\Omega_{eff}t}{2}\right)\nonumber\\
		&\ \ \ \ \qquad\quad+i c_1(0)\dfrac{\Omega }{\Omega_{eff}}\sin\dfrac{\Omega_{eff}t}{2} \Bigg]e^{-i\Delta t/2}\\[18pt]
		&c_1(t)=\Bigg[c_{1}(0) \left(\cos\dfrac{\Omega_{eff}t}{2} +\dfrac{i\Delta}{\Omega_{eff}}\sin\dfrac{\Omega_{eff}t}{2}\right)\nonumber\\
		&\ \ \ \ \qquad\quad+i c_2(0)\dfrac{\Omega }{\Omega_{eff}}\sin\dfrac{\Omega_{eff}t}{2} \Bigg]e^{i\Delta t/2}
	\end{align}
	\label{e31}
\end{subequations}
Where \( \Omega_{eff} = \sqrt{\Omega^2 + \Delta^2} \) is the generalized Rabi frequency, \( \Delta = \omega_P - \omega_{eg} \) is the single-photon detuning, and \( \Omega \) is the Rabi frequency.\\
Consider the following scenario, where the atom is initially in the ground state \( \ket{g} \), so it is clear that \( c_1(0) = 1 \) and \( c_2(0) = 0 \). Assuming the single-photon resonance condition (\( \Delta = 0 \)), we can then obtain the probability coefficients.
\begin{subequations}
	\begin{align}
		&c_1(t)=\dfrac{1}{2}\left( e^{i\Omega_{eff}t/2} + e^{-i\Omega_{eff}t/2}\right)\\[8pt]
		&c_2(t)=\dfrac{1}{2}\left( e^{i\Omega_{eff}t/2} - e^{-i\Omega_{eff}t/2}\right)
	\end{align}
	\label{e32}
\end{subequations}
Based on the calculated probability coefficients, the wavefunction in Eq.\eqref{e30} can be rewritten.
\begin{align}
	\ket{\psi(t)} =\dfrac{1}{2}&\Big( e^{-i(\omega_{g}+\Omega/2)t }\ket{g} + e^{-i(\omega_{g}-\Omega/2)t}\ket{g}  \nonumber\\[10pt]
	&+ e^{-i(\omega_{e}+\Omega/2)t }\ket{e} +  e^{-i(\omega_{e}-\Omega/2)t}\ket{e} \Big)
	\label{e33}
\end{align}
By carefully examining the above relation, we observe that four oscillatory frequencies and consequently, four different energy values have now emerged for the system.\\
Alternatively, we could have achieved the same result by examining the Hamiltonian of the system and calculating the eigenvalues, which would undoubtedly yield these four expressions as well.
\begin{figure}[h!]
	\centering
	\includegraphics[scale=0.5]{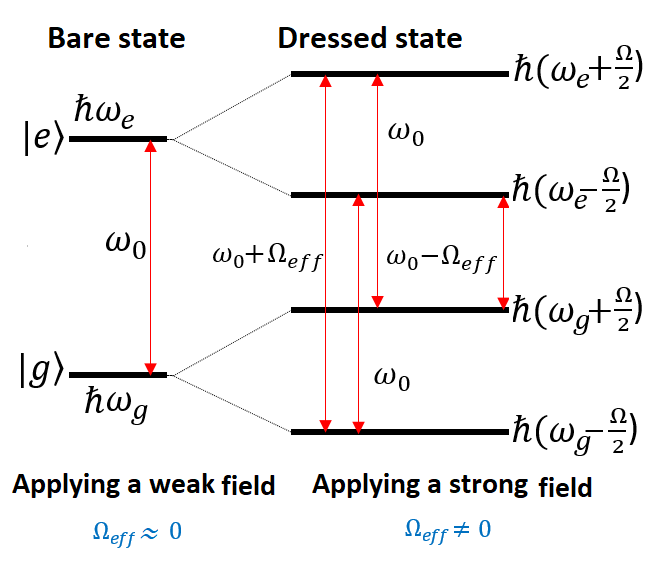}
	\caption{By applying an electromagnetic field in the single-photon resonance condition (\( \Delta = 0 \)), we observe the splitting of energy levels, which leads to the Autler-Townes effect.}
	\label{ss}
\end{figure}
This result indicates that the energy diagram of the system, as depicted in Fig.\ref{ss}, should split into four energy levels. This splitting is commonly called  the Autler-Townes effect.
\begin{align}
	&\omega^{\pm}_{g}=(\omega_{g}\pm \Omega/2) \\[10pt]
	& \omega^{\pm}_{e}=(\omega_{e}\pm \Omega/2) 
	\label{e34}
\end{align}
When a strong electromagnetic field is applied to an atom, the original atomic states, known as \textit{bare states} (e.g., \( \ket{g} \) and \( \ket{e} \)), evolve into what are called \textit{dressed states}. The states that emerge after applying the strong field to the atom (as shown on the right-hand side of Fig.\ref{ss}) are usually denoted by \( \ket{+} \) and \( \ket{-} \). These dressed states result from the interaction between the atom and the classical external field, leading to a new set of quantum states that reflect the combined properties of both the atom and the field.\\
Each bare state, such as \( \ket{g} \) or \( \ket{e} \), can be written as a superposition of these dressed states. For example, by neglecting the photon states for the  $\ket{e}$ level, we can express the excited state \( \ket{e} \) as a linear combination of the dressed states \( \ket{+} \) and \( \ket{-} \)
\begin{equation}
	\ket{e} = a\ket{+} + b\ket{-}
	\label{altw}
\end{equation}
where \( a \) and \( b \) are the probability amplitudes. When a strong field is applied to the system, it causes level splitting, as shown in Fig.\ref{ss}. Working directly with the bare states can often be challenging. For this reason, special new states are defined as eigenstates of the time-independent Hamiltonian derived from the final Hamiltonian, which includes interactions. These eigenstates, obtained by diagonalizing the final Hamiltonian, are called dressed states.\\
Dressed states are used because\cite{Barnett}
\begin{enumerate}
	\item These states and their energies are well-known.
	\item The system dynamics become simpler since the final states are superpositions of the dressed states.
	\item The probability of being in any of the dressed states is time-independent.
	\item The final states are superpositions of the uncoupled bare states.
\end{enumerate}
In fact, it is common to describe the atom-field interaction in terms of dressed states when a strong laser field drives the atom \cite{CARDOSO}. This approach reveals that a strong field causes each bare state to evolve into a superposition of dressed states, as shown in Eq.\eqref{altw}, leading to\textit{ interference effects} and level splitting that are more easily understood through the dressed state framework.\\
The Autler-Townes effect leads to various phenomena, including electromagnetically induced transparency and Mollow triplets.
\subsection{\label{CPT}Coherent Population Trapping}
Coherent Population Trapping (CPT), is an additional consequence of spontaneous emission observed by Alzetta et al in 1976 using a Lambda atom. In their experiment, the hyperfine levels of sodium played the role of the $\ket{g}$ and $\ket{s}$ levels. They showed that under conditions of two-photon resonance, the fluorescence from the state  $\ket{e}$  is strongly suppressed, and the population of the level $\ket{e}$ effectively reduces to zero.\\
The applications of CPT extend beyond its fundamental scientific interest, playing a crucial role in the development of ultra-stable and precise atomic clocks, enabling groundbreaking advancements in timekeeping. CPT also finds applications in the burgeoning field of quantum information processing, offering a method to manipulate and store quantum information within atomic systems.\\

CPT can be understood in the context of quantum interference. There are two contributions to the amplitude probability for the population of state \( \ket{e} \): one from state \( \ket{g} \) and the other from state \( \ket{s} \). When the experimental conditions are such that the amplitude probabilities of states \( \ket{g} \) and \( \ket{s} \) are equal in magnitude but opposite in sign, quantum interference becomes completely destructive, and the population of state \( \ket{e} \) becomes zero. At this point, we observe that the system is in a coherent superposition of states \( \ket{g} \) and \( \ket{s} \)\cite{Swain}.\\

Therefore, this phenomenon shows a remarkable demonstration of atomic interference effects, leading to transfer of atomic population to a superposition of decoupled states (dark states) \cite{Butts}. After the system transitions to the dark state, it can no longer absorb or emit radiation.
The critical requirement for observing CPT is the application of two coherent fields with time-independent amplitudes to the system, combined with exact two-photon resonance ($ \delta=0 $). Consequently, CPT involves stable population trapping in the coherent superposition of two ground states (e.g., states \( \ket{g} \) and \( \ket{s} \) for a lambda system). These ground states are coupled by coherent fields to an intermediate state (as depicted in Fig.\ref{nores})\cite{xu}.\\

For a deeper insight into CPT and dark states, we solve an eigenvalue equation based on Eq.\eqref{e28} for a lambda system. By imposing the two-photon resonance condition ($\delta=0$) and assuming all Rabi frequencies are real ($\Omega^{*}_j=\Omega_j$), we can express this equation as follows
\begin{align}
	\hat{H}_{\Lambda} &=-\hbar\left(\begin{matrix}0&\Omega_P&0\\\Omega_P&\Delta_1&\Omega_C\\0&\Omega_C& 0 \\\end{matrix}\right)
	\label{e35}
\end{align}
We seek the eigenstates of this Hamiltonian, which are the atomic states dressed by two fields \( \omega_{C} \) and \( \omega_{P} \). We consider the eigenvalue equation as \( \hat{H}\ket{\psi_n(t)}=\hbar\lambda_n\ket{\psi_n(t)} \). \\

Due to the oscillating nature of the external perturbation, which are the electromagnetic fields driven to the atom, the quantities \( \lambda_n \) and \( \hbar\lambda_n \) represent the eigenfrequencies and eigenvalues of the energy, respectively. Consider the system's wavefunction in the basis of the atomic diabatic or bare states (i.e., \( (\ket{g},\ket{e},\ket{s}) \)) as follows
\begin{equation}
	\ket{\psi_{Di}(t)}=c_g (t)\ket{g} + c_e (t)\ket{e} + c_s (t)\ket{s}
	\label{Diaba12}
\end{equation}
In the above equation, $\ket{\psi_{Di}(t)}$ represents the diabatic or unperturbed state of the system, and the $c_g, c_e, c_s$ are also called the diabatic probability coefficients.
We can readily obtain the eigenvalues.
\begin{equation}
	\lambda_0=0 ,\qquad \lambda_\pm=-\dfrac{\Delta}{2}\pm \bar{\Omega}
	\label{e36}
\end{equation}
Where \( \bar{\Omega}=\sqrt{\Omega_P^2 +\Omega_C^2 +(\Delta/2)^2} \), with \( \Omega_P \) and \( \Omega_C \) represent the Rabi frequencies of two driving fields interacting with the atom. These frequencies quantify the intensity of interaction between the fields and the atomic transitions. \( \Delta \) represents the single-photon detuning, indicating the energy difference between the photon and the atomic transition. When the detuning is zero (i.e., photon energy matches transition energy), the system is said to be in resonance, and the interaction between the field and the atom is strongest. Thus, the corresponding eigenstates are given by\cite{peter}.
\begin{subequations}
	\begin{align}
		&\lambda_0 \ :  \quad \ket{D}=\dfrac{1}{\Omega_{rms}}\Big(\Omega_C \ket{g}-\Omega_P \ket{s}\Big)  \label{e37-1}\\[10pt]
		&\lambda_\pm \ : \quad \ket{B_{\pm}}=\dfrac{1}{{N_{\pm}}}\Big(\Omega_P \ket{g}-\lambda_\pm \ket{e}+\Omega_C \ket{s}\Big)  \label{e37-2}
	\end{align}
	\label{e37}
\end{subequations}
Here, $\Omega_{rms}=\sqrt{\Omega_P^2 +\Omega_C^2} $ and $N_{\pm}=\sqrt{\Omega_{rms}^2 +\lambda_\pm ^2}$. The states $\ket{D}$ and $\ket{B_\pm}$ 
are called the \textbf{dark state} and the \textbf{bright states}, respectively. These adiabatic states are the eigenstates of the Hamiltonian given in Eq.\eqref{e35}.
Since the states $\ket{D}$ and $\ket{B_{\pm}}$ explicitly involve the fields, they are dressed states. Therefore, they form a suitable dressed basis, in which expansions are based\cite{Eberly_1995}.
\begin{equation}
	\ket{\varphi _{Ad}(t)}=a_-(t)\ket{B_-}+a_0(t)\ket{D}+a_+(t)\ket{B_+}
	\label{Adia}
\end{equation}
The state \( \ket{\phi_{Ad}(t)} \) represents the adiabatic state of the system, and \( a_i(t) \)s (where \( i = 0, \pm \)) are also referred to as the adiabatic probability coefficients. 
We obtained these eigenstates under the condition of zero damping factors, such as spontaneous radiation (\( \Gamma = 0 \)) \cite{stenh}.
From Eq.\eqref{e37-1}, we observe that the eigenstate corresponding to zero energy ($\lambda_0=0$), i.e., $\ket{D}$, does not include the intermediate state $\ket{e}$.
By defining the mixing angle as $tan{\theta}={\Omega_P}/{\Omega_C}$, the dark state can be written as
\begin{equation}
	\ket{D}=\cos{\theta}\ket{g}-\sin{\theta}\ket{s}
	\label{ee38}
\end{equation}
The results presented for \( \ket{D} \) are highly remarkable. We see that when two fields with Rabi frequencies \( \Omega_C \) and \( \Omega_P \) are driven to a lambda system under the specified conditions, it will be in a \textbf{coherent superposition} of the form \( \ket{D}=\cos{\theta}\ket{g}-\sin{\theta}\ket{s} \). In the absence of perturbing environmental factors, the atom will remain in this state. In this situation, the population of each level is given by\cite{peter}
\begin{equation}
	\begin{array}{lr}
		\abs{c_g}^2=cos^2{\theta}=\dfrac {\Omega^2_C}{\Omega_{rms}^2}  
		\vspace{0.1 cm}\\
		\abs{c_e}^2=0
		\vspace{0.1 cm}\\
		\abs{c_s}^2=sin^2{\theta}=\dfrac {\Omega^2_P}{\Omega_{rms}^2}  
	\end{array}
	\label{e38}
\end{equation}
The state \( \ket{D} \) is called a dark state because it does not include the intermediate state, and it is immune to spontaneous emission and population decay due to the intermediate state. Thus, it can maintain its coherent superposition state. In this coherent atomic superposition, the system will neither absorb radiation from the applied fields nor emit any radiation. The population stays in this state until decoherence, such as spontaneous emission, occurs. Consequently, state \( \ket{D} \) establishes a coherent population trap for the system under investigation. To prepare the system in the dark state, we need to turn on the field slowly enough so that the system is transferred adiabatically from the \( \ket{g} \) to the \( \ket{D} \) state.

Therefore, CPT requires the formation of a dark state, which occurs in \( \Lambda \)-type three-level systems. These systems possess two long-lived lower levels that enable efficient dark state creation. Conversely, CPT typically does not occur in V-type systems, where the presence of two short-lived upper levels hinders the formation of a dark state \cite{Singh}.
However, exceptions exist. Experimental observations of CPT in V-type systems have been achieved, such as in the two-electron atom $^{174}$Yb \cite{Singh}. Additionally, numerical simulations suggest its possibility in a V-type $^{87}$Rb system \cite{SHARABY}. In these cases, the dark state is independent of $\ket{g}$ and is a superposition of the levels $\ket{e}$ and $\ket{s}$ \cite{Gong}.
\begin{equation}
	\ket{D}=\cos{\theta}\ket{s}-\sin{\theta}\ket{e}
	\label{eee38}
\end{equation}
Similar to \( \Lambda \)-type systems, $\Xi$-type atoms (as shown in Fig.\ref{ladd}) also possess two long-lived levels, denoted as $\ket{g}$ and $\ket{s}$. This enables the formation of a dark state analogous to the \( \Lambda \)-type case. To achieve a dark state in $\Xi$-type atoms for CPT, under a two-photon resonance condition described by $\delta = \Delta_1 + \Delta_2 = 0$ \cite{vit, Snigirev2012, Xu2016}, the effective Hamiltonian for the $\Xi$-type system in Eq.\eqref{ee29} becomes mathematically equivalent to that of the \( \Lambda \)-type system in Eq.\eqref{e35}. Consequently, the dark state for the $\Xi$-type system can be expressed using Eq.\eqref{ee38}, and CPT will emerge \cite{Xu2016, Robinson, Suptitz:97, Garcia, Snigirev2012, Mukherjee, Ruth}.

CPT is an example of the STIRAP phenomenon and is a specific case of EIT, where slowly varying driven fields can be used to effectively manipulate the population and coherence of atoms \cite{vit}. An instructive review of CPT and dark states, along with their applications in spectroscopy, can be found in Ref.[\onlinecite{ARIMONDO}].
\subsection{Stimulated Raman Adiabatic Passage}
In Sec.\ref{CPT}, we assumed that the Rabi frequencies \( \Omega_{P,C} \) do not vary with time. However, phenomena such as Stimulated Raman adiabatic passage (STIRAP) require the presence of time-varying fields. In this situation, since the Rabi frequencies are time-dependent, the Hamiltonian and the system's state functions (dark and bright states) will also be time-dependent. For a time-dependent perturbation, if a physical system is in one of its instantaneous eigenstates at some instant of time, at a later time other instantaneous eigenstates will acquire population due to the transitions induced by the time dependence of the perturbation.\\
The phenomenon of STIRAP was introduced in 1990 by Gaubatz et al. in a paper titled "Population transfer between molecular vibrational levels by stimulated Raman scattering with partially overlapping laser fields" \cite{Gaubatz}. Quantum adiabatic passage techniques, such as STIRAP, are widely used for achieving quantum control in quantum information processing. In recent years, STIRAP has been employed for quantum computations and communications in superconducting circuits \cite{Zheng2022,Gaubatz,berg,vit}.\\
STIRAP efficiently transfers population between two discrete quantum states by coupling them via an intermediate state \cite{vit}. It finds applications in various fields, including atomic and molecular physics, such as laser cooling and trapping, quantum state control, and quantum information processing.\\
We will explore how time-varying fields, acting as perturbations, can drive atoms and enable the observation of STIRAP and its fascinating consequences arising from atomic interference effects during atom-field interactions.
Readers interested in a deeper understanding of STIRAP can refer to Refs. [\onlinecite{Gaubatz,berg,Bergmann_2019,vit,Shore2017,RevModPhys.70.1003}]. Here, we will focus on specific phenomena that illustrate atomic superposition and interference within STIRAP, while others will only be introduced.

\subsubsection{Mechanism of STIRAP}
\begin{figure}[h!]
	\centering
	\includegraphics[scale=0.5]{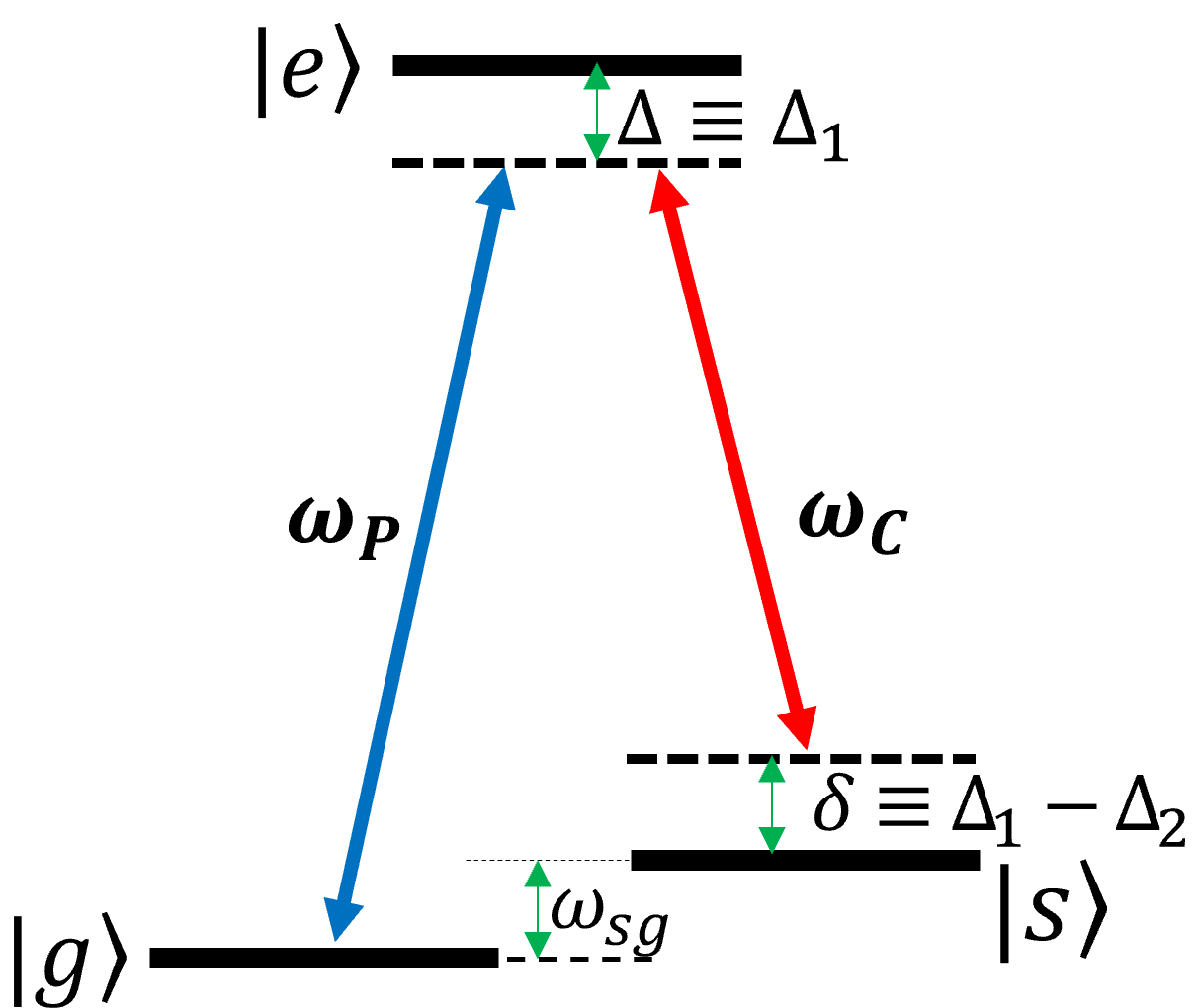}
	\caption{The $\Lambda$-type atom configuration, when the frequencies of the driving fields are in the non-resonant state.}
	\label{stirap}
\end{figure}

\begin{figure}[h!]
	\centering
	\includegraphics[scale=0.15]{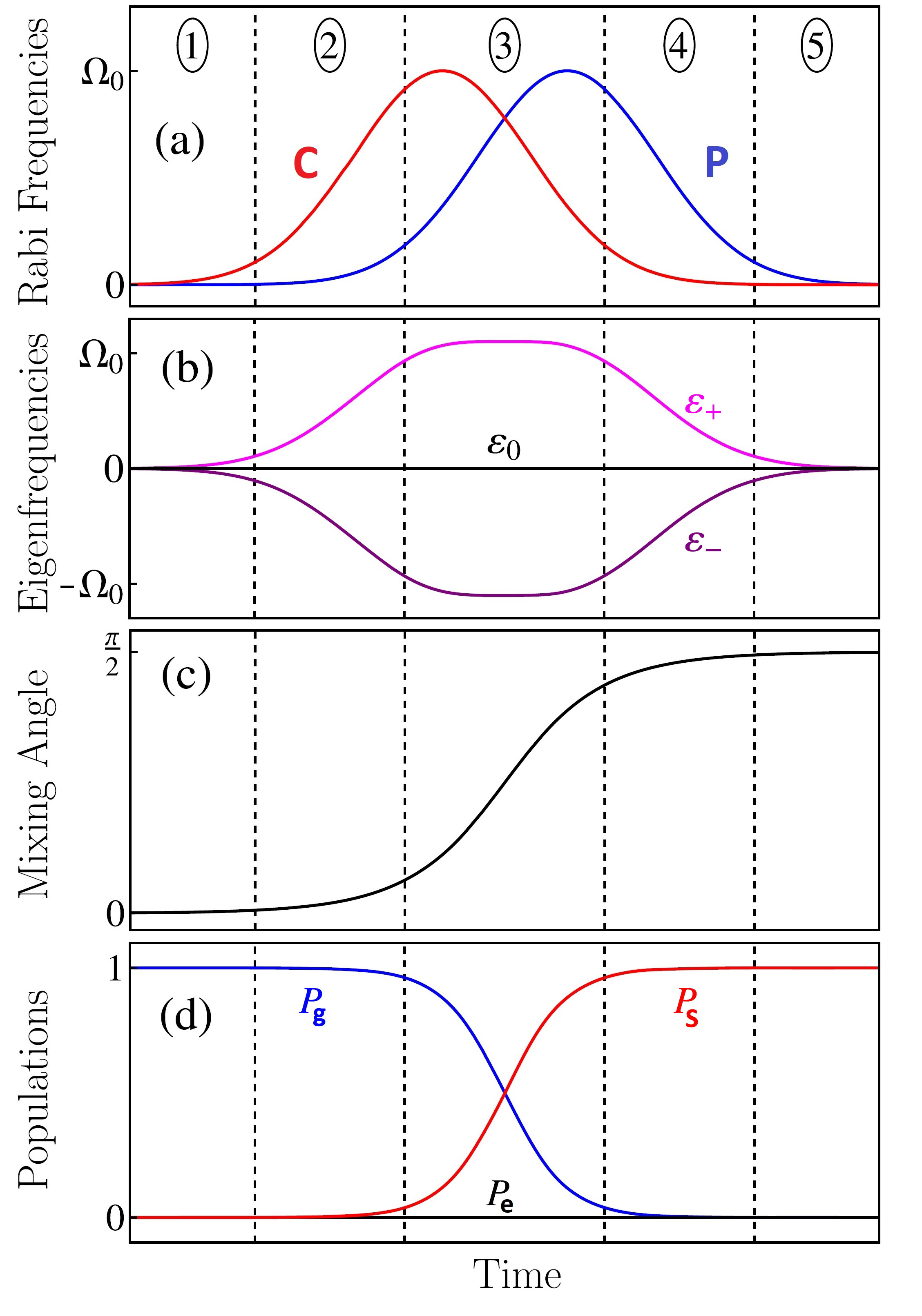}
	\caption
	{
		Example of STIRAP induced by Gaussian pulses with peak value $\Omega_0$ under single-photon resonance ($\Delta=0$) conditions:
		(a)  Shows the time dependence of the two Rabi frequencies $P$ and $C$.
		(b) Shows the time evolution of the eigenvalues of the adiabatic energy expression given in Eq.\eqref{e36}.  It is observed that whenever either of the Rabi frequencies is non-zero, the corresponding eigenergy $ \lambda_{\pm}(t) $ is eliminated (Autler-Townes effect), while the value of $ \lambda_0(t) $ is always zero.
		(c) Shows the time evolution of the mixing angle $ \theta(t) $.
		(d) Shows the time evolution of the populations $ P_n(t) $ for $ n=g,e,s $. We see that the population of level $ P_e $ is always zero.
		The vertical dotted lines indicate the five stages of STIRAP in the system. Figure adapted from [\onlinecite{VITANOV200155}].
	}
	\label{stirapstep}
\end{figure}

The most remarkable characteristic of STIRAP is the elimination of spontaneous emission from the excited state \( \ket{e} \) during the population transfer process \cite{Bergmann_2019}. The population transfer using STIRAP is notable for the following reasons\citep{vit}\\
	1. It is immune to losses due to spontaneous emission from the intermediate state.\\
	2. It is robust against small experimental variations (such as laser intensity and pulse shape).\\
While the $\Lambda$-type system is the typical configuration for STIRAP due to its efficient population transfer, the $\Xi$- system has also been explored \citep{Shore2017,Garcia}. In the most basic version of STIRAP, as depicted in Fig.\ref{stirap}, two coherent laser fields couple the intermediate state $\ket{e}$ of a $\Lambda$-type atom to the ground state $\ket{g}$ and the excited state $\ket{s}$.
Then we will have a complete population transfer from the initial population state $ \ket{g}$ to the final state $ \ket{s}$.\\
For the generation of STIRAP, the following three conditions must be met.\\
1. The amplitudes of the driving fields should be time-dependent\cite{peter,vit}.\\
2. Two-photon resonance \( \delta = 0 \)  must also be satisfied. For a $\Lambda$-type atom, based on Eq. \eqref{e27}, we can write \citep{vit}. 
\begin{align}
	\delta=\Delta_1-\Delta_2 \quad \xrightarrow{\delta=0} \ \ \ \Delta_1=\Delta_2\equiv \Delta
	\label{e40}
\end{align}
However, this condition holds true for other atom configurations as well, like $\Xi$-type\cite{Xu2016}.\\
In most cases, STIRAP works better for \( \Delta = 0 \), but since the formation of the dark state is independent of \( \Delta \), STIRAP can be independent of the single-photon detuning in the adiabatic limit\cite{berg,vit}.\\
3. The driving fields must satisfy the adiabatic evolution condition. Adiabatic following of the system's wavefunction with the dark state (adiabaticity) requires that the changes in the angle \( \theta(t) \) be sufficiently slow (adiabatic). More precisely, the rate of change of the angle \( \theta(t) \) must be much smaller than the eigenvalues difference of the adiabatic states. Therefore, the pulse durations of \( C \) and \( P \) should be tuned such that when the rate of change of \( \theta(t) \) is at its largest, the eigenvalues splitting is also at its maximum \cite{vit}. The condition for adiabatic transformation in STIRAP can be expressed as follows\cite{kuk}
\begin{align}
	\Omega_{rms}(t)\gg\dfrac{d\theta(t)}{dt}=\dfrac{\abs{\Omega_C(t)\dot{\Omega}_P(t)-\Omega_P(t)\dot{\Omega}_C(t)}}{\Omega^2_P(t)+\Omega^2_C(t)}
	\label{e43}
\end{align}
The above equation demonstrates that the smoothness of the driven pulses is a critical requirement for the adiabatic condition. When the adiabatic conditions are satisfied, the efficiency of STIRAP becomes less sensitive to small variations in laser intensity, pulse duration and delay, and even changes in the transition dipole moments.\\
In the STIRAP method, the timing of applying pulses to the atom must be carefully calculated. For instance, in one common method for implementing STIRAP, at the beginning of the process, the condition \( |\Omega_C(t=-\infty)| > 0 \) is ensured, while \( \Omega_P(t=-\infty) = 0 \) or \( |\Omega_C(-\infty)| \gg |\Omega_P(-\infty)| \). Based on the definition of the mixing angle \( \theta \), where \(tan\theta = {\Omega_P (t)}/{\Omega_C (t)} \), it's observed that the angle's value at the beginning of the process is \( \theta = 0 \). Furthermore, at the end of the process, the Rabi frequencies should have an opposite phase relative to the initial state. Thus, the following equations illustrate this concept\citep{vit}
\begin{align}
	\abs{\Omega_C(t=\infty)}<<\abs{\Omega_P(t=\infty)}  \ \Rightarrow \ \theta=\dfrac{\pi}{2}
	\label{e41}
\end{align}
In intermediate times, the two Rabi frequencies will have approximately equal magnitudes, meaning $\abs{\Omega_C(t)}\simeq\abs{\Omega_P(t)}$. Again, it is emphasized that for satisfying adiabatic evolution, the changes in Rabi frequencies must be smooth.\\
By dividing the interaction between the atom and the laser field into five distinct stages, we can gain a deeper understanding of the STIRAP mechanism. In Fig.\ref{stirapstep}, these five stages are illustrated\cite{berg,vit}.

\textbf{Stage 1:Field C induces the Autler-Townes phase.}In this stage, only pulse \( C \) is present, which connects the states \( \ket{e} \) and \( \ket{s} \). This pulse induces the Autler-Townes splitting, where the energy levels of the adiabatic states exhibit splittings with values \( \lambda_\pm \) as seen in Eq.\eqref{e36}. In this case, since the population is completely in state \( \ket{g} \), then \( |c_g( -\infty )|^2 = 1 \), and the other levels have no population. Therefore, considering \( \theta( -\infty ) \approx 0 \) and using Eqs.\eqref{ee38} and \eqref{e38}, for the initial time (\( t = -\infty \)) of the process, we can write
\begin{equation}
      \ket{D(-\infty)}=\ket{g}
	\label{e38e}
\end{equation}
In the bare state basis, we observe that all coefficients are zero except for \( |c_g|^2 = 1 \). This allows us to deduce the initial condition \( a_0(-\infty) = 1 \). Consequently, for the system's wavefunction in the dressed state or adiabatic state, based on Eq.\eqref{Adia}, we obtain
\begin{equation}
	\ket{\varphi_{Ad}(-\infty)}=\ket{g} 
	\label{e42}
\end{equation}

\textbf{Stage 2: The field C induces the CPT effect.} While pulse \( C \)  remains strong, we slowly turn on pulse \( P \), which is much weaker than pulse \( C \) at this stage. The state vector \( \varphi \) initially remains close to the ground state \( \ket{g} \). The \( P \) field does not induce transitions to the excited state \( \ket{e} \) because \textbf{destructive interference} —similar to the mechanism behind electromagnetically induced transparency, as discussed in Sec.\eqref{sec:EIT}—suppresses these transitions. This interference eliminates the transition rate from \( \ket{g} \) to the Autler-Townes states created by the strong Stokes field coupling \( \ket{e} \) and \( \ket{s} \), keeping the system predominantly in \( \ket{g} \).\\
However, by the  end of this stage, as the Pump field strengthens relative to the start, a small fraction of the population begins to transfer to state \( \ket{s} \)  (see Fig.\ref{stirapstep}\textcolor{blue}{(d)}). This small transfer marks the weakening of destructive interference and prepares the system for the adiabatic passage phase (the next stage), although most of the population still resides in \( \ket{g} \) at this point. 

\textbf{Stage 3: Adiabatic passage phase.} In the previous step, since pulse \( P \) had just entered the system, it was not yet as strong as pulse \( C \). However, in this stage, over time, both fields become strong. It's worth noting that pulse \( P \) is still increasing while pulse \( C \) is decreasing. The angle \( \theta(t) \) increases from zero to \( \pi/2 \) in this stage, and the system's state vector remains in a linear combination of \( \ket{g} \) and \( \ket{s} \). As a result, the intermediate state will be depopulated.

\textbf{Stage 4:Induced CPT by field \( P \).} This stage is similar to the second stage, with the difference that both fields are decreasing, and the intensity of field \( P \) is still greater than that of field \( C \). Now, the population has completely transferred to state \( \ket{s} \). Because pulse \( P \) has coupled states \( \ket{g} \) and \( \ket{e} \), the weaker pulse \( C \) cannot induce transitions to state \( \ket{e} \).
In this situation, the associated Autler-Townes splitting protects the population in the state \( \ket{s} \). Similar to the second step, where the laser \( C \) protected the population in \( \ket{g} \), here, we will also have an induced CPT by the \( P \) pulse in the system.

\textbf{Stage 5: Pulse \( P \) induces the Autler-Townes effect.} Now, there is no longer a pulse $C$, and the induced Autler-Townes splitting by pulse $P$ gradually decreases to zero. The state vector of the system becomes equal to $\ket{s}$, and the population is completely transferred from state $\ket{g}$  to state $\ket{s}$ . The STIRAP process is now complete. Thus, we have successfully transferred all the population initially in state $\ket{g}$  to state $\ket{s}$ through these five stages.

While STIRAP is typically implemented in $\Lambda$-type systems, it can be effectively applied in $\Xi$-type systems as well, provided that the pulse duration is shorter than the lifetime of level $\ket{s}$\cite{Garcia,Sangouard}. Observations of STIRAP in $\Xi$-type systems have been demonstrated\cite{Snigirev2012, Mukherjee, Ruth,Smith:92,Sangouard,Broers,Antti}, notably observed experimentally with nanosecond pulses in rubidium atoms\cite{Suptitz:97}.\\
In V-type systems, the direct application of STIRAP, as commonly used in \(\Lambda\)-type or cascade-type configurations, is not straightforward. This is because the mechanisms of coherence and population transfer that rely on a dark state do not have a direct analogue in V-type configurations. A V-type system consists of a single ground state coupled to two excited states, and achieving adiabatic passage requires strategies tailored to handle the interference effects and unique coupling dynamics of this setup.\\
Despite these challenges, there has been progress in using V-type systems to implement STIRAP. Recent experimental work on superadiabatic quantum driving has demonstrated population transfer in a three-level solid-state spin system\cite{Gong}. This approach leverages the V-type level structure of the electronic ground-state triplet of the NV-center spin in diamond. As discussed in Sec.\eqref{CPT}, the dark state in a V-type system is a superposition of the two excited states \(\ket{e}\) and \(\ket{s}\). 
When the system is initially in state \(\ket{s}\), the corresponding dark state is \(\ket{D(-\infty)} = \ket{s}\). Through adiabatic evolution, the system remains in the dark state, avoiding transitions to other eigenstates. At the final time, the system evolves to the state \(\ket{D(+\infty)} = \ket{e}\), bypassing the intermediate state \(\ket{g}\). Notably, the dark state contains no component of the intermediate state \(\ket{g}\) at any point during the evolution \cite{Gong}.
\subsubsection{B-STIRAP}
\begin{figure}[h]
	\centering
	\subfigure[Intuitive pulse.]
	{
		\includegraphics[width=0.19\textwidth]{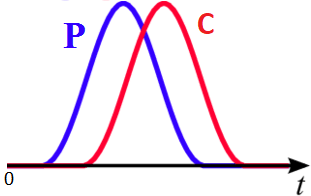}
		\label{i-pulse}
	}
	\hspace*{0.5 cm}
	\subfigure[Counterintuitive pulse.]
	{
		\includegraphics[width=0.18\textwidth]{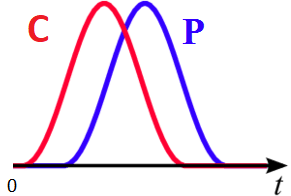}
		\label{c-pulse}
	}
	\caption{Example of intuitive and counterintuitive pulses.}
	\label{pccp}
\end{figure}

So far, we have utilized a counterintuitive pulse sequence (CP), where pulse \( C \) is applied to the system first, followed by pulse \( P \), to achieve STIRAP. In quantum optics, counterintuitive phenomena such as electromagnetically induced transparency, Autler-Townes splitting, and coherent population trapping play a significant role in the precise control of the optical properties of a medium\cite{mahana2023}. However, these phenomena are highly sensitive to the shapes and types of the applied pulses. By applying two different pulse sequences, PC and CP, to the system as depicted in Fig.\ref{pccp}, significantly different results can be observed, especially when considering the single-photon detuning \( \Delta \). In STIRAP, the counterintuitive pulse sequence CP completely transfers the population to state \( \ket{s} \). This is achieved by the dark state, independent of \( \Delta \).
In the intuitive pulse sequence (PC) , where pulse \( P \) precedes \( C \), population transfer occurs via bright states, which depend strongly on the single-photon detuning \(\Delta\). Conversely, in the counterintuitive sequence (CP) , with \( C \) applied before \( P \), the population transfers fully to the dark state \(\ket{s}\), independent of \(\Delta\). For a \(\Lambda\)-type atom, this distinction arises from the \(\Delta\)-dependence of bright states\cite{Bergmann_2019,peter,vit,VITANOV200155}.  
\begin{subequations}
	\begin{align}
		&\ket{B_{+}(t)}=\sin\theta(t)\sin\varphi(t) \ket{g}+ \cos\varphi(t)\ket{e}\label{e48}\\
		&\qquad\qquad+\cos \theta(t)\sin\varphi(t) \ket{s}\nonumber\\[8pt]
		&\ket{D(t)}=\cos \theta(t) \ket{g}- \sin\theta(t) \ket{s}\label{e49}\\[10pt]
		&\ket{B_{-}(t)}=\sin\theta(t)\cos\varphi(t) \ket{g}-\sin \varphi(t) \ket{e}\nonumber\\
		&\qquad\qquad+\cos \theta(t) \cos \varphi(t) \ket{s}\label{e50}
	\end{align}
	\label{eee}
\end{subequations}
\hspace*{-0.18cm}where \(\tan 2\varphi(t) = {\Omega_{\text{rms}}(t)}/{\Delta}\), \(\Omega_{\text{rms}}(t) = \sqrt{\Omega_P^2(t) + \Omega_C^2(t)}\), and \(\tan\theta(t) = {\Omega_P(t)}/{\Omega_C(t)}\).\\
Let's consider two distinct cases to investigate atomic interference \cite{vit}.\\

\textbf{1-  Single-photon  detuning:}
Under single-photon resonance (\(\Delta = 0\)), \(\varphi = \pi/4\), and both bright states (\(\ket{B_\pm}\)) are populated. The interference between pathways leads to generalized Rabi oscillations in the final population of state \(\ket{s}\). The population distribution at \(t \to +\infty\) is calculated using the diabatic basis transformation:
\begin{align}
	\mathbf{c}(t) =\hat{\mathbf{R}}(t)\mathbf{a}(t)
	\label{e55}
\end{align}
where \(\hat{\mathbf{R}}(t)\) relates adiabatic and diabatic amplitudes. For the intuitive pulse sequence PC, the mixing angles \(\varphi(t)\) and \(\theta(t)\) simplify the adiabatic amplitudes, yielding
\begin{align} 
	&a_{+}(+\infty)=\dfrac{1}{\sqrt{2}}e^{-iA}\nonumber\\
	&a_{0}(+\infty)=0	\label{e66}\\
	&a_{-}(+\infty)=\dfrac{1}{\sqrt{2}}e^{+iA}\nonumber
\end{align}
with \(A = \int_{-\infty}^\infty \Omega_{\text{rms}}(t) \, dt\). Substituting these into the diabatic basis, the final level populations are:
\begin{align}
	P_{s}(+\infty)&=\cos^2 A\nonumber \\  
	P_{g}(+\infty)&=0 \label{e69} \\
	P_{e}(+\infty)&=\sin^2 A \nonumber
\end{align}
Thus, under single-photon resonance and an intuitive pulse sequence, the system reaches a  \textbf{superposition}  of \(\ket{e}\) and \(\ket{s}\) at \(t \to +\infty\).\\

\textbf{2- Single-photon detuning:} For $\Delta \neq 0$, the angle $\varphi(t)$ is defined as
\begin{equation}
	\tan 2\varphi(t) = \frac{\Omega_{rms}(t)}{\Delta} \nonumber
	\label{e70}
\end{equation}
As $t \to \pm \infty$, $\varphi(\pm \infty) = 0$. Although the fields $P$ and $C$ remain in two-photon resonance ($\delta = 0$), adiabatic evolution enables complete population transfer from $\ket{g}$ to $\ket{s}$. This occurs through the adiabatic state $\ket{B_-(t)}$, which connects $\ket{g}$ and $\ket{s}$ \cite{vit}. Initially, $\theta(-\infty) = \pi/2$, $\varphi(-\infty) = 0$, and the population resides in $\ket{B_-(-\infty)} = \ket{g}$
\begin{align}
	a_+(-\infty) &= 0, \ a_0(-\infty) = 0, \ a_-(-\infty) = 1
	\label{e73}
\end{align}
At $t \to +\infty$, the adiabatic state evolves to $\ket{B_-(+\infty)} = \ket{s}$, resulting in:
\begin{equation}
	\ket{g} \xleftarrow{\text{$t \to -\infty$}} \ket{B_-(t)} \xrightarrow{\text{$t \to +\infty$}} \ket{s}
	\label{e74}
\end{equation}
This process, termed bright STIRAP (B-STIRAP), involves transient population of $\ket{e}$ during the evolution, as the system transitions via $\ket{B_-(t)}$ \cite{sten}.

\begin{figure}[h!]
	\centering
	\includegraphics[width=0.98\linewidth]{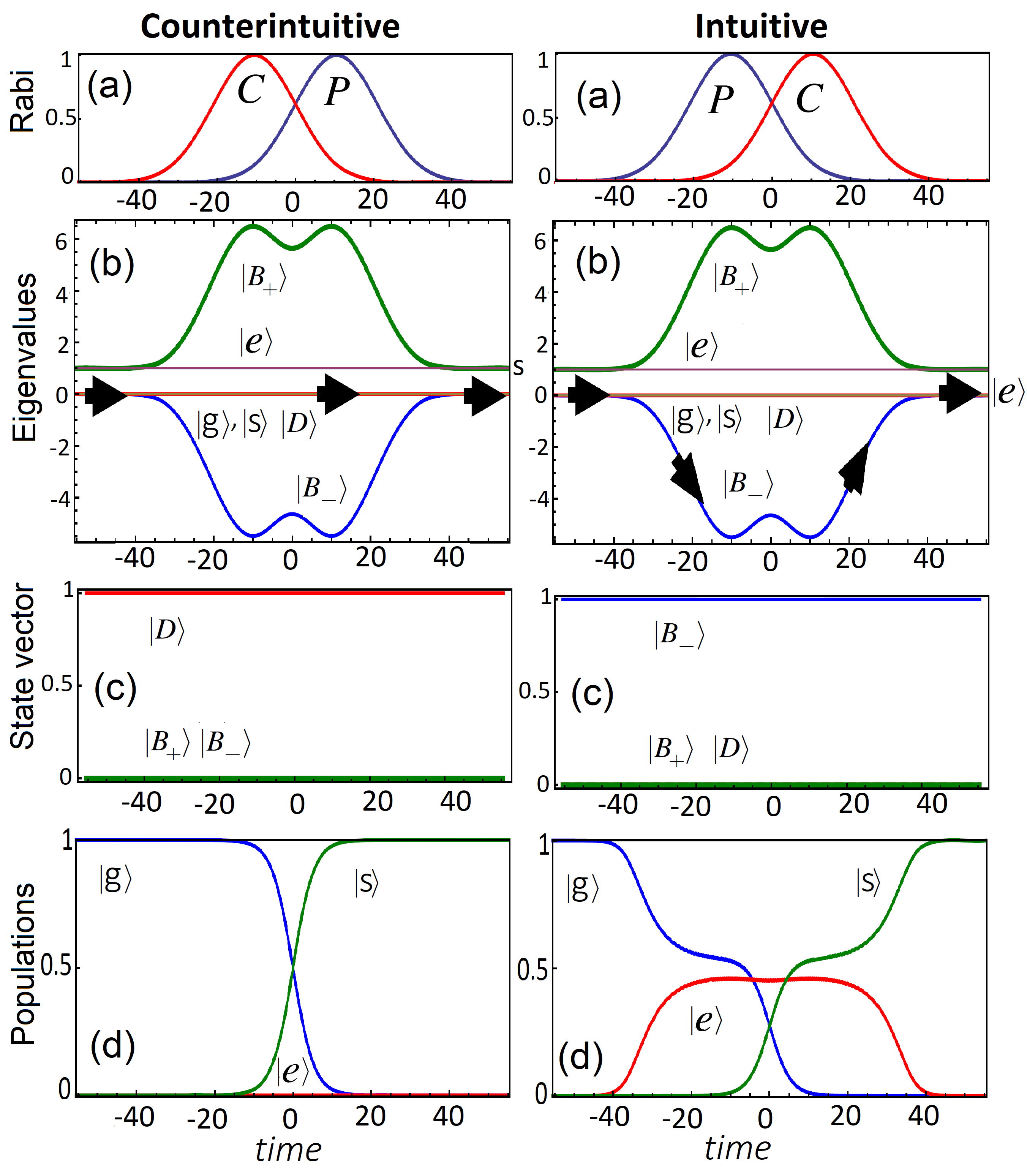}
	\caption
	{An example of STIRAP Using Intuitive (Right) and Counterintuitive (Left) Pulses in the Presence of Two-Photon Resonance and Single-Photon Detuning:
		(a) Rabi frequencies.
		(b) Represents the eigenvalues of the adiabatic states $ \ket{D}, \ket{B_\pm} $ and the diabatic states $\ket{g}, \ket{e}, \ket{s} $. The thick arrow illustrates the trajectory of the system's quantum state in the adiabatic basis during the STIRAP process. State $\ket{e} $ is positively detuned, while states $\ket{g} $ and $\ket{s}  $, corresponding to the adiabatic state $\ket{D}$ with zero eigenvalue, are degenerate.
		(c) The adiabatic components of the state vectors are shown.  For the counterintuitive pulse sequence, the state vector remains aligned with $\ket{D}$, while for the intuitive pulse sequence, the state vector is aligned with $\ket{B_-}$.
		(d) Populations of each level based on the type of applied pulse (intuitive or counterintuitive). Figures adapted from [\onlinecite{Shore2017}]
		}
	\label{pp}
\end{figure}

In contrast, for the counterintuitive sequence $CP$, the population transfers via $\ket{D(t)}$, avoiding $\ket{e}$ entirely. Thus, $CP$ achieves direct transfer to $\ket{s}$, while $PC$ populates $\ket{e}$ transiently. Fig.\ref{pp} illustrates this distinction. During B-STIRAP, $\ket{e}$ and $\ket{s}$ form a superposition
\begin{equation}
	P_e(t) = \sin^2 \varphi(t) = \frac{1}{2} \left( 1 - \frac{\Delta(t)}{\sqrt{4\Omega_{rms}^2(t) + \Delta^2(t)}} \right)
	\label{e75}
\end{equation}
Successful B-STIRAP requires the lifetime of $\ket{e}$ to exceed the pulse duration \cite{vit}.
\\
 For further study on B-STIRAP, one can refer to the sources Refs.[\onlinecite{Grigoryan, berg, vit, Shore2017}].
 \subsubsection{\label{F-ST}F-STIRAP}
 Shortly after the discovery of STIRAP, it was realized that when the ratio of the two Rabi frequencies remains constant, the mixing angle $\theta(t) = \tan[-1]({\Omega_P(t)}/{\Omega_C(t)} )$ will also remain constant. In this case, the system's state vector (\(\Lambda\) and \(\Xi\)-types) becomes frozen in a coherent superposition of states  $\ket{g}$ and  $\ket{s}$. However, this phenomenon occurs only if the mixed Rabi frequencies satisfy the following condition\cite{vit,Sangouard}.
 \begin{equation}
 	\lim_{t\to -\infty} \dfrac{\Omega _P(t)}{\Omega _C(t)}=0 \  \  , \  \  \lim_{t\to +\infty} \dfrac{\Omega _P(t)}{\Omega _C(t)}=e^{i\alpha}\tan \Theta
 	\label{e76}
 \end{equation}
 in which $ \Theta=\theta(+\infty) $ and $ \alpha $ is a constant phase, that appears due to the assumption that the fields $ C $ and $ P $ have phases $ \phi_C $ and $ \phi_P $, respectively. Therefore, if we have \cite{shore}.
 \begin{equation}
 	c_g(t)=\cos \theta(t) \  \  , \  \   c_s(t)=-e^{i\alpha} \sin \theta(t)
 	\label{e77}
 \end{equation}
 Then, at $ t\rightarrow +\infty $, the angle $ \theta(t) $ can be frozen at an arbitrary value such as $ \Theta $. The final state will be a coherent superposition of probability coefficients. Since $c_g(+\infty)=\cos\Theta$ and $c_s(+\infty)=e^{i\alpha}\sin \Theta$, we can use either Eq.\eqref{ee38} and Eq.\eqref{e38} or Eq.\eqref{e49} to see the form of the state vector that represents this coherent superposition.
 \begin{align}
 	\ket{D(+\infty)}= \cos\Theta\ket{g}- e^{i\alpha}\sin \Theta\ket{s}
 	\label{e78}\
 \end{align}
 Therefore, instead of STIRAP, we have F-STIRAP (Fractional STIRAP) or partial STIRAP, in which only a controlled fraction of the population is transferred to level $\ket{s}$.
 \begin{figure}[h!] 
 	\centering
 	\includegraphics[scale=0.095]{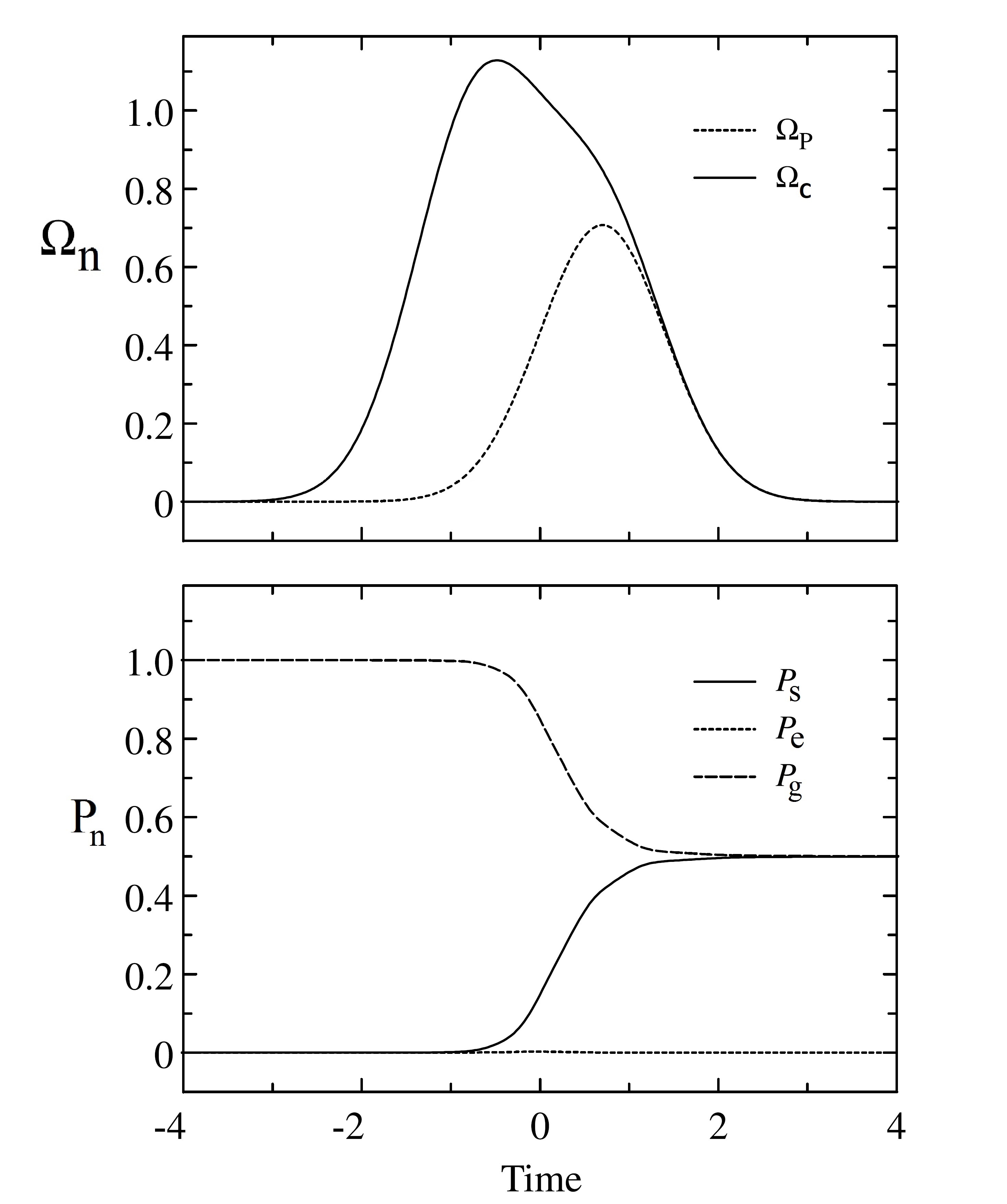}
 	\caption{Time evolution of population for F-STIRAP with single-photon resonance ($\Delta=0$). Figure adapted from [\onlinecite{Vit99}]. }
 	\label{f-stirap}
 \end{figure}
 In F-STIRAP, it is important to note that the applied pulses are counterintuitive. This is because only the adiabatic state $\ket{D(t)}$ is considered in the calculations for this phenomenon. For F-STIRAP, we consider a specific case where $\alpha=0$ and $\Theta={\pi}/{4}$.
 \begin{equation}
 	\ket{D(+\infty)}=\ket{\psi}=\dfrac{1}{\sqrt{2}} \left( \ket{g} - \ket{s} \right)
 	\label{e83}
 \end{equation}
 This superposition corresponds to a Hadamard transform for a quantum bit \cite{Vit99}. Extending F-STIRAP may be advantageous for imaging, sensing, and detection due to its ability to generate stronger signals and maintain signal stability during propagation through a medium\cite{Chathanathil}.\\ 
 In atomic optics, counter-propagating pulses $C $ and $P$ can create a coherent superposition accompanied by a momentum transfer of $ 2\hbar k  $ for half of the atoms. Consequently, F-STIRAP acts as a coherent beam splitter ($ BS $) in this scenario.
 The use of counter-propagating fields $C$ and $P$ to create F-STIRAP has turned this phenomenon into a popular tool for the production of atomic beam splitters\cite{Vit99}.\\
 Two-photon resonance ($ \delta=0 $) is usually a necessary condition for observing STIRAP. This condition is certainly valid when the peak frequencies of the Rabi fields are approximately equal. However, both experimental\cite{Dupont,Srensen} and theoretical \cite{Boradjiev,PhysRevA.76.062321} studies have shown that when the fields $C$ and $P$ are significantly different, the population transfer profile (with respect to two-photon resonance) becomes asymmetric. Such conditions often arise in applications of STIRAP involving interactions between fields originating from different sources \cite{vit}. Examples of such occurrences include\\
 1. When the applied fields originate from different sources.\\
 2. In the Vacuum STIRAP phenomenon.
 
 Originally, STIRAP was used in atoms for coherent momentum transfer, aiming to build coherent beam splitters and mirrors for atomic interferometers. We know that coherent atomic excitation involves the absorption and emission of photons, hence it is always accompanied by the transfer of momentum from photons to atoms. This momentum change is the basis of laser cooling for atoms.\\
 Coherent momentum transfer facilitated by optical beams plays a pivotal role in the design of mirrors, beamsplitters, and atomic interferometers, within the realm of atomic optics. An atomic beamsplitter divides the atomic wavefunction into a macroscopic coherent superposition of two wavepackets traveling in distinct spatial directions. Subsequently, an atomic mirror reflects these wavepackets, allowing them to converge and interfere. Through the combination of beamsplitters and mirrors, we can assemble an atomic interferometer. However, achieving this interference necessitates coherence between the atomic beamsplitters and mirrors. Therefore, leveraging properties like coherent momentum and population transfer without any losses, STIRAP emerges as a valuable tool in atomic interferometry\cite{vit}.
 \subsection{Some Other Interference Phenomena}
 So far, we have examined STIRAP and several examples of its subsets. We observed how this technique induces atomic coherent superposition and consequently atomic interference. The variety of STIRAP phenomena arises from the method and type of pulses driven into the atoms. These differences result in various superpositions between different levels, as discussed in this section. In the following, we intend to briefly introduce some other types of STIRAP and the phenomena that lead to the transfer of population from the initial state to the final state, without going into details.
 \subsubsection{V-STIRAP}
 In the realm of quantum optics, inducing transitions in a system necessitates the application of external fields. Laser fields provide the necessary conditions to achieve this, playing a key role in stimulated Raman scattering. However, it is crucial to recognize that vacuum fields can also produce measurable effects. To illustrate this point, let us consider the case of an atom confined within a cavity.
 We know that the vacuum field of the cavity is always present. Therefore, this field, along with the electric dipole moment associated with the transition $ \ket{e}\leftrightarrow \ket{s} $ (denoted as $ \mu_{es} $), creates a constant vacuum Rabi frequency $ C $. Now, by applying a laser field $ P $ (the pump) to this system, we can produce a pseudo-STIRAP transition and observe that the mixing angle $ \theta(t) $ changes in response to the increase in the field $ P $. Thus, in this case, which is called Vacuum STIRAP (V-STIRAP), the vacuum field inside the cavity replaces the laser field $ C $ \citep{shore}. To observe the superposition resulting from this new situation, we examine the dark state vector. In V-STIRAP, the dark state energy is given by $ E_n = \hbar n\omega $, and the state vector is equal to\cite{PhysRevA.51.1578,Shore2017}
 \begin{equation}
 	\ket{D(t)}=\cos\theta(t) \ket{g,n}- \sin\theta(t)\ket{s,n+1}
 	\label{v-sti}
 \end{equation}
 where $ n $ represents the number of photons in the cavity. Therefore, under these conditions, we observe an atomic superposition. For further study of this phenomenon, we can refer to the Refs.[\onlinecite{PhysRevLett.126.113601,Shore2017,vit,Parkins,Vasilev_2010,PhysRevA.51.1578,PhysRevLett.85.4872}].
 \subsubsection{C-STIRAP}
 In a typical STIRAP setup, pulses \( P \) and \( C \) interact with a three-level system, resulting in complete and adiabatic population transfer from the initial to the final state. Complete population transfer to the final state occurs in the presence of two-photon resonances, and for non-zero values of \(\delta\), the population evolution is not adiabatic. However, when pulses \( P \) and \( C \) are chirped laser pulses, due to the characteristics of this type of field, even in the presence of a two-photon detuning (\(\delta \neq 0\)), population can be adiabatically transferred from the initial to the final level \cite{Chathanathil}. This phenomenon is known as Chirped STIRAP (C-STIRAP).
 The pulse $P$ or $C$ in C-STIRAP can be written as follows, considering a Gaussian envelope\cite{Chathanathil}
 \begin{equation}
 	\mathbf{E}_{_{P,C}}(t)=\mathbf{E}_{_{P_0,C_0}} \ \ e^{-(\frac{t-t_{_{P,C}}}{\tau_{_{P,C}}})^2}\cos[\omega_{_{P,C}}(t-t_{_{P,C}})+\frac{\alpha_{_{P,C}}}{2}(t-t_{_{P,C}})^2]
 \end{equation}
 where \(\omega_{_{P,C}}\) denotes the field frequencies, \(\mathbf{E}_{_{P_0,C_0}}\) the peak amplitudes, \(\tau_{_{P,C}}\) the pulse durations, \(\alpha_{_{P,C}}\) the chirp rates, and \(t_{_{P,C}}\) the central times of \(E_{_{P,C}}(t)\).
 Based on what we have seen in the previous sections, calculating the Hamiltonian and Rabi frequencies using this field will not be a very complicated task. Considering the deformation of the applied fields on the system, the form of the single-photon and two-photon detunings should also be modified, and new definitions may need to be written for them\cite{Chathanathil}.
 \begin{subequations}
 	\begin{align}
 		&\Delta(t)=\Delta-\alpha_{_{P}}(t-t_{_P})\\[4pt]
 		&\delta(t)=-\delta+\alpha_{_{C}}(t-t_{_C})-\alpha_{_{P}}(t-t_{_P})
 	\end{align}
 	\label{dsdsokc11}
 \end{subequations}
 Considering the Eqs.\eqref{dsdsokc11}, which define the single-photon and two-photon detunings, we can still observe population transfer by carefully choosing the values of the chirp rates. In this process, the chirp rates must be selected such that their sum and difference with $\delta$ result in the vanishing of $\delta(t)$. Then, we will observe that the state vector describing the system will be the dark state $\ket{D(t)}$. Therefore, we have a superposition of the initial and final states, and the population will be completely and adiabatically transferred from the initial state to the final state\cite{Chathanathil}.
 \subsubsection{CHIRAP and RCAP}

Adiabatic passage is a widely-used technique for achieving nearly 100\% efficiency in population transfer within multi-level quantum systems. However, STIRAP, despite its advantages, cannot be applied to cascade excitation processes of molecular or atomic vibrational levels—commonly referred to as vibrational ladder climbing—when chirped laser pulses are involved. This limitation arises from the challenge of maintaining precise two-photon resonance across the vibrational ladder due to synchronization issues \cite{Chelkowski}. \\
Without two-photon resonance, each adiabatic eigenstate acquires a component of the excited state $\ket{e}$, which is subsequently lost through spontaneous emission \cite{berg}. To overcome this challenge, Chirped Rapid Adiabatic Passage (CHIRAP), also known as Raman Chirped Adiabatic Passage (RCAP), has been developed as a robust alternative to traditional three-state adiabatic passage. CHIRAP utilizes chirped laser pulses to adiabatically transfer populations between quantum states by sweeping through the resonance frequencies of the transitions \cite{Band}. \\
This method supports either complete or selective population transfer from an initial state to a final state \cite{Magnes, Melinger}. Unlike C-STIRAP, which requires precise timing and a time delay between laser pulses, CHIRAP involves the simultaneous application of pulses, simplifying the experimental setup \cite{berg, Chathanathil}.  \\
The concept of rapid adiabatic passage in the optical domain was first explored by Loy and Grischkowsky, who studied population inversion using chirped laser pulses \cite{Xihua, Chang}. A typical CHIRAP setup involves the use of a chirped pulse laser (laser P) and a monochromatic laser field (field C) interacting with a $\Lambda$-type atomic system. This configuration ensures efficient population transfer while maintaining adiabatic conditions, enabling the creation of coherent superpositions between the initial and final states \cite{Chelkowski, Malinovsky2001}.  

RCAP, a variation of CHIRAP, employs two chirped pulses in an off-resonance Raman configuration, improving both the efficiency and robustness of the transfer process against experimental variations \cite{Malinovsky2001}. This makes RCAP particularly suitable for applications in quantum control and the manipulation of atomic and molecular states \cite{Chelkowski}. Furthermore, RCAP enables the observation of coherent superpositions of the initial and final states, which is essential for certain quantum applications. \\
RCAP offers several notable advantages \cite{Chelkowski} \\ 

\textbf{1. Relaxed Frequency Tuning:} Unlike other methods requiring precise tuning of individual transition frequencies, RCAP achieves near-perfect efficiency by slowly sweeping through resonances. This eliminates the need for meticulous adjustments. \\ 
\textbf{2. Precise Excitation Control:} By halting the chirp at a specific frequency, RCAP enables precise state selection, making it a versatile tool for selective population transfer.\\  
\textbf{3. Rapid Dissociation: }RCAP facilitates rapid dissociation processes before vibrational energy redistribution occurs, which is ideal for studying ultrafast molecular dissociation. \\ 
\textbf{4. Low Fragment Energy Dissociation:} With sufficiently strong chirps, RCAP disrupts system coherence, resulting in bond dissociation where the remaining fragments exhibit very low vibrational energy.\\  

In addition to these advantages, CHIRAP is especially useful in systems where STIRAP requires far-infrared laser pulses for population transfer. However, CHIRAP typically demands higher laser powers than STIRAP, often approaching the ionization threshold. Thus, it is most effective in systems with strong Raman transitions. Despite these requirements, experiments with CHIRAP have demonstrated interference effects and coherent population transfer, similar to STIRAP \cite{Krish12, Bandrauk}. \\
 For further reading on this phenomenon, you can refer to the sources mentioned in this section.
 \subsubsection{TCC}
 Quantum control (QC) can be expressed as a problem of finding methods to achieve complete population transfer from an arbitrary initial wavepacket to a target wavepacket. This can be accomplished using a set of coherent optical fields or, equivalently, a set of customized laser pulses\cite{Thanopulos}.
 In recent years, QC has been realized using various methods. Some of these methods include:\\
 \textbf{1. Coherent Control (CC):} A prevalent approach to achieving QC is coherent control, which involves manipulating the phases and other parameters of optical fields so that the selectivity of the final state can be achieved using controlled laser interference between two or more quantum paths.\\
 \textbf{2. Optimal Control (OC):} A general method for optimizing laser pulses using pulse shaping techniques to achieve the desired outcome.\\
 \textbf{3. Adiabatic Passage:} This involves a method for effectively transferring population between selected states and recently, their coherent superpositions.\\
 In this section, we will focus on the coherent control (CC) technique. Coherent control in population transfer from an initial state to a target state and the creation of desired coherent superpositions between two states have attracted significant attention due to their applications. Numerous methods have been proposed and employed for this purpose, including three widely used strategies: \textit{STIRAP}, \textit{CHIRAP}, and Temporal Coherent Control (\textit{TCC}). While STIRAP and CHIRAP have been discussed previously, let's delve into the concept of TCC.
 
 Coherent control is a method for manipulating dynamic processes using light, aiming to control quantum interference phenomena. It's typically achieved by shaping the phase of laser pulses \cite{Shapiro,Kohler}. Several theoretical approaches to quantum control have been proposed and investigated, but optimal control, which involves shaping laser pulses to achieve a well-defined target, is mostly inaccessible to experimenters except in very limited cases \cite{Aziz,Kohler}.
 In coherent control, quantum interference between several quantum pathways is employed to modulate a specific channel. Each quantum pathway arises from the interaction between a laser field and an atomic or molecular system. Adjusting the relative phases of the laser fields leads to either constructive or destructive interference between the quantum pathways. Consequently, the probability of exciting a process is coherently controlled \cite{Aziz}.
 Control of interference between different pathways can be achieved experimentally by adjusting the relative phase between the laser modes used for excitation. However, achieving this control is impeded by two factors: the combined effects of partial coherence in light sources and the rapid decoherence of quantum processes in materials due to scattering. Experience shows that extensive control remains achievable as long as the separation of the light pulses is shorter than the decoherence rates, despite these challenges\cite{Thanopulos}.\\
 Coherent control can be achieved by combining a laser's fundamental frequency (\( \omega \)) with one of its harmonics (\( n\omega \)), where the outcome depends on whether \( n \) is odd or even. For odd harmonics (\( n = 3, 5, \dots \)), the quantum pathways from \( \omega \) and \( n\omega \) converge to the same excited state, allowing interference to modulate the total cross-section (overall process probability). In contrast, even harmonics (\( n = 2, 4, \dots \)) direct the pathways to degenerate states with opposite parity, enabling control over the differential cross-section (spatial distribution of outcomes). By adjusting the relative phase, amplitude, or timing of the frequencies, this method exploits quantum interference to steer both the intensity and directionality of quantum processes, with applications in selective reactions, spectroscopy, and quantum technologies\cite{Charron,Aziz}. \\ 
  In TCC, a sequence of two time-delayed pulses is employed to trace two quantum pathways \cite{Aziz,Jones}. The phase interference is associated with the time delay between the two pulses. A change in the time delay leads to interferograms exhibiting high-frequency oscillations modulated by a slow envelope arising from the movement of the wavepacket in the excited state. However, for process control, the time delay must be stabilized with a precision much better than the optical period to enable selection between constructive or destructive interference leading to enhancement or suppression of the total cross-section of the process\cite{Aziz}.\\
 To achieve population transfer via TCC, the form of the total applied field to the system will be \(\mathbf{E}(t) = \mathbf{E}_{P}(t) + \mathbf{E}_{C}(t)\), where each of the fields \(P\) and \(C\) can be written as follows \cite{Aziz}.
 \begin{subequations}
 	\begin{align}
 		&\mathbf{E}_{_{P}}(t)=\mathbf{E}_{_{P_0}}(t) e^{-i\omega_{_{L}}t}\label{TCC1}\\[5pt]
 		& \mathbf{E}_{_{C}}(t)=\beta\mathbf{E}_{_{P}}(t-\tau) =\beta\mathbf{E}_{_{P_0}}(t-\tau)e^{-i\omega_{_{L}}(t-\tau)}\label{TCC2}
 	\end{align} 
 	\label{TCC}
 \end{subequations}
 In Eq.\eqref{TCC}, it is assumed that the central frequency is the same for both fields, denoted as \(\omega_{_{L}}\). However, the field \(P\) is applied to the system with a time delay \(\tau\) relative to the field \(C\). In Eq.\eqref{TCC2}, the parameter \(\beta\) is a real quantity representing the ratio between the amplitudes of the two pulses. The quantity \(\mathbf{E}_{_{P_0}}\) denotes the envelope or maximum value of the driven field. Therefore, by applying these pulses to the system under conditions detailed in Ref.[\onlinecite{Aziz}], we will observe a \textbf{coherent superposition of the initial and final states}.\\
 Beyond using individual methods, combining two or three coherent population transfer techniques can achieve quantum coherent control. For example, in 1994, Band proposed a selective and efficient scheme for a four-level lambda or ladder system. This scheme integrated STIRAP and CHIRAP techniques for coherent population transfer control \cite{Band,Magnes}. In another work in 2007, Petr Král et al. suggested a method combining STIRAP and TCC to achieve selective and complete population transfer. This approach is called coherently controlled adiabatic passage\cite{Thanopulos}.
 As a final example, Yang et al. in 2010 demonstrated that by integrating the techniques of STIRAP, TCC, and CHIRAP, it is possible to achieve selective and robust complete population transfer from the initial state to any of the final states. They showed that efficient population control in the selected excited state can be realized in a multilevel system \cite{Xihua}.\\
 For further details on the topics discussed in this section, please refer to the following Refs.[\onlinecite{Aziz,Salour,Zare1998,Zhenrong,Chu2002,Ngoko,Thanopulos,Jones,Charron,Kohler,Shapiro,Xihua}].
 \subsection{\label{sec:EIT}Electromagnetically Induced Transparency}
 Optical pumping has enabled physicists to manipulate atomic populations using optical fields. Electromagnetically Induced Transparency (EIT) expands this control to the domains of coherent interference and quantum states, allowing for the simultaneous manipulation of light by atoms. It presents a novel approach to achieve atom-light coupling through enhanced coherence \cite{PhysRevLett.84.5094}.
 In contrast to the standard quantum electrodynamics (QED) approach for single atoms, which requires strong coupling of a single atom to a photonic mode, here strong coupling arises due to collective enhancement provided by a large ensemble of identical atoms \cite{RevModPhys.75.457}. In this scenario, the resulting collective atomic state exhibits high coherence and can preserve the quantum information initially carried by the probe field.\\
 EIT is a powerful technique for rendering a material system transparent to resonant laser radiation, while preserving the nonlinear optical properties associated with the system's resonant response. Essentially, in this phenomenon, the field propagates through the medium as if it were absent \cite{Swain}. Therefore, using EIT, we can eliminate the environment's effect on electromagnetic radiation. Making an opaque medium transparent is achieved using quantum interference \cite{RevModPhys.75.457}. In fact, many of the important properties of EIT arise from the nature of quantum interference in a material that was opaque at the beginning of the process \cite{RevModPhys77}.\\
 In 1989, Imamo\v{g}lu and Harris proposed a groundbreaking concept based on their studies of dressed states (long-lived excited states). They suggested that the energy level structure required for quantum interference could be manipulated using an external laser field \cite{Imamo}. This concept opened new avenues for controlling light-matter interactions. Soon after, in 1990, Harris et al. expanded upon this idea. They demonstrated how to incorporate frequency conversion within a four-wave mixing scheme involving atomic bound states, significantly enhancing the frequency conversion process.
 The key to achieving this effect, now known as Electromagnetically Induced Transparency or $EIT$, lies in two phenomena:
  \begin{enumerate}
 	\item Neutralization of the linear susceptibility (absorption) at resonance.
 	\item Amplification of the nonlinear susceptibility through constructive interference
 \end{enumerate}
 The term "EIT" first appeared in a landmark paper by Harris et al. (1990) to describe this phenomenon. In EIT, destructive interference in a laser medium cancels out the linear response\cite{Harris1990}. 
 They demonstrated that when a strong coupling field ($ \omega_C $) is used for a resonant transition  in a three-level system, the absorption of the probe field $ \omega_P $ can be reduced or even eliminated\cite{Harris,abraham}. This strong coupling ($ \omega_C $) implies that Rabi frequency is comparable to or larger than the spontaneous emission rate of the system. EIT can be used under more restricted conditions to eliminate self-focusing, defocusing, and improve the passage of laser beams through inhomogeneous refractive gases\cite{Harris}.
 When a field is applied to the system, we encounter an optical response from the atomic medium. This response that is caused by the induced coherence of atomic states due to laser radiation, leading to quantum interference between excitation pathways and controlling the optical response. This allows us to eliminate absorption and dispersion at the resonant frequency of a transition \cite{RevModPhys77}. For a historical overview of the process that led to the emergence of EIT, one can refer to the Ref.[\onlinecite{RevModPhys77}].\\
 EIT and CPT share a common underlying physical principle for neutralizing the absorption of an applied field to the system \cite{RevModPhys77}. However, there are also differences between them. The distinction between CPT and EIT lies in the assumption made about the initial state of the atom. In CPT, it is assumed that the atom starts in a dark state and is prepared in this state. In contrast, in EIT, the assumption is that the atom is injected into a dark state through a combined action using the coupling field ($ \omega_C $) \cite{zub}.
 Note that one method for observing CPT is modifying the absorption profile for a field in the presence of another field \cite{steck}, a condition provided by EIT. Furthermore, by carefully studying the physics of EIT and STIRAP phenomena, we can observe both commonalities and differences between these two phenomena \citep{vit}.\\
 1. Both phenomena utilize  the consequences of interference in probability amplitudes for transitions between quantum system states.\\
 2. EIT primarily occurs due to field propagation in high-density media, while STIRAP is often utilized in low-density environments to achieve precise control or manipulation of population distribution across quantum states.\\
 3. In EIT, the ratio of weak to strong Rabi fields $ \left({\Omega_P}/{\Omega_C}\right) $ remains constant or undergoes very small variations. Consequently, a steady dark state emerges. However, in STIRAP, it is essential for the dark state to evolve over time, as the ratio $ {\Omega_P}/{\Omega_C} $ changes from zero to infinity during the process.\\
 4. Both processes are resistant to small changes in field intensities.\\
 5. The combination of these two processes is utilized for light stopping, meaning the transfer of properties of a light pulse to a medium for storage and readout.\\
 6. Both processes are the physical foundation of \textbf{slow light}, in which a strong field alters the refractive index and consequently changes the group velocity of the weak field.\\
 \subsubsection{Mechanism of EIT}
EIT investigations in ladder-type systems are suitable for fundamental research, as EIT transitions become easily accessible when low-power coupling diode lasers are applied to elements such as rubidium. However, in lambda-type systems, EIT holds the greatest promise for commercial applications due to the easier attainment of complete transparency in the rubidium \(D_1\) line \cite{abraham}. 
V-type systems offer another configuration for EIT with distinct advantages\cite{Qin_2022}. Here, the EIT process avoids population trapping, allowing researchers to distinguish different coherent processes\cite{Marangos}. Additionally, V-type excitation utilizes counter-propagating beams for velocity discrimination\cite{Lazoudis} and optical pumping effectively suppresses residual absorption for high-precision spectroscopy\cite{Acosta}.\\

 In EIT, we apply two fields with frequencies different from the Raman transition. By applying these two fields, the electrons in the sample become stationary. Since the electrons no longer move, they cannot contribute to the dielectric constant. This immobility occurs when, for each applied frequency, two sinusoidal forces with opposite phases are applied to the electrons\cite{Harris}.
 While the previous explanation provided a classical picture, a more accurate description requires quantum mechanics, where we deal with probability amplitudes and expectation values of electron coordinates. From a quantum perspective, we consider a lambda-type system. The  probability amplitude of the excited state, denoted by $ \ket{e} $, is driven by two terms with equal magnitudes but opposite signs. One driving term is proportional to the probability amplitude of the ground state, \( \ket{g} \), while the other driving term is proportional to the probability amplitude of the state \( \ket{s} \), with an opposite phase.
 Both driving terms have the same frequency, denoted by $ \omega_P $, and the same amplitude. This ensures that the probability amplitude of the excited state, $ \ket{e} $, and the expectation value of the electron's motion at each applied frequency remain zero.\\
 In EIT, the $P$ field is tuned close to the resonance frequency between the ground state, $ \ket{g} $, and the excited state, $ \ket{e} $. This field serves to measure the absorption spectrum of the transition.
However, the control field, denoted by \(\omega_C\), is significantly more powerful than the probe field, \(\omega_P\), i.e., \(\omega_C \gg \omega_P\). The control field is tuned close to the resonance frequency of the states \(\ket{s}\) and \(\ket{e}\). It is important to note that in the absence of the control field, the atomic medium will be opaque to the probe field, \(\omega_P\), as photons are absorbed by atoms and re-emitted spontaneously\cite{Harris}.
 \subsubsection{Investigative Approach}
\begin{figure}[h!] 
 	\centering
 	\subfigure[]
 	{
 		\includegraphics[width=0.15\textwidth]{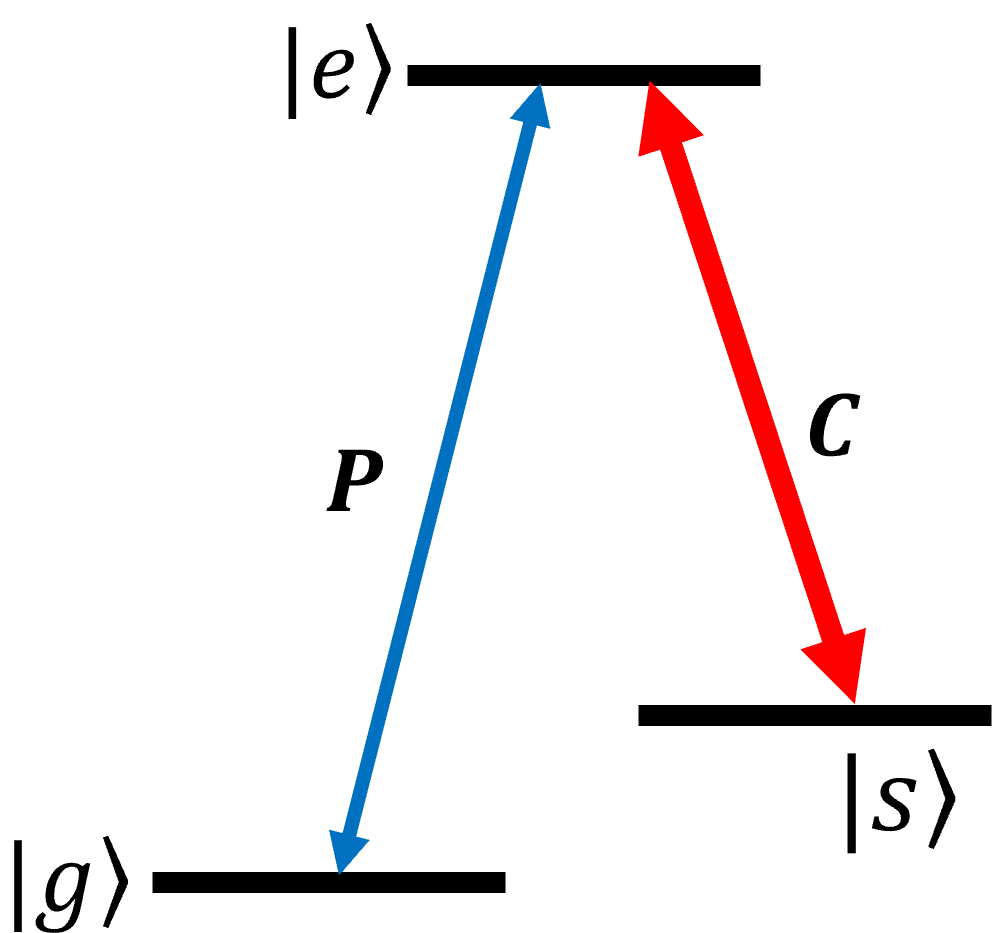}
 		\label{EIT-approche1}
 	}
 	\subfigure[]
 	{
 		\includegraphics[width=0.22\textwidth]{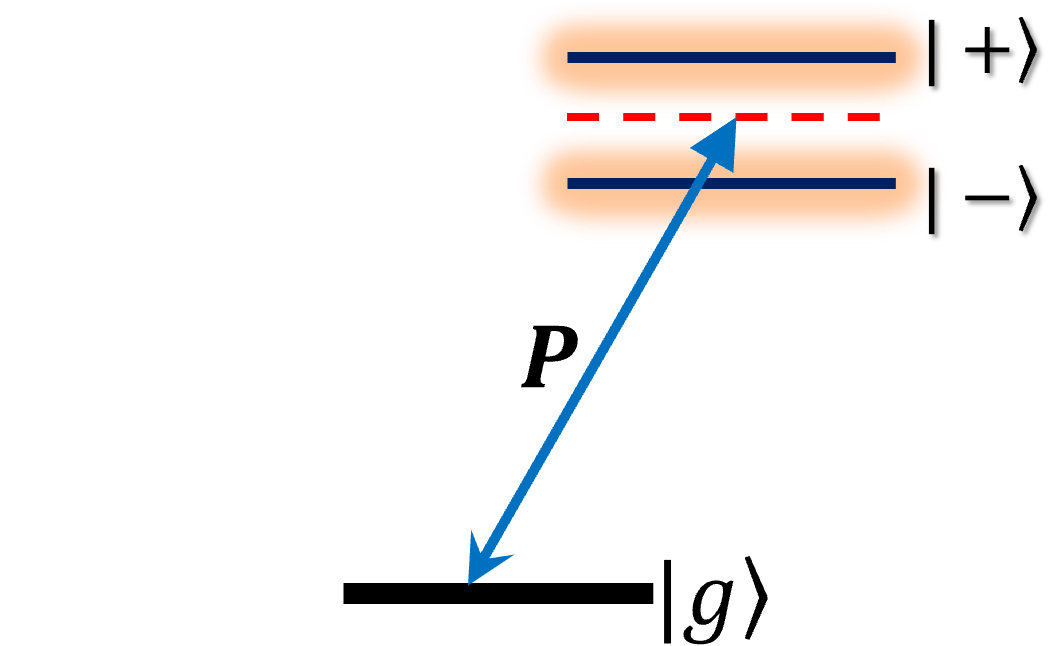}
 		\label{EIT-approche2}
 	}
 	\caption{Energy levels in an atomic system: (a) Lambda-type atom in the bare states picture. (b) $\Lambda$-type three-level atom in the dressed states picture. }
 	\label{EIT-app}
 \end{figure}

 In 1991, Boller et al., in their discussion of the first experimental observation of EIT in strontium vapor, pointed out that there are two equivalent approaches and physical pictures for observing EIT and understanding quantum interference in it. These are the dressed-state picture and the bare-state picture\cite{Boller,RevModPhys77,yong}.\\ 
 \textbf{Bare state picture:} In this approach, EIT can be achieved through different pathways between bare states. Here, the probability amplitudes between the states can destructively interfere and prevent absorption. For example, in the $\Lambda$-type atom, absorption of the $ \omega_P $ field occurs via the transition from state $ \ket{g} $ to $ \ket{e} $. In this three-level system, as shown in Fig.\ref{EIT-approche1}, the fields can directly transfer population from $ \ket{g} $ to $ \ket{e} $ or distribute it indirectly via the $ \ket{g}-\ket{e}-\ket{s}-\ket{e} $ pathway.
 Since the coupling field is stronger than the probe field, the decay rate of state \( \ket{s} \) is low, and the indirect pathway has a probability amplitude comparable in magnitude to the direct pathway. If the fields are in resonance, coupling from \( \ket{e} \) to \( \ket{s} \) leads to a phase shift of \( \pi/2 \), and coupling from \( \ket{s} \) to \( \ket{e} \) also results in a phase shift of \( \pi/2 \).
 Therefore, the probability amplitude of the indirect excitation pathway has a phase shift of $\pi$ compared to the direct excitation pathway, which leads to destructive interference and consequently causes the transitions to be canceled out. Now, if state $ \ket{s} $ has a relatively long lifetime, it leads to the emergence of a fully transparent window within the absorption line $ \ket{g}-\ket{e} $\cite{RevModPhys77,yong}.\\
 \textbf{Dressed states picture:} 
 The basis of this picture is the work of  Imamo\v{g}lu and Harris in 1989, where dressed states can be considered as two closely spaced resonances that effectively decaying to a continuum, as shown in Fig.\ref{EIT-approche2} for a $\Lambda$-type atom . The field $ \omega_C $ is driven to the system, causing the states \( \ket{e} \) and \( \ket{s} \)  to couple and resulting in two dressed states that effectively decay to the same continuum.
 If the field \( \omega_P\) exactly matches the transition frequency between \( \ket{e} \) and \( \ket{g} \), then the linear responses arising from these two dressed states will be equal  in magnitude but with opposite signs. Consequently, the response at this frequency will be canceled out due to Fano-like interference\cite{RevModPhys77,yong}.
 In this picture, EIT can be understood as a combination of Autler-Townes splitting and Fano interference between dressed states. When the strong coupling field is turned on, it dresses the excited state $\ket{e}$ and splits its energy level into two parts (the magnitude of this splitting is determined by the Rabi frequency between the two states $\ket{g}$ and $\ket{e}$). Now, the field $\omega_P$ sees two possible transitions for the two dressed states, and the transition coefficients from state $\ket{s}$ to these two dressed states, which are driven by $ \omega_P $, have different phases. Exactly at their midpoint, a destructive interference will occur. As a result, when $ \omega_P $ is tuned exactly to resonance, the excitation probability is zero, and the atomic medium becomes transparent, which is the physical basis of EIT.\vspace*{-0.5 cm}
 \subsubsection{Absorption and Dispersion}
 We analyze EIT by solving the master equation in the bare-state picture, enabling us to derive the system's response function. This approach connects the system's absorption ($\alpha$) and dispersion ($\beta$) coefficients, derived from the real and imaginary parts of the first-order susceptibility (\( \chi^{(1)} \)), to the medium's optical properties influenced by destructive interference\cite{abraham}.
 \begin{equation}
 	\chi^{(1)} = \chi' + i\chi''
 	\label{e84}
 \end{equation}
The absorption and dispersion coefficients are related to the first-order susceptibility as
 \begin{subequations}
 	\begin{align}
 		&\alpha = \frac{\omega_P n_0}{c} \chi'' = \frac{\omega_P n_0}{c} \Im{\chi^{(1)}}\\[6pt]
 		&\beta = \frac{\omega_P n_0}{2c} \chi' = \frac{\omega_P n_0}{2c} \Re{\chi^{(1)}}
 	\end{align}
 	\label{e85}
 \end{subequations} 
 \( n_0 \)  is the background refractive index, and  $c$  is the speed of light.  The susceptibility \(\chi^{(1)}\) is derived from the material's polarization, which depends on the density matrix elements describing the quantum state of the system. For a \(\Lambda\)-type atom, when driven by a field given as  $ E_P(z,t)=(\varepsilon_P e^{-i(\omega_pt-kz)}+c.c.) /2$, where \(\varepsilon_P\) is the amplitude of the driven field and \(\omega_P\) is its frequency, the susceptibility \(\chi^{(1)}\) is expressed as \cite{Xiao} 
 \begin{equation}
	\chi^{(1)} = -\frac{2N |\mu_{eg}|^2}{\hbar \epsilon_0 \Omega_P} \rho_{eg}
	\label{e98}
\end{equation}
Where, \( N \) represents the density of atoms, and the Rabi frequency associated with transitions \(\ket{g} \leftrightarrow \ket{e}\) is defined as \(\Omega_P = -{\mu_{eg} \varepsilon_P}/{\hbar}\). This formulation provides a precise framework for calculating absorption and dispersion phenomena related to electromagnetically induced transparency (EIT). It offers key insights into the optical response of the medium\cite{abraham, boyd, zub}. 
 To compute the system's response, it suffices to evaluate the density matrix element \( \rho_{eg} \) using the von Neumann equation
 \begin{equation}
 	\frac{d\hat{\rho}}{dt} = -\frac{i}{\hbar} [\hat{H}_{int}, \hat{\rho}] - \frac{1}{2} \{ \hat{\Gamma}, \hat{\rho} \}
 	\label{e99}
 \end{equation}
 where $\hat{\Gamma} $ represents spontaneous emission, with matrix elements defined as $\langle m|\Gamma_{mn}|n \rangle = \Gamma_m \delta_{mn} $ \cite{zub}. Considering the specific form of the driving field, we must modify some details in Eq.\eqref{e28} to accurately represent the Hamiltonian of a $\Lambda$-type atom. The modified Hamiltonian is given by
 \begin{equation}
 	H_\Lambda = -\frac{\hbar}{2} \left\{ 2(\Delta\hat{\sigma}_{ee} + \delta\hat{\sigma}_{ss}) + \frac{1}{2} [\Omega_P\hat{\sigma}_{ge} + \Omega_C\hat{\sigma}_{es} + \text{h.c.}] \right\}
 	\label{e102}
 \end{equation}
 where $\Delta = \omega_P - \omega_{eg}$ and $\delta = (\omega_P - \omega_C) + \omega_{es} - \omega_{eg}$. Solving Eq.\eqref{e99} in the steady-state regime yields the expression for $\rho_{eg}$ \cite{zub}.
 \begin{align}
 	\rho_{eg} = \frac{1}{2Z} \bigg[ &-\Omega_P (\delta^2 \Delta + \gamma_3^2 \Delta - \frac{\delta}{4}\Omega_C^2) \nonumber \\
 	&+ i\Omega_P (\delta^2 \gamma_1 + \gamma_1^2 \gamma_3^2 + \frac{\gamma_3}{4}\Omega_C^2) \bigg]
 	\label{e113}
 \end{align}
  where \( Z = (\delta\Delta - \gamma_1 \gamma_3 - {\Omega_C^2}/{4})^2 + (\gamma_3 \Delta + \gamma_1 \delta)^2 \),  \( \gamma_1 = (\Gamma_g + \Gamma_e)/2 \) and \( \gamma_3 = (\Gamma_g + \Gamma_s)/2 \). Substituting \( \rho_{eg} \) into the relevant equations allows calculation of absorption and dispersion for a \(\Lambda\)-type atom driven by \( \omega_P \) \cite{abraham, zub}.
  \begin{figure}[h!] 
 	\centering
 	\includegraphics[width=1\linewidth]{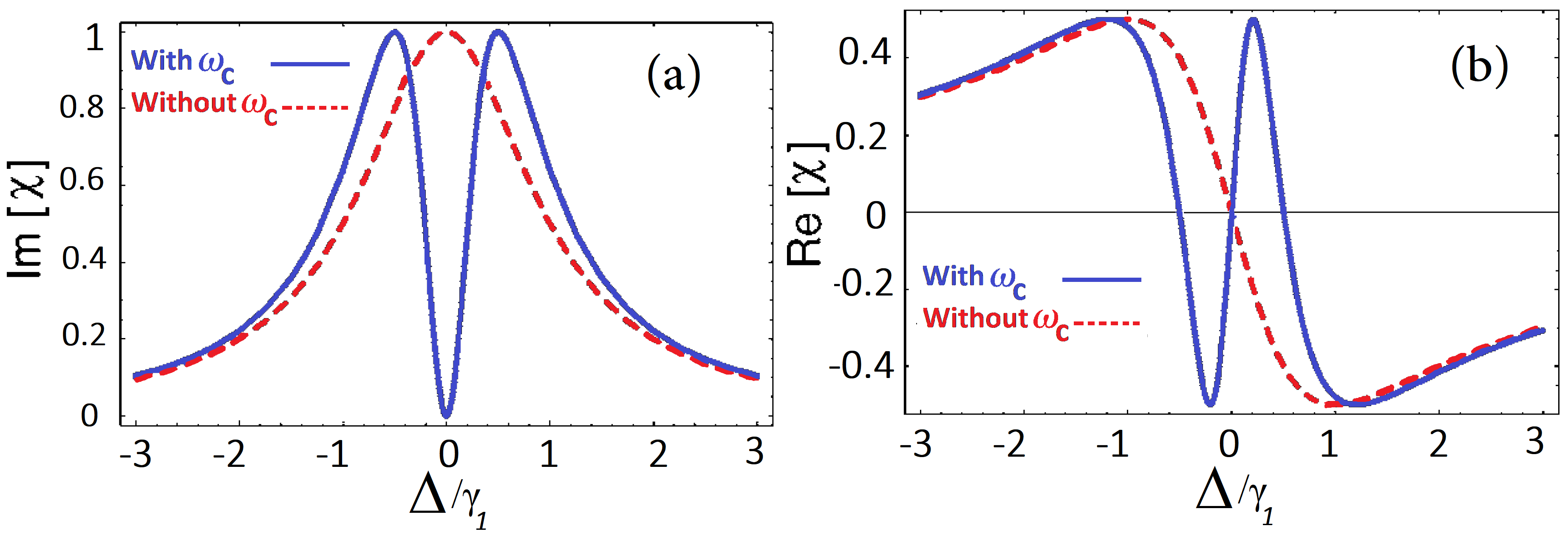}
 	\caption{(a) Absorption and (b) Dispersion as a function of the single-photon detuning, plotted without the coupling field (red line) and with the resonant coupling field (blue line). It is clear that when the coupling field is turned on, we observe the EIT effect. In fact, the plots shown with red lines appear when the strong coupling field $ \omega_C $ is zero. Otherwise (shown by the blue lines in the plots), EIT creates a transparency window in the absorption region, causing both the real and imaginary parts of the linear susceptibility to vanish. This phenomenon demonstrates the key role of destructive interference, allowing the probe field to pass through the medium without any absorption. Figure adapted from [\onlinecite{RevModPhys77}].}
 	\label{EIT-ab}
 \end{figure}
 \begin{subequations}
 	\begin{align}  
 		&\alpha=-\dfrac{N\abs{\mu_{eg}}^2n_0 \omega_P}{\hbar \epsilon_0 cZ}(\gamma_1 \delta^2 + \gamma_1^2 \gamma_3^3+ \dfrac{\gamma_3}{4}\Omega_C^2)	\label{e114-1}\\[6 pt]
 		&\beta=\dfrac{N\abs{\mu_{eg}}^2n_0 \omega_P}{2\hbar \epsilon_0 cZ}(\delta^2 \Delta +\gamma_3^2 \Delta -\dfrac{\Omega_C^2}{4}\delta)	\label{e114-2}
 	\end{align}
 	\label{e114}
 \end{subequations}
Eq.\eqref{e114} and Fig.\ref{EIT-ab} showcase a remarkable phenomenon known as Electromagnetically Induced Transparency. Under ideal conditions, where the decay rate \( \gamma_3 \) is zero and perfect two-photon resonance (\( \delta = 0 \)) is achieved, the real and imaginary parts of the linear susceptibility vanish. This complete cancellation highlights the crucial role of \textbf{destructive interference}. With no net response from the medium, the probe field (\( \omega_P \)) experiences perfect transparency. Interestingly, modifying the Rabi frequency (\( \Omega_C \)) of the coupling laser only affects the absorption profile, not the overall transparency effect driven by destructive interference \cite{zub}.\\
We can observe this destructive interference and consequent transparency in \(\Xi\) and V-type systems. For example, the coefficients for absorption (\(\alpha\)) and dispersion (\(\beta\)) in a \(\Xi\) or  V-type system are given by\cite{Marangos, Harris1990}:
\begin{subequations}
	\begin{align}
		&\alpha = -\frac{N|\mu_{eg}|^2 n_0 \omega_P}{\hbar \epsilon_0 c Y} \Big[8\delta^2 \Gamma_e + 2\Gamma_s(|\Omega_C|^2 + \Gamma_e\Gamma_s)\Big] \label{ee114-1}\\[12pt]
		&\beta = \frac{N|\mu_{eg}|^2 n_0 \omega_P}{2\hbar \epsilon_0 c Y} \Big[-4\delta(|\Omega_C|^2 - 4\delta\Delta_2) + 4\Delta_1\Gamma_s^2 \Big] \label{ee114-2}
	\end{align}
	\label{ee114}
\end{subequations}
where \( Y = (4\delta\Delta_1 - \Gamma_e\Gamma_s - |\Omega_P|^2)^2 - 4(\Gamma_s\Delta_1 + \Gamma_e\delta)^2 \) and \(\delta = \Delta_1 + \Delta_2\) is the two-photon detuning for the \(\Xi\)-type atom. Also, \(\Delta_{1,2}\) are the single-photon detunings defined in Eq.\eqref{e27}.  With attention to these relationships, it is observed that in a ladder-type atomic system, the absorption at exact two-photon resonance (\(\delta = 0\)) reaches a vanishing value due to destructive interference\cite{Marangos}. This phenomenon is a characteristic feature of EIT, where the presence of a coupling laser induces quantum interference between different excitation pathways.\\
For a comprehensive exploration of EIT in \(\Xi\)-type atoms, refer to Refs. [\onlinecite{boyd, Marangos, Harris1990, Julio, abraham, Pittman, ZHANG20152495, Su_2019}].

$\Lambda$- and  \(\Xi\)-type atomic configurations are more commonly employed for achieving narrow linewidths in EIT compared to V-type systems. This preference arises because the EIT linewidth in V-type systems can be comparable to the inherent atomic linewidths of the transitions involved. This similarity makes it challenging to isolate the contribution arising from the EIT-induced coherence effect [Hyun]. Consequently, identifying the presence of EIT in V-type systems becomes more difficult\cite{Hoshina14}. Crucially, the excited-state lifetime plays a key role in determining EIT linewidths, with longer lifetimes leading to narrower linewidths.\\
However, for a V-type atomic system, we can calculate absorption (\(\alpha\)) and dispersion (\(\beta\)) coefficients from the density matrix equations. A V-type system (Fig.\ref{tls2}) consists of two upper states \(|e\rangle\) and \(|s\rangle\) coupled to a common ground state \(|g\rangle\) by probe and coupling lasers, respectively.\\
The absorption and dispersion formulas are typically derived from the imaginary and real parts of the coherence term \(\rho_{ge}\) in the density matrix (as observed from Eq.\eqref{e85} and Eq.\eqref{e98}). Specifically, the matrix element responsible for probe absorption is \(\rho_{ge}\). After solving the master equation in the steady-state regime, we find \cite{Hoshina14,Zhao2002,Kang:17}
\begin{equation}
\rho_{ge}=\dfrac{\Omega _P}{2 K_2 \left(4 K_0 K _1-\Omega _C^2\right)}\Big[{\rho _{ss} \Omega _C^2}+{\rho _{gg} \left(4 K_0 K _2-\Omega _C^2\right)}\Big]
\end{equation}
where the steady-state populations are given by
\begin{align}
		\rho_{ss}= \dfrac{s}{K_3} ,\ \
		\rho_{gg}= \dfrac{2 + s + {2\Gamma_t}/{\Gamma_s}}{K_3}
	\end{align}
and the other coefficients are
\begin{subequations}
	\begin{align}
		K_0 &= \Delta_P - \Delta_C + i\gamma_{es} \\[10pt]
		K_1 &= \Delta_P + i\gamma_{ge} \\[10pt]
		K_2 &= \Delta_C - i\gamma_{gs}\\[10pt]
		K_3 &=2 \left[1 + \frac{\Gamma_t}{\Gamma_s} + s \left(1 + \dfrac{\gamma}{2}\right)\right]\\[10pt]
		s &= \dfrac{\Omega_C^2 / (\gamma_{gs} \Gamma_s)}{1 + (\Delta_C / \gamma_{gs})^2} \\[10pt]
		\gamma &= \dfrac{\Gamma_s - \Gamma_{sg}}{\Gamma_t}.
	\end{align}
\end{subequations}

In the above relations, single-photon detunings are represented by \(\Delta_P = \omega_P - \omega_{ge}\) and \(\Delta_C = \omega_P - \omega_{gs}\), where \(\omega_P\) is the probe field frequency and \(\omega_{ge}\) and \(\omega_{gs}\) are the transition frequencies for a V-type atom. Decay rates for the excited states \(|e\rangle\) and \(|s\rangle\) are denoted by \(\Gamma_e\) and \(\Gamma_s\), respectively, while the ground state decay rate (\(\Gamma_g\)) is assumed to be zero. The system can be either open or closed. Here, \(\Gamma_{ig}\) (\(i = e, s\)) represents the decay rate from state \(|i\rangle\) to the ground state \(|g\rangle\). Additionally, the transverse decay rate between states \(|i\rangle\) and \(|j\rangle\) (\(i, j = g, e, s\)) is given by \(\gamma_{ij} = {\Gamma_i + \Gamma_j}/{2}\).\\
The transit time decay rate, \(\Gamma_t\), accounts for the finite time atoms spend interacting with the laser beam. This parameter is employed when solving the density matrix equations in the steady-state regime\cite{Kang:17,Zhao2002}. \\
As a result, for a V-type atom, EIT occurs, which can be observed from the equations described. The destructive interference reduces the absorption of the probe laser at the resonance frequency. This interference between the transitions driven by the probe and coupling lasers is essential for EIT.\\
For further study on EIT in V-type atoms, one can refer to Refs.[\onlinecite{Kang:17,Zhao2002,Hoshina14,Lazoudis,Qin_2022,Fulton,Dong2023,Welch1998,Echaniz,Kumar,Boon,Zhao2002,Higgins,Bharti}].
\subsection{Interference with Giant Atoms}
  The traditional paradigm in quantum optics assumes that atoms interact with light as point-like entities. However, recent breakthroughs in the field have shattered this long-held belief. Artificial atoms, meticulously constructed from superconducting circuits, exhibit a revolutionary property: the ability to couple with an electromagnetic field at multiple points separated by a full wavelength of light. This unique characteristic, absent in conventional quantum optics with ordinary atoms, paves the way for fascinating interference effects.
  This section delves into these groundbreaking systems, aptly named "giant atoms," and explores their profound implications for light-matter interactions\cite{Anton2020}.
  
  Quantum optics traditionally relies on the dipole approximation, which treats atoms as point-like entities when interacting with light at optical frequencies.\\
  This simplification stems from the significant size disparity between atoms (radius \( \approx 10^{-10} \) m) and the wavelength of light (\( \lambda \approx 10^{-6} - 10^{-7} \) m)~\cite{Leibfried2003}. Even for larger Rydberg states (radius \( \approx 10^{-8} - 10^{-7} \) m), quantum optics experiments typically involve microwave radiation with a much longer wavelength (\( \lambda \approx 10^{-2} - 10^{-1} \) m)~\cite{Haroche2013}. This vast difference in scales justifies the dipole approximation (\( r \ll \lambda \)) within theoretical frameworks, facilitating the description of light-matter interactions\cite{Anton2020}.\\
  However, recent exploration of artificial atoms in quantum optics has challenged this long-held assumption. Artificial atoms, encompassing engineered systems like quantum dots~\cite{Hanson2007} and superconducting qubits (qubits)~\cite{Xiang2013, GU2017}, emulate key properties of natural atoms. While the circuits of superconducting qubits can reach larger sizes (radius \( \approx 10^{-4} - 10^{-3} \) m), they remain dwarfed by the wavelength of the microwave fields they interact with.
  
  A groundbreaking experiment in 2014~\cite{Gustafsson2014} necessitated a re-evaluation of the dipole approximation. This experiment involved coupling a superconducting transmon qubit to surface acoustic waves (SAWs)~\cite{Anton2020,Koch2007}.\\
  Due to the slower propagation velocity of SAWs, their wavelength (\( \lambda \approx 10^{-6} \) m) became comparable to the size of the qubit. 
  
 Additionally, the qubit's interdigitated capacitance layout, consisting of a series of interleaved conductive fingers, was specifically designed to function as an effective transducer for SAWs. This configuration provided the necessary electric field coupling to the propagating acoustic waves and enabled interaction at multiple distinct points along the wave’s path. These coupling points were precisely spaced at intervals of \( \lambda/4 \), leveraging the periodic structure to maximize constructive interference and ensure coherent coupling between the qubit and the SAWs. This arrangement allowed the qubit to experience the wave’s spatially varying field at well-defined intervals, a critical factor for observing frequency-dependent interference effects characteristic of systems where quantum emitters couple at multiple points to a propagating mode\cite{Anton2020}.\\
 
\begin{figure}[h!]
  		\includegraphics[width=8.5cm]{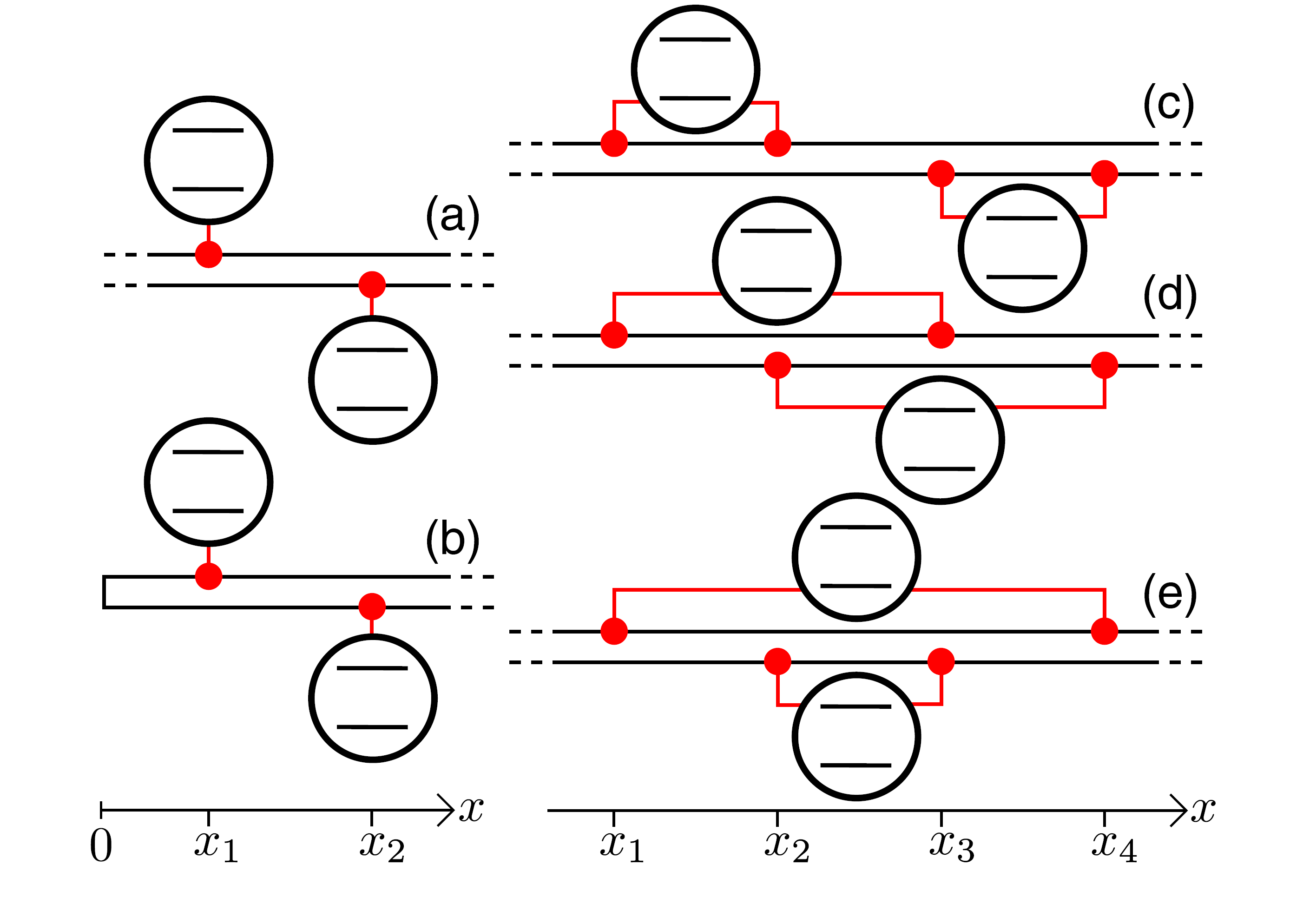}
  		\centering 
  		\caption{Illustration of atomic coupling configurations to a waveguide (transmission line): (a) two small atoms coupled to an open transmission line, (b) two small atoms coupled to a semi-infinite transmission line, (c) two separate giant atoms with independent coupling points, (d) two braided giant atoms with interwoven coupling points, and (e) two nested giant atoms where one atom’s coupling points are fully contained within the other’s. Red circles mark the coupling points, with the leftmost atom labeled '\textbf{a}' and the other '\textbf{b}'. Red lines depict the connections between the  transmission line and the coupling points. Two lines within the circles symbolize the internal structure of the atoms. Figure adapted from [\onlinecite{PhysRevLett.120.140404}]. }
  			\label{giant1}
  		\end{figure}

  Unlike small atoms, which couple to a waveguide at a single point due to their compact size, giant atoms couple at multiple points, with significant distances between these points relative to the wavelength of the waveguide modes. This wavelength, \( \lambda \), is determined by the atomic transition frequency \( \omega_a \) and the propagation velocity \( v \) in the waveguide, given by \( \lambda = 2\pi v / \omega_a \) \cite{Anton2020}.
  This groundbreaking experiment has inspired theoretical investigations into a new paradigm of giant atoms\cite{Kockum2014}. The defining feature of giant atoms is their multiple coupling points, which give rise to novel interference effects \cite{Anton2020}.
  \begin{figure}[h!]
  	\includegraphics[width=8cm]{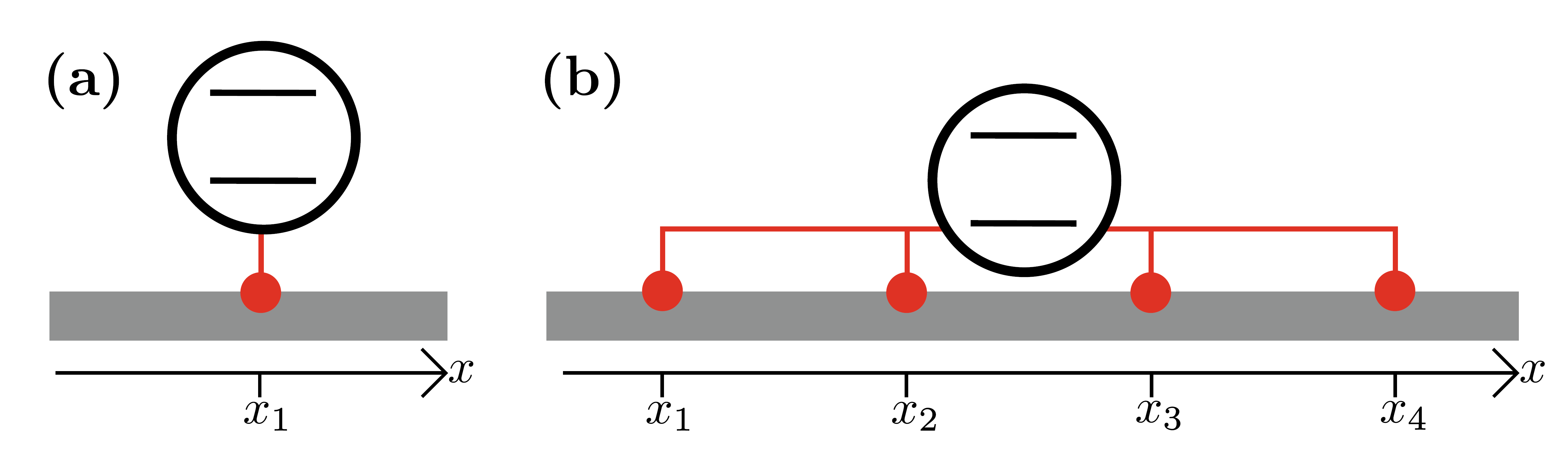}
  	\centering 
  	\caption{Illustration and comparison of coupling mechanisms in a 1D waveguide for a small atom and a giant atom. (a) A small atom couples to the waveguide, represented by a gray surface, at a single point marked by a red line (\(x_1\)). The coupling is localized, and the atom can be treated as a point-like object. (b) A giant atom couples to the waveguide at multiple points, each marked by a red line (\(x_k\)), with separations \( |x_k - x_n| \) comparable to the wavelength of the waveguide modes. These distributed coupling points introduce phase shifts and interference effects, resulting in frequency-dependent behaviors and novel interference phenomena. The internal structure of the atoms is symbolized by two lines within the circles. Figure adapted from [\onlinecite{Anton2020}].}
  		\label{giant0}
  	\end{figure}
  	As shown in Fig.\ref{giant0}, small and giant atoms exhibit distinct waveguide coupling behaviors. Small atoms, modeled as point-like objects, couple at a single point, while giant atoms, with extended spatial profiles, couple at multiple points separated by distances significant relative to the wavelength \( \lambda \) of the waveguide modes.  Fig.\ref{giant1} illustrates various atom-waveguide configurations: two small atoms coupled to an open transmission line, two small atoms coupled to a semi-infinite transmission line, two separate giant atoms, two braided giant atoms, and two nested giant atoms\cite{Kockum2014,PhysRevLett.120.140404}.
  For the observation of interference effects, we analyze the master equation for a giant atom. Giant atoms are typically implemented in waveguide QED, where a continuum of bosonic modes propagates within a one-dimensional waveguide and interacts coherently with atoms coupled to this waveguide\cite{Anton2020,GU2017}.\\
  We begin by considering the total system Hamiltonian (using units where \( \hbar = 1 \))
\begin{align}
  H = H_{\rm a} + H_{\rm wg} + H_{\rm I},
\end{align}
  where
  \( H_{\rm a} \) is the atomic Hamiltonian, representing the internal energy levels
\begin{align}
  H_{\rm a} = \sum_m \omega_m \ketbra{m}{m},
  \end{align}
  with \( m \) labeling atomic levels and \( \omega_m \) their corresponding energies.
  \( H_{\rm wg} \) is the waveguide Hamiltonian, describing the bosonic modes
\begin{align}
  H_{\rm wg} = \sum_j \omega_j \left( a^\dag_{{\rm R}j} a_{{\rm R}j} + a^\dag_{{\rm L}j} a_{{\rm L}j} \right),
\end{align}
  where \( j \) indexes the waveguide modes, \( \omega_j \) are their frequencies, and \( a^\dag \) and \( a \) are the creation and annihilation operators for right-moving (R) and left-moving (L) modes.
 \( H_{\rm I} \) is the interaction Hamiltonian, describing the coupling between the atom and waveguide modes
  \begin{align}
  	H_{\rm I} &= \sum_{j, k, m} g_{jkm} \left(\hat{ \sigma}_-^{(m)} +\hat{ \sigma}_+^{(m)} \right) \\
  	&\quad \times \left( a_{{\rm R}j} e^{-i \omega_j x_k / v} + a_{{\rm L}j} e^{i \omega_j x_k / v} + a^\dag_{{\rm R}j} e^{i \omega_j x_k / v} + a^\dag_{{\rm L}j} e^{-i \omega_j x_k / v} \right)\nonumber
  \end{align}
  The key difference for giant atoms lies in this interaction Hamiltonian, which includes multiple coupling points between the giant atom and the waveguide, leading to interference effects.\\
  The atomic levels \( m = 0, 1, 2, \ldots \) have energies \( \omega_m \), and are connected through lowering and raising operators \(\hat{ \sigma}_-^{(m)} = \ketbra{m}{m+1} \) and \(\hat{ \sigma}_+^{(m)} = \ketbra{m+1}{m} \). The phase factors \( e^{\pm i \omega_j x_k / v} \) in the interaction Hamiltonian, absent for small atoms, are responsible for these interference effects\cite{Anton2020}.\\
  \begin{figure}[h!]
  	\includegraphics[width=8cm]{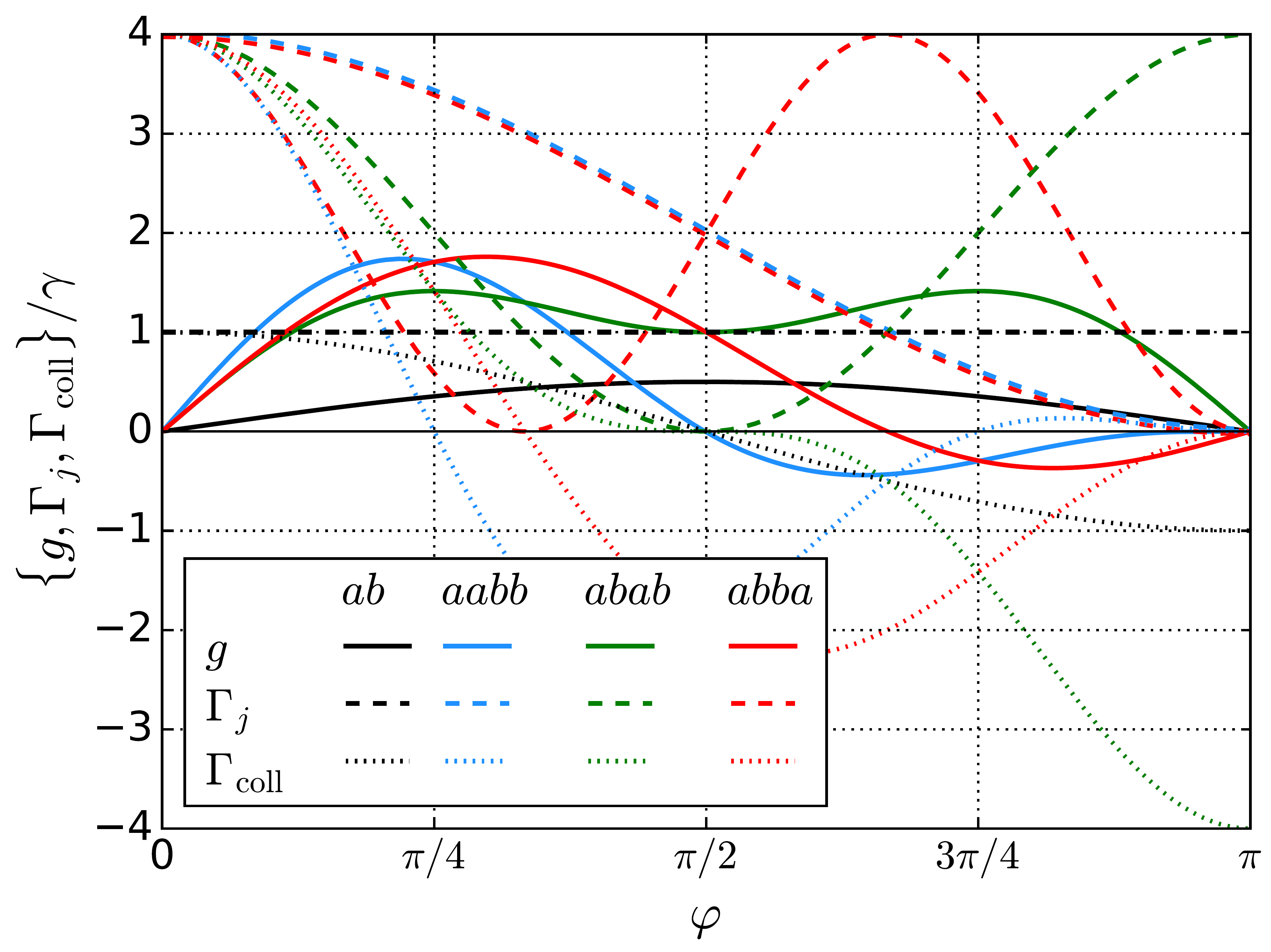}
  	\centering 
  	\caption{Exchange interaction (solid lines) and decoherence rates (individual: dashed lines; collective: dotted lines) as functions of \( \phi \) for the setups illustrated in Fig.\ref{giant1}, where \( \phi = k |x_{j+1} - x_j| \) represents the phase acquired when traveling between consecutive connection points. Here, \( k = \omega_a / v \) is the wave number, with \( v \) as the velocity of the modes in the waveguide. This phase factor allows interference effects between points, which can suppress decoherence under certain conditions, creating decoherence-free subspaces. Labels denote the ordering of connection points for the two atoms:  \textbf{ab}  (small atoms, black),  \textbf{aabb}  (separate giant atoms, blue),  \textbf{abab}  (braided giant atoms, green), and \textbf{abba} (nested giant atoms, red). The case of small atoms in a semi-infinite waveguide [Fig. \ref{giant1}\textcolor{blue}{(b)}] is not shown separately, as it is qualitatively similar to the nested giant atoms. Note that two red dashed lines represent \( \Gamma_a \) and \( \Gamma_b \) individually. Figure adapted from [\onlinecite{PhysRevLett.120.140404}]}
  	\label{giant2}
  \end{figure}
  
	Kockum et al. (2018) investigated a setup where multiple giant atoms couple at discrete points along a one-dimensional (1D) waveguide (see Ref.[\onlinecite{PhysRevLett.120.140404}]), demonstrating that interference is essential to achieving decoherence-free interactions. \\
	Their results (Fig.\ref{giant2}) show that braided giant atoms, in particular, exploit interference effects to control relaxation rates and coupling strengths (see Fig.\ref{giant1}). The phase \( \phi \) acquired between connection points allows for non-zero exchange interactions while suppressing decoherence, an effect unachievable with conventional small atoms.\\
	
	In this setup, phase-sensitive interference between coupling points leads to the cancellation of emissions, suppressing decay channels and protecting the atomic system from energy loss to the waveguide. Key parameters include the exchange interaction strength  $\ket{g}$, the phase \( \phi \), and the relaxation rates \( \Gamma_j \) and \( \Gamma_{\text{coll}} \), both of which can vanish under certain configurations, further enhancing decoherence-free conditions through destructive interference.\\
	
	In the braided configuration, interference permits non-zero  $\ket{g}$ with \( \Gamma_j = 0 \) (where \( \Gamma_j \) is the individual relaxation rate for atom \( j \)) when \( \phi = (2n + 1) \pi \) (for integer \( n \)). This phase condition causes emissions from one atom’s connection points to interfere destructively, while allowing coherent absorption at the other atom’s points. This engineered interference effectively isolates the system from decoherence while enabling stable quantum interactions, positioning braided giant atoms as a powerful tool for robust quantum operations\cite{PhysRevLett.120.140404}.\\
	
  For more details about giant atoms, refer to Refs.[\onlinecite{Anton2020,Gustafsson2014,Kockum2014,GU2017,Joshi2023,PhysRevLett.120.140404}].
\section{\label{IAWT}Interference of atoms with themselves}
Interferometry using atoms and even molecules has been known for a long time. Atomic interferometry was introduced in 1973 by Franz and Altschuler, and since then, it has been widely studied\cite{Berman}. Atom interferometry is the art of coherently manipulating the translational motion of atoms. Motion here refers to the displacement of the center of mass, and coherence relates to the phase of the de Broglie wave representing this motion. The primary result of this coherence is interference, which is most effectively utilized in interferometers. In these devices, atom waves can travel through two or more alternate paths, producing an interference pattern that is scientifically valuable\cite{RevModPhys.81.1051}.\\
In this section, we will explore the phenomenon of atomic self-interference, where atoms exhibit wave-like properties even when treated as individual particles. This phenomenon is prominently observed in atom interferometry experiments, where the resulting interference pattern arises from the superposition of different quantum states of the atoms. Atomic interferometry provides extensive information about the internal structure of atoms and their properties, such as mass, magnetic moment, and absorption frequencies\cite{Berman}.

We can catalog interferometers according to their features, such as internal state changing interferometers, time domain vs. space domain interferometers, atom traps and waveguides, and others\cite{RevModPhys.81.1051}. Here, to explore the interference of atoms with themselves, we will focus on internal state changing interferometers. These interferometers use beam splitters that change an atom's internal state, similar to a polarizing beam splitter in optics, with stimulated Raman transitions playing a key role in this process.

Atomic interferometry forms the basis for a new generation of quantum sensors, enabling ultra-precise measurements of fundamental physical constants such as the fine structure constant ($\alpha$) and the gravitational constant ($G$). For instance, in 2009, Müller et al. utilized atomic interferometry to perform a highly precise measurement of gravitational redshift, demonstrating a deviation of \(7 \times 10^{-9}\) from general relativity's predictions\cite{redshift}. This technique's exceptional sensitivity was further highlighted by Asenbaum et al. (2020), who achieved a deviation of \(10^{-12}\) in their test of the equivalence principle\cite{Equivalence}.\\
Atomic interferometry is also highly effective as an internal sensor in accelerometers and gravimeters. Because atoms have mass, atom interferometers are significantly more sensitive to external forces than optical interferometers of comparable arm length, generating significant interest in using atoms for precision measurements.\\
Advances in laser cooling and trapping now allow for the routine preparation of atomic ensembles at microkelvin temperatures. At these temperatures, both the internal and motional states of the atoms can be precisely controlled using microwave and optical manipulation techniques\cite{Ravi2020}. For precision measurement purposes, such as atomic clocks, this requires very slow (cold or ultracold) atoms\cite{Seidel2007}. \\
We will then discuss cold atom interferometers, which can be classified into two basic types based on different Raman pulse sequences: the Ramsey-Bordé interferometer and the Mach-Zehnder interferometer. Other atom interferometers are considered modifications of these two types\cite{Wang_2015}.\\
Unlike optical interferometry, which employs mirrors and beam splitters, atomic  interferometry manipulates the trajectories of atoms using interactions between atoms and laser light. This technique shares core principles with other interferometers and follows a five-step process\cite{RevModPhys.81.1051}.\\
\textbf{1.\hspace*{0.15cm}Preparation (including Laser Cooling): }The initial state of the atom cloud is carefully prepared. This often involves laser cooling the atoms to ultracold temperatures (microkelvin range) to achieve high coherence and minimize thermal noise\cite{Chu98, Wang_2015,Seidel2007}.\\
\textbf{2.\hspace*{0.15cm}Splitting (using Stimulated Raman Transitions):} The atomic wavefunction is coherently divided into multiple paths. A common technique for splitting utilizes stimulated Raman transitions. This process, illustrated in Fig.\ref{pp3} using a three-level atom model, involves single-mode laser fields that couple two ground states and an excited state, inducing a coherent transition and effectively splitting the wavefunctions\cite{Wang_2015}. \\
\textbf{3.\hspace*{0.15cm}Interaction:} The split wavefunctions experience potentially different interactions due to their distinct spatial locations. These interactions may be influenced by external factors like gravity, magnetic fields, or specific light fields\cite{RevModPhys.81.1051}.\\
\textbf{4.\hspace*{0.15cm}Recombination:} The separated wavefunction components are coherently recombined\cite{RevModPhys.81.1051}.\\
\textbf{5.\hspace*{0.15cm}Detection:} The phase shift of the resulting interference pattern is measured. By analyzing this phase shift, we can extract information about the external influences that affected the split wavefunctions\cite{Weiss1993,RevModPhys.81.1051}.\\
Crucially, during the splitting and recombination stages (steps 2 and 4), the atomic wavefunction maintains its coherence. This allows the matter waves associated with the split paths to interfere with each other when recombined, creating observable patterns with regions of both constructive and destructive interference. Analyzing this interference pattern is the heart of atomic interferometry, revealing information about the external influences the split wavefunctions experienced during step 3.\\
In this section, we will examine two types of atom interferometers commonly used to observe quantum interference: the Mach-Zehnder interferometer and the Ramsey-Bordé interferometer. For a comprehensive list of atomic interferometers, please refer to Ref.[\onlinecite{RevModPhys.81.1051}].
\subsection{\label{mach}Mach-Zehnder Interferometer}
\begin{figure}[h!]
	\includegraphics[width=8cm]{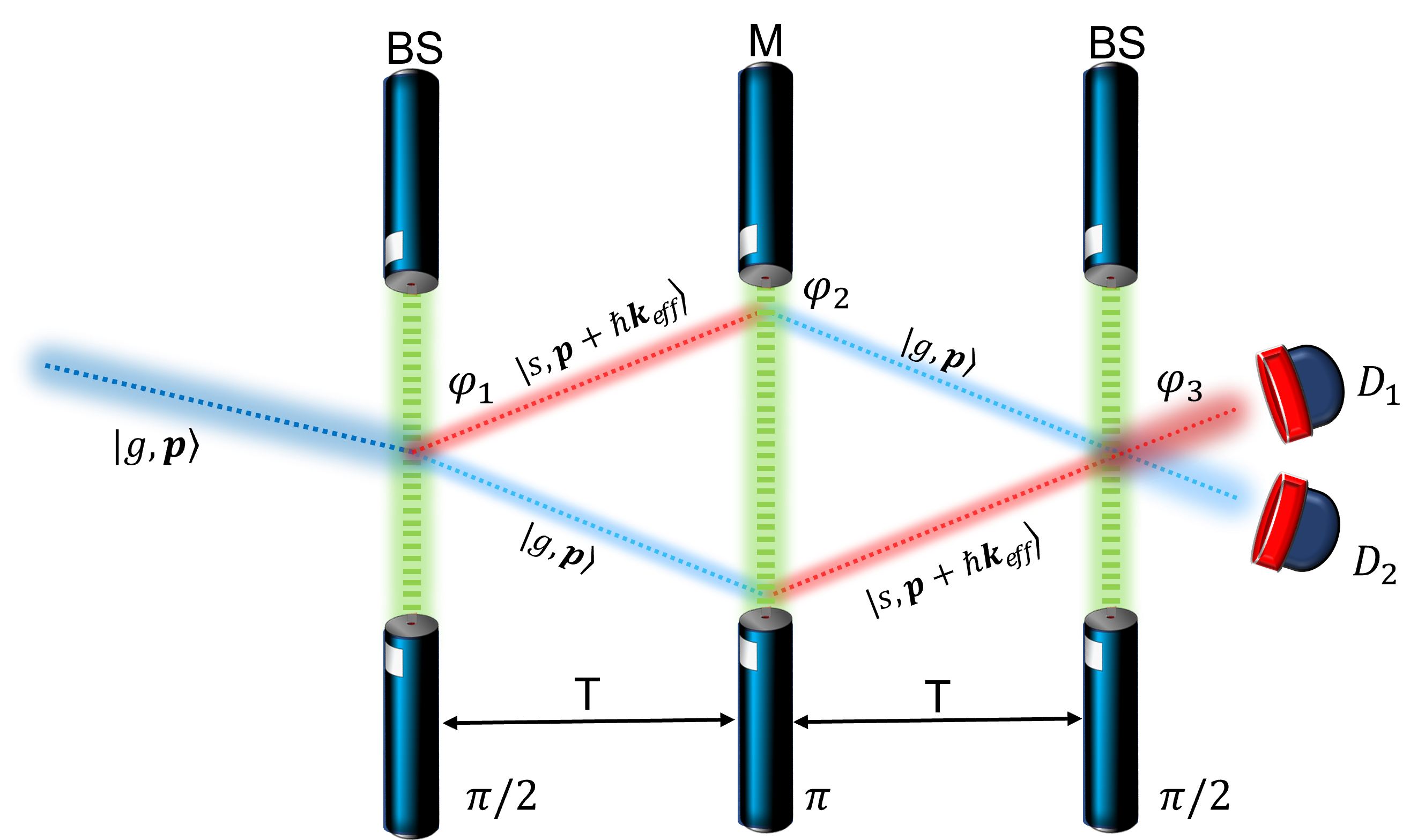}
	\centering 
	\caption
	{\textbf{An illustration of an atomic Mach-Zehnder interferometer:} Atoms, initially prepared in the ground state $\ket{g,\boldsymbol{p}}$, enter the interferometer. Passing through three $\pi/2 - \pi - \pi/2$ laser pulses, each applied to the atoms with a time interval of $T$, results in the creation of atomic interference. This interference is subsequently detected by detectors $D_1$ and $D_2$.
	}
	\label{pp5}
\end{figure}

The Atomic Mach-Zehnder Interferometer (AMZI) is an advanced instrument that leverages the principles of quantum mechanics and the wave nature of atoms to perform highly precise measurements. Modeled after the optical MZI, which uses light, the atomic version employs atoms instead of photons. We employ \(\Lambda\)-type atoms for interferometry and will discuss the formation of atomic interference in these atoms and the governing relationships in more detail. The mechanism of creating atomic interference in the AMZI can be expressed as follows\\
\begin{figure}[h!]
	\includegraphics[width=3.5cm]{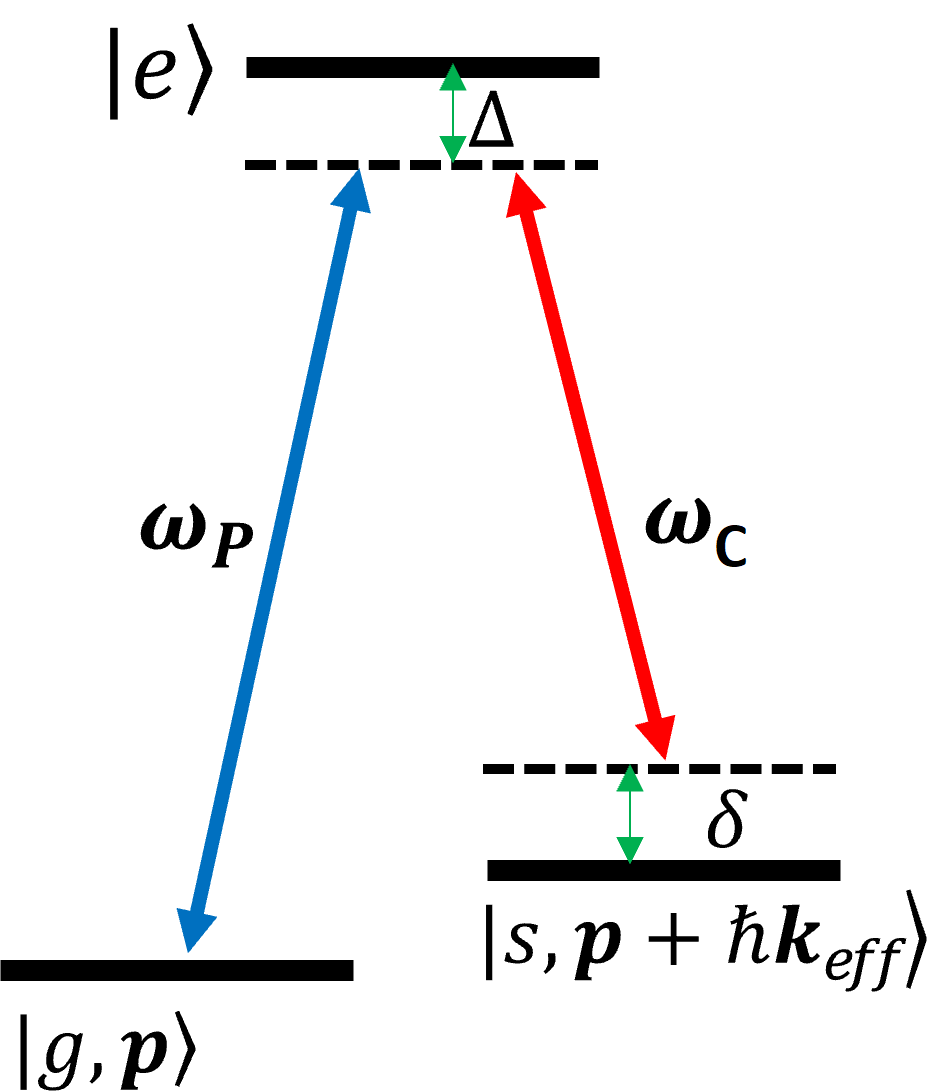}
	\centering 
	\caption
	{\textbf{Schematic of a three-level $\Lambda$-type atom:} This atom consists of two ground states$\ket{g, \textbf{p}}$ and $\ket{s ,\textbf{p}+\hbar\textbf{k}_{\text{eff}}}$ , and one excited state  $\ket{e}$. In this figure, $\Delta$ is the single-photon detuning and $\delta$ is the two-photon detuning. The atom is irradiated by two lasers with frequencies $\omega_{_\text{C}}$ and $\omega_{_\text{P}}$.
	}
	\label{pp3}
\end{figure}
\begin{figure}[h!]
	\includegraphics[width=8cm]{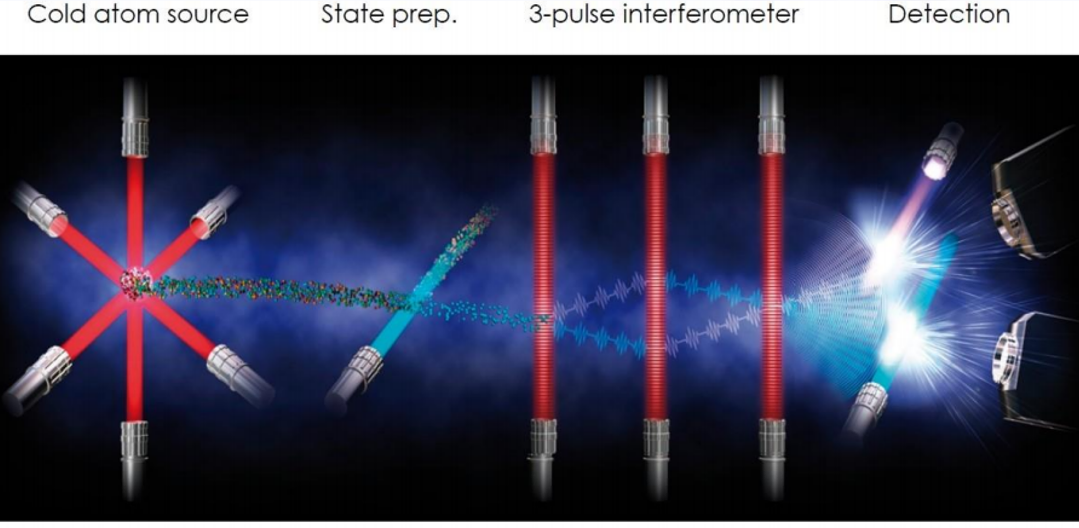}
	\centering 
	\caption
	{\textbf{A conceptual diagram of an atomic interferometer}, where atoms, after cooling and filtering, enter the Mach-Zehnder interferometer.Figure adapted from [\onlinecite{Cold}].
	}
	\label{pp4}
\end{figure}
\textbf{1.}\hspace*{0.15cm}We prepare a source of three-level atoms (Fig.\ref{pp3}) where the intermediate level can be adiabatically eliminated.\\
\textbf{2.}\hspace*{0.15cm}Laser cooling techniques (Fig.\ref{pp4}) cool the atoms\cite{Cold}.\\
\textbf{3.}\hspace*{0.15cm}A laser (Fig.\ref{pp4}) eliminates atoms not in the desired ground state\cite{Cold}.\\
\textbf{4.}\hspace*{0.15cm}Remaining atoms in the ground state ($\ket{g,\textbf{p}}$) enter the interferometer.\\
\textbf{5.}\hspace*{0.15cm}Two $ {\pi}/{2} $ laser pulses act as a beam splitter, dividing the atomic beam into two paths. One path is specific to ground-state atoms ($\ket{g,\textbf{p}}$), the other for excited-state atoms ($\ket{s ,\textbf{p}+\hbar\textbf{k}_{eff}}$). Laser pulses act as a rotation operator (e.g., rotation along y-axis with $\ket{s}=(0 \ 1)^T$ and $\ket{g}=(1 \ 0)^T$)\cite{Cold,fox,Hu2018}.\\
\begin{align} 
	R_y(\theta)=
	\begin{bmatrix}
		\cos \theta/2 & -\sin\theta/2 \\
		\sin \theta/2 & \cos \theta/2 \\
	\end{bmatrix}
	\label{zzz23}
\end{align}
If ${\theta=\pi/2}$, then 
\begin{align} 
	R_y(\dfrac{\pi}{2})\ket{g}=\dfrac{1}{\sqrt{2}}\big(\ket{g}+\ket{s}\big)
	\label{zz23}
\end{align}
After applying the pulse, the initial state is in a superposition of the ground and excited states. Each of these two states will traverse separate paths. Therefore, the $\pi/2$ pulse acts as a \textbf{50-50 beam splitter}. It is important to note that the laser pulse in an atomic Mach-Zehnder interferometer induces a phase to the state \(|s\rangle\), as shown in Fig.~\ref{pp5}. Therefore, we should multiply \(|s\rangle\) by \(e^{-i\phi_1}\) in Eq.~\eqref{zz23}.\\
 In this process, we will have a stimulated Raman transition, which plays a very key role in understanding atomic interferometers \cite{steck}.
The quantity $\textbf{p}$ in $\ket{g, \textbf{p}}$ represents the initial momentum of the atom. As discussed in Sec.\eqref{ram}, after applying the first laser pulse, the atom absorbs a photon with frequency $\omega_1$ (or momentum $\hbar \textbf{k}_1$) and eventually emits it stimulatedly with frequency $\omega_2$ (or momentum $\hbar \textbf{k}_2$).
It should be noted that the momentum directions $\hbar \textbf{k}_1$ and $\hbar \textbf{k}_2$ are opposite to each other because laser beams are applied in opposite directions. Therefore, the atom's momentum in the state $\ket{s}$ must be $\textbf{p}+\hbar (\textbf{k}_1 -\textbf{k}_2)$, and the final state should be $\ket{s,\hbar \textbf{k}_{eff}}$, where $\textbf{k}_{eff}=\textbf{k}_1 -\textbf{k}_2$ \cite{Garrido}.
Thus, by driving the atom with two $\pi/2$ laser pulses in opposite directions, we put the atom in a superposition of two states $\ket{g,\textbf{p}}$ and $\ket{s , \textbf{p}+\hbar\textbf{k}_{eff}}$. After exiting the laser pulses, the state $\ket{g,\textbf{p}}$ almost continues its initial path (which it entered the interferometer with). However, the excited state $\ket{s , \textbf{p}+\hbar\textbf{k}_{eff}}$, after receiving a kick due to the momentum $\hbar\textbf{k}_{eff}$, changes its state. Thus, we see how the stimulated Raman transition plays a prominent role in atomic interferometry \cite{Cold}.
After a time interval $T$, two new laser pulses, called $\pi$ pulses, are applied to the atom(s). We saw that pulses act as rotation operators. Let's examine how the $\pi$ pulse affects each of the states.
\begin{align} 
	R_y(\pi)\ket{g}=\ket{s} \  , \ R_y(\pi)\ket{s}=-\ket{g} 
	\label{z23}
\end{align}
We observe that after applying the $\pi$ pulses, the state \(|g\rangle\) changes to \(|s\rangle\) and the state \(|s\rangle\) changes to \(|g\rangle\). Therefore, we conclude that the $\pi$ pulses act like perfect \textbf{mirrors}, causing a complete reversal of the atomic state. Here, we should use \(e^{i\phi_2}\) to the states.\\
\textbf{6.}\hspace*{0.15cm}In the final step, after a time interval $T$ from the $\pi$ pulse, we apply another set of ${\pi}/{2}$ pulses to the atom(s). These pulses put each of the states $\ket{g}$ and $\ket{s}$ into a superposition, resulting in atomic interference. To calculate the probability of finding an atom in the excited state $\ket{s}$, we need to sum the probability amplitudes of the state $\ket{s}$ in the upper and lower paths and calculate the probability as follows \cite{Berman}.
\begin{align} 
	P_s=\vert C^{^{up}}_{s,\textbf{p}+\hbar \textbf{k}_{eff}}+C^{^{down}}_{s,\textbf{p}+\hbar \textbf{k}_{eff}}\vert^2=\dfrac{1}{2}[1- \cos (\Delta \varphi_L)]
	\label{z24}
\end{align}
The quantities $ \Delta\varphi_L=\varphi_1 -2\varphi_2 +\varphi_3 $ represent the phases associated with the applied Raman lasers at different times. As shown in Fig.\ref{pp5}, these phases correspond to the time range from $ t=0 $ (with phase $ \varphi_1 $) to $ t=2T $ (with phase $ \varphi_3 $)\cite{Berman}. We can express $ \Delta\varphi_l=k_{eff}aT^2 $ and, considering the relationship between phase and position according to the equation $\varphi=kx$, by determining the probability \( P_s \), we can calculate the gravitational acceleration.
\subsection{Ramsey-Bordé interferometer}
\begin{figure}[h!]
	\includegraphics[width=8cm]{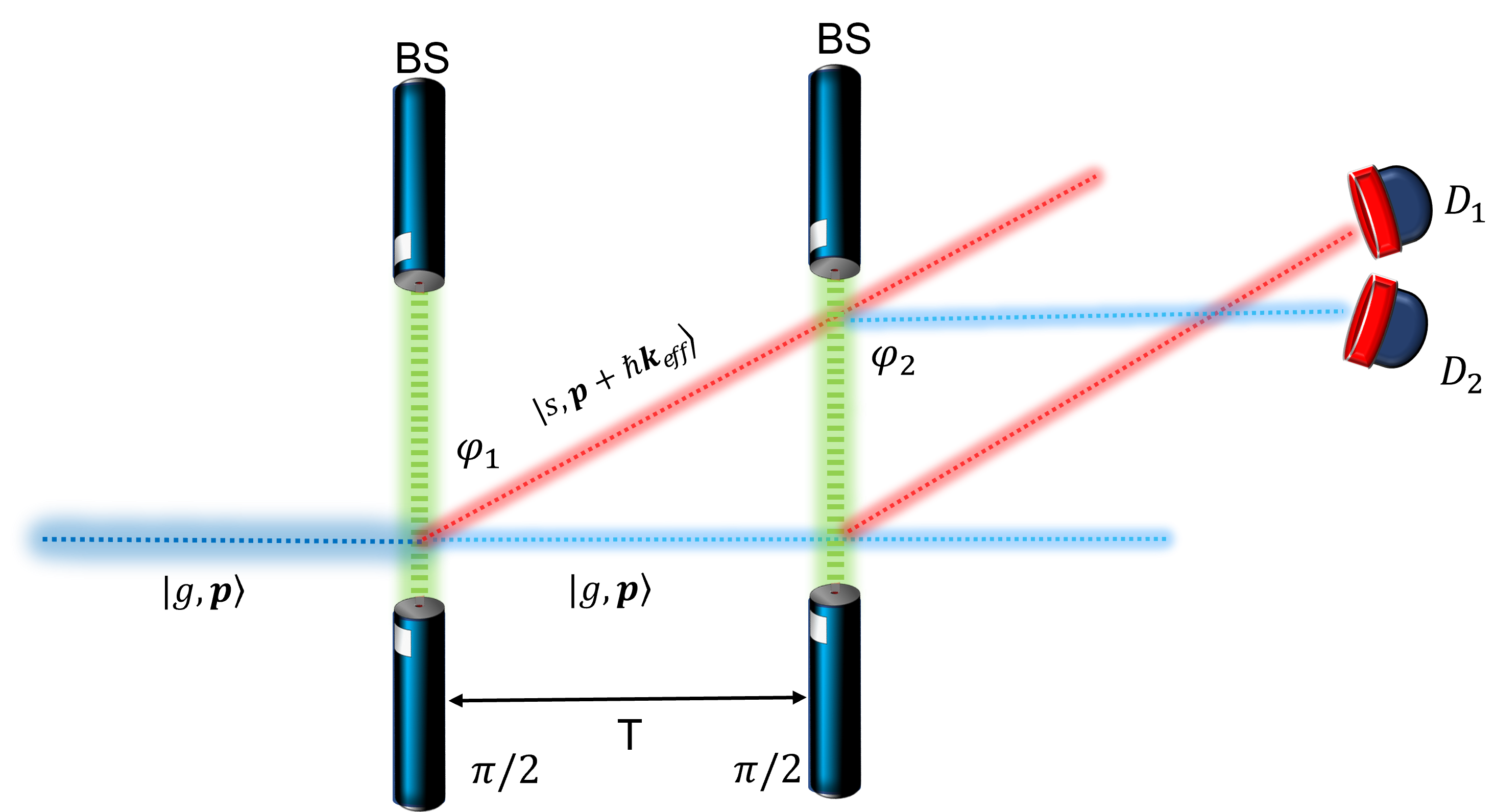}
	\centering 
	\caption
	{\textbf{An illustration of a Ramsey-Bordé atomic interferometer:} Atoms are initially prepared in their ground state, denoted by $\ket{g,\boldsymbol{p}}$. The atoms enter the interferometer and pass through two pairs of $\pi/2 - \pi/2$ laser pulses. These pulse pairs are separated by a precisely controlled time interval, denoted by T. The specific sequence of these laser pulses creates atomic interference, which is then detected by detectors $D_1$ and $D_2$.
	}
	\label{pp6}
\end{figure}

Atom interferometry based on a sequence of light pulses is inspired by Ramsey’s separated oscillatory field methods, introduced around 1950 to improve the stability of atomic clocks\cite{Ramsey,Cadoret2009}. In 1989, Christian Bordé extended Ramsey's principles to the domain of laser interferometry\cite{BORDE198910}, resulting in what is now often called a Ramsey–Bordé interferometer. This interferometer closely resembles the AMZI and can be constructed using a method based on two $\pi/2$ Raman pulse sequences, as shown in Fig.\ref{pp6}.\\
The process is similar to AMZI, but the details are different, which we will discuss.

\textbf{1.\hspace*{0.15cm}Preparing the Stage:} A cloud of cold atoms is prepared in a specific internal state, typically the ground state of the atom $\ket{g,\boldsymbol{p}}$, before being sent to the interferometer. Herein, we use a $\Lambda$-type atom (Fig.\ref{pp3}), where the intermediate energy level of these atoms can be adiabatically eliminated.\\

\textbf{2.\hspace*{0.15cm}Inducing Superposition:}  The first ($\pi/2$) laser pulse pair creates a superposition of the atom's ground and excited states, similar to a Hadamard gate in quantum computing. This process introduces a relative phase shift, $\phi_{1}$, resulting in a rotation of the original state vector.
 For further details, refer to step 5 of Sec. \ref{mach}.\\

\textbf{3.\hspace*{0.15cm}Free Evolution:}  During free evolution time \( T \), external forces can induce a phase difference between superposition paths, impacting the interference pattern. After time \( T \), the state vector is a combination of phase-shifted ground and excited states. The phases \(\phi_g\) and \(\phi_e\) for these states are given by \(\phi_i = {T E_i}/{\hbar}\), where \( E_i \) represents the energy levels\cite{Cadoret2009}.\\

\textbf{4.\hspace*{0.15cm}The Second  Pair of Pulses:}  The second pair of laser pulses (\(\pi/2\) pulses) is applied after the free evolution zone, acting as a second beam splitter that further splits and mixes the atomic wave packets \(\ket{g,\textbf{p}}\) and \(\ket{s,\textbf{p}+\hbar\textbf{k}_{\text{eff}}}\). This pulse induces \(\phi_2\) and also recombines the superposition of states. However, the phase difference accumulated during free evolution will affect this coherence\cite{Cadoret2009}.\\
	
\textbf{5.\hspace*{0.15cm}Detection and Interference:} The atoms undergo a state-dependent detection process, enabling the measurement of the relative populations of the final states. This measurement reflects the resulting interference pattern.  The probability \(P_s\) of finding the atom in the excited state after passing through the interferometer is\cite{Cadoret2009}
\begin{equation}
	P_s = \dfrac{1}{2}\Big[1+\cos \left( {\phi_e - \phi_g + \phi_1 - \phi_2}\right)\Big]
\end{equation}
The interference pattern in a Ramsey-Bordé interferometer depends on the total phase difference between two atomic paths after a free evolution period. In an undisturbed state, recombined atoms show equal populations in final states, resulting in no signal. However, external forces that shift this phase produce a detectable signal proportional to the phase shift's magnitude and direction. The interferometer achieves high coherence by carefully controlling laser pulses and timing, enabling it to detect subtle external influences with great precision. This setup demonstrates atomic self-interference, using atoms' wave-like nature to facilitate high-precision measurements of forces, accelerations, and fundamental constants.
\section{\label{Iqt}Interference in quantum technologies}
At the forefront of the ongoing revolution in quantum technologies lies a fundamental and intriguing phenomenon: quantum interference. Unlike the deterministic framework of classical physics, quantum mechanics allows particles to exist in superpositions, where a quantum bit, or qubit, can occupy multiple states (0 and 1) simultaneously. This principle of superposition, intertwined with wave-particle duality, is central to understanding quantum interference\cite{Nielsen_2010,Zubairy-Q,dirac}.\\The ability to control and exploit quantum interference underpins many advanced quantum technologies, enabling revolutionary progress in various domains. These include:

1. \textbf{Quantum Computing:} Interference enables quantum algorithms to explore and solve complex problems more efficiently than classical algorithms. By manipulating the phases of qubit states, quantum computers can perform parallel computations and achieve significant speedups for specific tasks \cite{Shor1997, Grover1996, Montanaro2016, Arute2019}.

2. \textbf{Quantum Cryptography:} Interference plays a pivotal role in securing quantum communication channels.  Quantum Key Distribution (QKD) use the principles of interference to detect eavesdropping and ensure secure key exchange \cite{Bennett1984, Ekert1991, Gisin2002, Scarani2009, Boaron2018}.

3. \textbf{Quantum Metrology and Imaging:} In quantum sensing and metrology, interference is harnessed to achieve unprecedented measurement precision. Devices like Superconducting Quantum Interference Devices (SQUIDs) rely on quantum interference to detect minute changes in magnetic fields, while advanced spectroscopic techniques use interference to analyze material properties with high accuracy \cite{Clarke2004, Caves1981, schnabel2017, Backes2021}.

These applications showcase the transformative power of quantum interference. It is not just a theoretical curiosity; it is a practical tool that drives technological advancements and deepens our understanding of the quantum world. Given the vast potential of quantum interference, this paper explores its role in three key areas  described in the following sections.
 \subsection{Quantum Computing}
Quantum interference is a key mechanism enabling the manipulation of qubit states through precise control over their phase relationships. Quantum algorithms exploit interference to explore a vast computational space in parallel. For instance, Shor’s algorithm, which factors large integers, and Grover’s algorithm, designed for unstructured search, utilize constructive interference to amplify correct solutions while using destructive interference to suppress incorrect ones\cite{Shor1997, Grover1996, Montanaro2016, Arute2019}. This capability promises exponential speedups over classical algorithms, with profound implications for cryptography, complex system simulations, and optimization problems \cite{Ladd2010, Preskill2018}.\\
In a conventional computer, bits are classical objects, like voltages, that represent "0" or "1." Quantum bits (or qubits), however, can exist in two quantum states,  $\ket{0}$ and $\ket{1}$, and can be realized in various physical systems, such as the polarization states of photons or the energy states of atoms. A computer using qubits is called a quantum computer\cite{Zubairy-Q}.\\
Qubits have unique properties not found in classical bits\\
1. \textbf{Quantum entanglement:}  Multiple qubits can become entangled, causing the state of one qubit to depend on the state of another, even when separated by large distances\cite{Zubairy-Q}. Entanglement is not the primary focus of this paper and is therefore not explored in detail.\\
2. \textbf{Coherent superposition}: Qubits can exist in a superposition of states, meaning they can be in a combination of $\ket{0}$ and $\ket{1}$ simultaneously. For any qubit, the state can be represented as\cite{Zubairy-Q,Nielsen_2010,Barnett2009}
\begin{equation}
	\ket{\psi}=c_0 \ket{0} + c_1 \ket{1}
\end{equation}
where \( c_0 \) and \( c_1 \) are complex numbers that satisfy \( |c_0|^2 + |c_1|^2 = 1 \). As noted in previous sections, quantum interference relies on the existence of superposed states. Without superposition, there are no multiple probability amplitudes to interfere.\\
Our quantum computation system comprises three key components\cite{peter,Barnett2009,Loft2020,Nielsen_2010}\\ 
1. A finite collection of qubits: These qubits can be atoms, photons, trapped ions, or other quantum systems.\\
2. Quantum gates: These operations manipulate the states of the qubits.\\
3. Measurement: We measure the final state of individual qubits to extract the desired information.\\
We are particularly interested in exploring the application of quantum interference in quantum computation. Quantum interference plays a crucial role in the construction of quantum gates. For example, the F-STIRAP technique utilizes interference effects to create a Hadamard gate for a single qubit\cite{Vit99}, as shown in Eq.\eqref{e83}. Sec.\eqref{F-ST} details how F-STIRAP leverages atomic interactions to achieve this.\\
Similarly, interferometers like AMZI or Ramsey-Bordé can be used to construct quantum gates depending on the applied laser pulse. A $\pi/2$ pulse, as described by Eq.\eqref{zz23}, can be used to construct a Hadamard gate, while a $\pi$ pulse (see Fig.\ref{pp5} for an example involving AMZI) can create a bit-flip gate (X-gate) \cite{Zubairy-Q,Ekert_1998,Barenco1995}.\\ 
In essence, quantum interference allows for precise control over the quantum states of the qubits, enabling the construction of various quantum gates. This ability to manipulate qubits through interference is fundamental to the power and versatility of quantum computation.\\

Quantum algorithms are specifically designed to operate on quantum computers, exploiting principles from quantum mechanics to potentially solve certain problems more efficiently than classical algorithms on classical computers. Here, we briefly review the application of quantum interference in some key quantum algorithms. However, it must be noted that quantum interference and superposition are nearly always present in all quantum algorithms.
\paragraph{\textbf{Deutsch's algorithm:}}
In 1985, David Deutsch introduced a foundational problem that illustrates the potential of quantum computers. Although it is a theoretical construct with limited practical applications, Deutsch's algorithm is one of the earliest and simplest quantum algorithms. It effectively demonstrates the potential advantage of quantum computing over classical computing for specific types of problems\cite{Zubairy-Q}.\\
Quantum computers can leverage parallelism to process data simultaneously. However, to obtain a result, a measurement must be performed, which collapses the quantum state to a single outcome from all possible results. To observe and utilize all potential outcomes, specialized algorithms are necessary. Deutsch's algorithm addresses this challenge by combining quantum parallelism with a fundamental property of quantum mechanics known as interference\cite{Nielsen_2010}.\\
\begin{figure}[h!]
	\includegraphics[width=7.5cm]{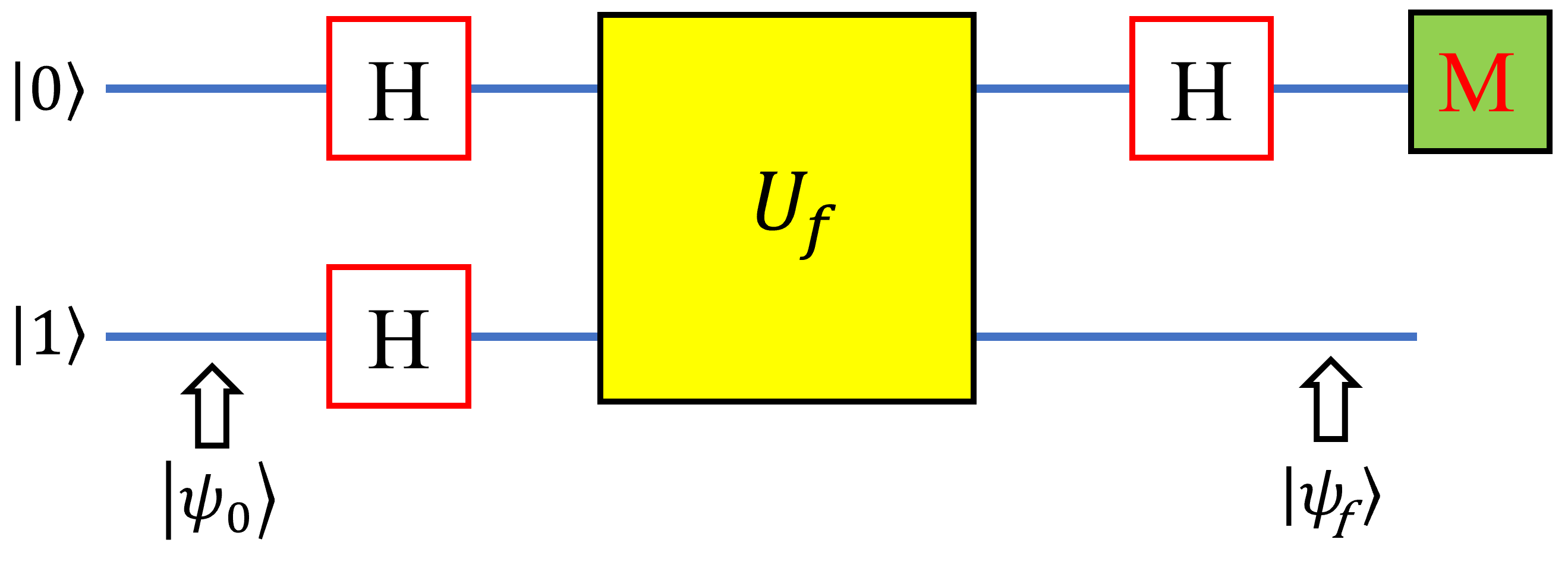}
	\caption{\textbf{Quantum circuit implementing Deutsch’s algorithm:} In this circuit, \(H\) is the Hadamard gate, \(U_f\) is the Oracle, and \(M\) denotes a measurement at the end of the process. Here, \(\ket{\psi_0}\) and \(\ket{\psi_f}\) are the initial and final states, respectively.}
	\label{deu}
\end{figure}
Suppose we are given a binary function $f(x)$, where x can only be 0 or 1. The function's output, $f(x)$, can also be either 0 or 1. In other words\\
$f(0)$ can be 0 or 1\\
$f(1)$ can be 0 or 1\\
Deutsch's algorithm can determine whether this unknown function $ f(x)$ is constant (always outputs the same value) or balanced (outputs 0 and 1 with equal probability) using only one evaluation with a quantum computer. For use of this algorithm, start with two qubits in the state $=\ket{0}\ket{1}$. Then the initial state of the two qubits is $\ket{\psi_0} =\ket{0}\ket{1}$. \\
Deutsch’s algorithm, depicted in Fig.\ref{deu}, begins with the initial state passing through two Hadamard gates applied to target qubits, followed by the application of \(U_f\). Here, \(U_f\) acts as a black-box oracle, transforming the state according to \(U_f \ket{x}\ket{y} = \ket{x}\ket{y \oplus f(x)}\), where \(\oplus\) denotes the bitwise XOR operation. The final state, modified by the last Hadamard gate applied to the qubit from \(U_f\), is represented as\cite{Nielsen_2010,Barnett2009}
\begin{equation}
	\ket{\psi_f} = \pm \ket{f(0) \oplus f(1)} \frac{1}{\sqrt{2}} ( \ket{0} - \ket{1} )
\end{equation}
Measurement of the data qubit reveals \(\ket{f(0) \oplus f(1)}\), yielding \(\ket{0}\) if \(f(0) = f(1)\) \([f(0) \oplus f(1) = 0]\), or \(\ket{1}\) if \(f(0) \neq f(1)\) \([f(0) \oplus f(1) = 1]\). This demonstrates the power of quantum superposition, enabling the determination of \(f(0) \oplus f(1)\) with a single evaluation of \(f(x)\), compared to the classical requirement of at least two evaluations.\\
This example underscores the distinction between quantum parallelism and classical randomized algorithms. In classical computing, states like \(|0\rangle|f(0)\rangle + |1\rangle|f(1)\rangle\) are mutually exclusive. Quantum computing, however, leverages interference among such states to reveal global properties of functions like \(f\). This interference is facilitated by operations such as the Hadamard gate, as exemplified in Deutsch’s algorithm\cite{Nielsen_2010,Barnett2009}.\\
Deutsch's algorithm was extended by David Deutsch and Richard Jozsa in 1992, resulting in the Deutsch-Jozsa algorithm.
\paragraph{\textbf{Grover's search algorithm:}}
Database searching is fundamental in information processing. When dealing with a database of \( N \) entries, the goal is to locate a specific record within an unsorted list. This involves finding a particular entry in the database identified as the target, known to exist exactly once. Grover's algorithm, developed in 1996, is a quantum algorithm designed for this purpose and offers a quadratic speed-up over classical algorithms for searching unsorted databases\cite{Grover97,Gregg,peter}.\\
Grover's algorithm relies on two core principles of quantum computing: superposition and interference. It leverages the inherent quantum parallelism of superposition to evaluate multiple entries simultaneously. Additionally, interference is strategically used to amplify the probability amplitude of the target entry while reducing the amplitudes of non-target entries \cite{Abdulrahman2024,Bogatyrev2023}.\\
The detailed workings of Grover's algorithm are quite complicated, so we will present the gist of the process using a simple example. The algorithm’s goal is to find a specific item within an unsorted database of size \(N\) with a complexity of approximately \(\sqrt{N}\) iterations, compared to the linear \(O(N)\) complexity of classical algorithms. The algorithm requires a quantum register comprising \(n\) qubits (\(n = \log_2(N)\)). Each qubit is initialized in the state \(|0\rangle\)\cite{Abdulrahman2024,fox}.\\
Let’s review Grover’s algorithm step by step\\

	\textbf{Step 1: Initialization and Application of the Hadamard Transform}\\
	
    1. \textit{Initialization:} Initialize \( n \) qubits in the state \( |0\rangle^{\otimes n} \), where \( n \) is the number of qubits needed to represent \( N \) entries (i.e., \( n = \log_2 N \)).\\
    
    2. \textit{Apply Hadamard Transform:}  Apply the Hadamard gate to each qubit to create an equal superposition of all \( N \) possible states\cite{fox,peter, Mermin2007,Nielsen_2010, Barnett2009}
\begin{equation}
			H^{\otimes n} |0\rangle^{\otimes n} = \frac{1}{\sqrt{N}} \sum_{x=0}^{N-1} |x\rangle
\end{equation}
		Now, the system is in the state
\begin{equation}
			|\psi\rangle = \frac{1}{\sqrt{N}} \sum_{x=0}^{N-1} |x\rangle
			\label{Grpsi}
\end{equation}

\textbf{Step 2: Oracle Application}
 The oracle acts as a black-box function to identify the target item $w$ being searched for. It flips the phase of the amplitude of the target state \( |w\rangle \) while leaving the other states unchanged. The oracle function \( O \) operates as follows\cite{Nielsen_2010}
\begin{equation}
			O|x\rangle = (-1)^{f(x)} |x\rangle
\end{equation}
where \( f(x) \) is defined by
		\begin{align*}
			f(x) &= 1 \quad \text{if } x = w \\
			f(x) &= 0 \quad \text{otherwise}
		\end{align*}
		
\textbf{Step 3: Amplitude Amplification with Diffusion Operator}
Grover’s algorithm leverages constructive and destructive interference to amplify the probability amplitude of the marked state(s) and diminish the amplitude of non-marked states. This amplification process involves two main reflections: reflection about the mean and reflection about the marked state(s), facilitated by the Diffusion Operator.\\

	I. \textit{Reflection about the Mean:} Perform a phase inversion about the mean amplitude of all states, flipping the sign of the amplitude for each state relative to the mean amplitude.\\
	
	II. \textit{Reflection about the Marked State(s) using Diffusion Operator:} Apply the Diffusion Operator \( D \), which amplifies the marked states and reduces the non-marked states through constructive and destructive interference\cite{Nielsen_2010}
	\begin{equation}
				D =  2|\psi\rangle \langle \psi| - I
	\end{equation}
where $\psi$ is shown in Eq.\eqref{Grpsi}. These operations create constructive interference for the marked state \( |w\rangle \) (where the amplitudes reinforce each other) and destructive interference for the non-marked states (where the amplitudes cancel each other out to some extent).\\

\textbf{Step 4: Iteration}
Repeat Steps 2 and 3 approximately \( {\pi\sqrt{N}}/{4} \) times, assuming the problem has exactly one solution. Each iteration increases the amplitude of the marked state \( |w\rangle \) while decreasing the amplitudes of non-marked states\cite{Nielsen_2010,peter,Rieffel2011,fox} due to constructive and destructive interference, respectively.\\

\textbf{Step 5: Measurement}
After approximately \({\pi\sqrt{N}}/{4} \) iterations, the amplitude of the marked state \( |w\rangle \) is significantly higher than that of other states. Measure the quantum state. Due to the amplified amplitude of the marked state, the measurement of the quantum state will yield the target state \( w \) with high probability.\\

By carefully orchestrating constructive and destructive interference, Grover's algorithm efficiently amplifies the probability of finding the target item, providing a quadratic speedup over classical search algorithms. This makes it an invaluable tool for searching unsorted databases in various fields such as cryptography, drug discovery, and financial modeling.\\

For more details about quantum computing, its applications, and the observed role of interference in them, you can refer to Refs.[\onlinecite{Nielsen_2010, peter, Rieffel2011, fox, Abdulrahman2024, Bogatyrev2023, Grover97, Gregg, Barnett2009, Loft2020, Ekert_1998, Barenco1995}].
\subsection{Quantum Cryptography}
Cryptography is the art of encoding messages so that only the intended recipient can read them\cite{fox}. Quantum cryptography introduces groundbreaking methods for secure communication, distinct from traditional cryptography that relies on complex mathematical techniques to prevent eavesdropping. It harnesses the principles of quantum physics, where information is transmitted and stored using physical carriers like photons in optical fibers or electrons in electrical currents\cite{Pathak2023, alexander}.

A key application of quantum cryptography is Quantum Key Distribution (QKD). QKD uses quantum mechanics to securely distribute private information\cite{Ojha, Nielsen_2010}. It allows two parties to generate private key bits over a public channel, which can then be used in a classical cryptosystem for secure communication. The essential requirement for QKD is that qubits are transmitted with an error rate below a certain threshold. The security of the generated key is guaranteed by the unique properties of quantum information, grounded in the fundamental laws of physics\cite{Nielsen_2010}.

Quantum interference ensures secure communication in quantum cryptography. Interferometric QKD systems, known for their robustness against polarization variations in optical fibers, have become increasingly practical. These systems often employ sources that emit photon pairs, which traverse short (S) and long (L) paths in interferometers at distant stations.\\
The self-aligned design, incorporating Faraday mirrors, automatically compensates for environmental polarization transformations, eliminating the need for manual alignment and enhancing system stability \cite{Zbinden1998}. \\
These systems use quantum interference to detect potential eavesdropping. Coincidence detection ensures that photons taking the same path (both S or both L) result in specific correlated outcomes.

Quantum interference between these two probability amplitudes is crucial. It gives rise to non-local quantum correlations, meaning correlations stronger than any classical explanation, that violate Bell's inequality. This interference ensures that the probability amplitudes of the photons taking the same path (either both long or both short) combine to produce specific correlated outcomes. This phenomenon, key to securely distributing cryptographic keys, underpins protocols like BB84 and B92, as demonstrated in experimental setups leveraging constructive and destructive interference with Faraday mirrors \cite{Zbinden1998, alexander}.\\

Phase encoding is another method for generating secure keys in QKD, leveraging the interference of quantum states to encode and transmit information securely. The idea of encoding qubit values in the phase of photons was first proposed by Bennett in his 1992 paper introducing the two-state protocol\cite{Gisin2002}.\\
In phase encoding, information is encoded in the phase difference between two states, with the integrity of this encoded information relying heavily on phase coherence. Thus, the interference mechanism becomes essential for both detecting and decoding the phase-encoded information.

This method involves three steps:
\begin{enumerate}
	\item \textbf{Preparation:}
	\begin{itemize}[leftmargin=3pt]
		\item \textit{Superposition:} Quantum particles, such as photons or atoms, are placed in a superposition of states.
		\item \textit{Phase Modulation:} A phase shift $\phi$ is introduced between the states to encode information, created using a Mach-Zehnder interferometer.
	\end{itemize}
	\item \textbf{Transmission:} The phase-encoded states are transmitted through a channel while maintaining coherence.
	\item \textbf{Measurement:}
	\begin{itemize}[leftmargin=3pt]
		\item \textit{Interferometers:} At the receiver's end, the states are combined using interferometers to produce interference patterns.
		\item \textit{Phase Measurement:} The phase $\phi$ is extracted from the interference pattern to decode the key.
	\end{itemize}
\end{enumerate}

Combining quantum states creates interference patterns based on their phase difference ($\phi$). Constructive interference occurs when $\phi = 0$ or multiples of $2\pi$, where the waves perfectly reinforce each other, resulting in a strong signal. Destructive interference happens when $\phi = \pi$, where the waves cancel each other out, leading to a weak signal\cite{Gisin2002}. The receiver analyzes the interference pattern, determined by the phase difference, to decode the encoded information, such as a key bit in quantum cryptography.

In the BB84 protocol with phase encoding, interference patterns are crucial for secure communication. Quantum states are encoded using four distinct phases: $0$ and $\pi$ for one basis (Z basis), and ${\pi}/{2}$ and ${3\pi}/{2}$ for the other basis (X basis). Constructive and destructive interference play a key role: when the phase difference is $0$ or $2\pi$, constructive interference results in a strong signal; when the phase difference is $\pi$, destructive interference leads to a weak signal. The key bit is determined by these interference patterns: if the phase is $0$ or ${\pi}/{2}$, the key bit is 0; if the phase is $\pi$ or ${3\pi}/{2}$, the key bit is 1. Alice prepares and sends a photon with a specific phase, and Bob measures the phase after it interferes. They then publicly compare their basis choices, retaining measurements where their bases match. These retained measurements are used to generate key bits, with interference ensuring the correct phase determination. This method, along with error correction and privacy amplification, secures the key distribution by exploiting quantum interference and phase differences for encoding information\cite{Gisin2002, Zbinden1998, Qincheng, Pathak2023}.

Phase encoding in QKD exploits the interference of quantum states to securely transmit cryptographic keys. By encoding information in the phase difference between states and decoding it through interference patterns, this method ensures that any eavesdropping attempts are detectable due to disturbances in phase coherence. This technique leverages fundamental principles of quantum mechanics to maintain security and integrity in key distribution.
 \subparagraph{Quantum memory:}
In this paper, we investigate the role of atomic interference in various quantum phenomena. While previous discussions have primarily focused on photon interference in quantum cryptography and QKD, certain phenomena, such as quantum memory, underscore the significance of atomic interference in related applications. 

Quantum memory is pivotal in both quantum cryptography and QKD, significantly influencing security and functionality. In QKD, if an eavesdropper possesses quantum memory, they can store quantum states without immediate measurement, preserving entangled information for later analysis when additional classical information becomes accessible.\\
This capability poses a potential threat to QKD security, as it may delay the detection of eavesdropping attempts. However, the no-cloning theorem and the disturbance caused by measurement still provide robust defenses against such attacks. For legitimate users, quantum memory enhances QKD protocols by enabling advanced techniques like entanglement swapping and delayed error correction, leading to more accurate and secure key generation.\\
Additionally, quantum memory is essential for developing scalable quantum networks, as it facilitates the operation of quantum repeaters and the secure distribution of keys among multiple users without requiring direct quantum channels. Therefore, quantum memory plays a transformative role in QKD, serving both as a tool to strengthen cryptographic defenses and as a potential vulnerability if exploited by adversaries\cite{PhysRevLett.79.4034,mor1999quantum}.

In quantum memory systems, atoms interact with light to create a stable medium for temporarily holding quantum information. Such systems often use atomic ensembles—collections of many atoms that interact collectively with light. These ensembles absorb quantum information encoded in light pulses and store it in a collective atomic state. The large number of atoms enables strong interactions with light, which enhances memory efficiency and reliability\cite{RevModPhys.82.1041,Simon_article}. 

A widely used technique, EIT, involves arranging atoms in a $\Lambda$-type scheme to become transparent to a probe light in the presence of a coupling light. Through EIT, the atomic ensemble slows or halts light pulses, mapping the quantum state of light onto atomic states that can later retrieve the light pulse. This approach provides a long-lived atomic storage state, making it suitable for use in quantum memory and repeaters in quantum communication\cite{Simon_article}. Quantum memories have been developed using three main approaches: optically controlled memories, engineered absorption, and hybrid schemes. Systems relying on a $\Lambda$-type configuration are classified as optically controlled memories\cite{Heshami2016}.

Another approach to quantum memory involves Raman interactions, where off-resonant light interacts with atoms in a $\Lambda$-type scheme. This process enables quantum states of light to be mapped onto atomic states without requiring strong resonance, thereby minimizing spontaneous emission. However, it requires powerful pump light, which introduces noise, making it most suitable for applications where cost or simplicity is prioritized\cite{Lijun}. Atomic interference plays a critical role in these systems; coherent interactions between light and atomic states generate stable interference patterns, essential for preserving the fidelity and stability of quantum memory. These interference patterns—constructive and destructive—prevent decoherence, stabilizing the stored quantum information even amidst environmental disturbances like thermal noise in warm atomic vapors. Additional measures, such as magnetic shielding and optimized temperature, further enhance this coherence, making atomic interference pivotal to achieving high-fidelity quantum memory and robustness against noise and decoherence\cite{Lijun}.
\subsection{Quantum Metrology}
Quantum metrology, a branch of quantum physics, enhances measurement precision by leveraging the unique properties of quantum states, such as superposition and entanglement. Over the past five decades, remarkable advancements in laser physics and nanotechnology have empowered scientists to manipulate individual quantum entities—such as atoms, ions, electrons, and Cooper pairs. These breakthroughs form the basis of quantum metrology, which focuses on measuring discrete quanta (such as charge or magnetic flux quanta) rather than continuous variables, as seen in classical metrology\cite{Uwe2015,dowling2003}. One of its key goals is to study the fundamental limits of precision, allowing for high-precision measurements of physical quantities like length, time, frequency, and temperature\cite{dowling2003,Xiaoying,Rui}.\\
A crucial concept in quantum metrology is quantum interference, which arises from the superposition of quantum states. When particles such as photons or atoms exist in a superposition of states, their wavefunctions can interfere constructively or destructively. This interference pattern is highly sensitive to changes in phase, making it possible to detect minute variations in the quantity being measured\cite{Degen}. Numerous applications of quantum metrology benefit from quantum interference, demonstrating the practical advantages of quantum-enhanced measurements.\\
In particular, quantum metrology has recently been demonstrated to improve the sensitivity of some of the most sophisticated optical instruments currently available, such as large-scale interferometers for gravitational wave detection, which are otherwise limited by photon shot-noise. Other examples of promising quantum-enhanced measurement techniques include particle tracking in optical tweezers, sub-shot-noise wide-field microscopy, quantum correlated imaging, spectroscopy, displacement measurement, and remote detection and ranging\cite{Berchera2019}.\\

 \paragraph{\textbf{SQUID}:}
  This phenomenon stands as a remarkable invention, first demonstrated in 1964, just two years after Brian D. Josephson's groundbreaking theoretical work\cite{Cova1996, Uwe2015}. These devices combine two fascinating phenomena: flux quantization and Josephson tunneling, making them one of the oldest and most sensitive magnetic field sensors ever developed\cite{Clarke2004, Degen}.\\
In a SQUID, the fundamental building block is the Josephson junction, which consists of two superconductors separated by a thin insulating barrier.\\
SQUID's extraordinary sensitivity lies in its ability to exploit quantum interference. Imagine a superconducting loop carrying a perfectly synchronized flow of electrons, a delicate ballet disrupted by even the faintest magnetic nudge. This nudge alters the phase of the electron wavefunctions, leading to constructive and destructive interference within Josephson junctions, similar to how light waves interact \cite{Feynman}. These junctions, microscopic bridges between superconductors, are the stage for this quantum choreography. Cooper pairs, electron pairs, can tunnel through this thin barrier, and the current flowing through it depends on the phase difference between the superconducting wavefunctions on either side – a concept rooted in quantum mechanics\cite{Josephson, barone1982}.\\
The current \( I \) flowing through a Josephson junction depends on the phase difference \( \delta \) between the superconducting wavefunctions on either side of the junction. According to the Josephson relations, the current \( I \) can be expressed as \cite{Uwe2015, Clarke2004}
\begin{equation}
	I = I_c \sin(\delta)
\end{equation}
where \( I_c \) is the maximum supercurrent or critical current of the junction\cite{Clarke2004}.\\

The superconducting loop in a SQUID exhibits a crucial property known as magnetic flux quantization. This dictates that the total magnetic field threading the loop can only assume specific discrete values. This characteristic, described by the magnetic flux quantum (\( \Phi_0 \)), allows SQUIDs to leverage quantum mechanics to translate the subtle influence of a magnetic field into a measurable change in voltage or current\cite{Tesche1977, Jaklevic}.\\
There are two primary SQUID designs: DC SQUIDs and RF SQUIDs. Both types rely on quantum interference to function, but with slight variations. DC SQUIDs, equipped with two Josephson junctions, exhibit a periodic oscillation in voltage as the applied magnetic field changes\cite{Clarke2004}. RF SQUIDs, on the other hand, utilize a single junction and a radiofrequency signal to unveil the magnetic field's influence through quantum interference\cite{Clarke1993, Clarke2004}.\\

When a magnetic flux \( \Phi \) threads through the superconducting loop of a SQUID, it induces a phase difference \( \delta \) across the Josephson junctions. The relationship between the magnetic flux \( \Phi \) and the phase difference \( \delta \) is given by\cite{Jaklevic, Uwe2015, Clarke2004}
\begin{equation}
	\delta = \frac{2 \pi}{\Phi_0} \Phi 
\end{equation}
where \( \Phi_0 = {h}/{2e} \approx 2.07 \times 10^{-15} \, \text{Wb} \) is the magnetic flux quantum, \( h \) is Planck's constant, and  $\ket{e}$ is the elementary charge.\\

The phase difference \( \delta \) affects the current \( I \) through the Josephson junction. As the magnetic flux \( \Phi \) changes, the phase difference \( \delta \) changes accordingly, modulating the current \( I \) through the junction. This modulation is a manifestation of quantum interference. Quantum interference in SQUIDs occurs because the phase difference \( \delta \) alters the supercurrent flowing through the junction, leading to constructive and destructive interference effects.\\

The voltage \( V \) across a SQUID is related to the current \( I \) through the junctions. For a DC SQUID, the voltage \( V \) exhibits oscillatory behavior with the applied magnetic flux \( \Phi \) \cite{Clarke1993}
\begin{equation}
	V = V_0 \sin\left( \frac{\pi \Phi}{\Phi_0} \right) 
\end{equation}
where \( V_0 \) is the maximum voltage.
This equation shows that the voltage output of the SQUID is periodic with respect to the magnetic flux \( \Phi \), reflecting the interference pattern resulting from the phase difference \( \delta \) induced by the magnetic field.\\

This quantum interference within the SQUID has a distinct measurable signature. In a DC SQUID, the voltage across the device oscillates as a function of the applied magnetic field. This oscillation is a direct consequence of the supercurrents' constructive and destructive interference through the two junctions. Similarly, in an RF SQUID, the modulated RF signal reflects the interference affecting the inductance of the loop due to the magnetic field. By analyzing these periodic variations, scientists can determine the applied magnetic field with exceptional precision\cite{Fagaly, Tesche1977, Clarke1993, Clarke2004}.\\

Quantum interference lies at the very core of SQUID's remarkable functionality. By manipulating the behavior of supercurrents through this fascinating phenomenon, SQUIDs have become invaluable tools across various scientific disciplines. From probing the faint magnetic signals of the brain for medical imaging to uncovering the intricate magnetic signatures of material science, SQUIDs continue to push the boundaries of our ability to measure and understand the world around us\cite{Clarke2004}.\\

\paragraph{\textbf{Atomic Clocks}:}
Quantum metrology has revolutionized precision measurement by leveraging atomic interference to create ultra-stable measurement standards. Unlike traditional methods dependent on physical artifacts, quantum metrology uses the unchanging constants of nature to establish universally stable units essential for scientific and technological progress \citep{gobel2015quantum_metrology}. A prime example of this advancement is the atomic clock, where interference mechanisms enable reliable and precise timekeeping. Through methods such as CPT, atomic clocks use interference effects to generate stable quantum states that resist decoherence and maintain precise frequency standards, critical for applications in navigation, telecommunications, and other essential fields \citep{arimondo2010advances, merimaa2003all_optical}.

Atomic interference is fundamental to atomic clocks' exceptional precision and stability, as it establishes coherent superpositions of quantum states. This process creates ultra-stable "dark states" with narrow spectral linewidths, which are essential for consistent and accurate timekeeping. These non-absorbing dark states (see Sec.\ref{CPT}) stabilize the clock’s reference frequency, minimizing unwanted atomic transitions, reducing decoherence, and enhancing overall stability.

The principles of atomic interference allow atomic clocks to achieve high precision without requiring large, energy-intensive microwave cavities. Techniques such as CPT offer precise frequency standards using optical fields, enabling the development of more compact, energy-efficient clocks suitable for portable applications \citep{gobel2015quantum_metrology}.

\textit{Cold Atomic Clocks}: By using laser cooling to reduce atomic motion, cold atomic clocks minimize Doppler broadening, leading to increased accuracy. In these clocks, atomic interference enables the formation of "dark states," reducing energy loss and maintaining high precision. Additional techniques like magneto-optical trapping extend observation times and enhance stability, as demonstrated in fountain atomic clocks \citep{zhu2013double_lambda,liu2009transient_cpt,vanier2005review}.

\textit{Warm Atomic Clocks}: In contrast to cold atomic clocks, warm atomic clocks (or thermal atomic clocks) use atoms within vapor cells at ambient or moderately elevated temperatures, often using elements like cesium or rubidium. Rather than relying on laser cooling, these clocks stabilize frequency with CPT, simplifying their design by eliminating complex cooling systems \citep{Camparo2007,vanier2005review,FriedemannTheses}. This approach yields compact, cost-effective clocks ideal for robust timekeeping applications such as telecommunications, GPS, and network synchronization \citep{FriedemannTheses}.

Compact atomic clocks, such as chip-scale atomic clocks (CSACs), leverage warm vapor cell methods to offer compact, low-power, and reliable timekeeping solutions. Advances in atom-chip technology have enabled smaller and more stable designs, including magnetically trapped atom clocks using Ramsey-type spectroscopy for improved accuracy \citep{Knappe:05,Zhang16}. These clocks typically exploit hyperfine transitions in isotopes like \(^{87}\text{Rb}\) or \(^{133}\text{Cs}\)., with interrogation through microwave cavities or integrated waveguides for precise frequency references \citep{arimondo2010advances,FriedemannTheses}. Recent progress in optical lattice and vapor cell technology has further optimized compact atomic clocks for balanced performance and portability \citep{zhu2013double_lambda,Knappe:05,Zhang16,Svenja}.

Whether in compact warm atomic clocks or high-precision cold atomic clocks, interference effects like CPT enable the ultra-narrow spectral lines and stable dark states that define clock accuracy. This precision supports diverse fields, from navigation and telecommunications to quantum computing and fundamental physics, allowing atomic clocks to continually advance precision standards in quantum metrology and push the boundaries of science and technology \citep{gobel2015quantum_metrology}.\\
\paragraph{\textbf{Atomic gravimeter}:}
One of the phenomena that highlights the vital role of atomic interference is the atomic gravimeter. These are highly precise instruments used to measure gravitational acceleration by employing the principles of cold atom interferometry. In this process, atoms are cooled to near absolute zero to form coherent matter waves. This approach enables atomic gravimeters to detect even small changes in gravitational forces by manipulating these matter waves through laser pulses. The laser pulses, typically arranged in a sequence of $\pi/2$-$\pi$-$\pi/2$ pulses, split, redirect, and recombine the atomic wave packets, forming an interference pattern whose phase shift is directly influenced by gravity (see \ref{mach}). The phase shift observed in the interference pattern corresponds to the gravitational acceleration experienced by the atoms as they fall (see Eq. \ref{z24}), providing a highly sensitive means of measuring gravitational acceleration with unprecedented accuracy \cite{Biedermann2015, Zhou2011, Cold}.

Atomic gravimeters, such as the mobile Gravimetric Atom Interferometer (GAIN), utilize a vertical configuration in which rubidium atoms are launched upward, increasing the interrogation time and thereby enhancing the measurement sensitivity. This setup, combined with advanced vibration isolation and compensation for the Coriolis effect, allows atomic gravimeters to perform high-precision measurements outside laboratory conditions, which is essential for portable applications \cite{Hauth2014}. Cold atom interferometry significantly enhances the sensitivity and accuracy of these devices over classical gravimeters, which rely on mechanical moving parts that are subject to wear and require periodic calibration. In contrast, atomic gravimeters achieve an absolute measurement of gravitational acceleration, offering a more stable and reliable measure of gravity over time \cite{Biedermann2015}.

Due to their high sensitivity, atomic gravimeters are valuable in various applications, from geophysical studies and Earth observation to testing theories in gravitational physics, such as general relativity and the weak equivalence principle. Precision measurements from these devices support mapping gravitational fields for environmental monitoring, resource exploration, and the precise determination of Newton's gravitational constant. For instance, Rosi et al. (2014) demonstrated that cold-atom interferometry can measure Newton’s gravitational constant with a relative uncertainty of 150 ppm, significantly advancing the precision attainable with traditional methods \cite{Rosi2014}. Thus, atomic gravimeters represent a major advancement in precision metrology, leveraging quantum mechanics and laser manipulation techniques to measure gravitational forces with exceptional accuracy and adaptability across various field applications \cite{Tino2021, Sinha2011, Cold}.\\

This text explores some applications of quantum interference in quantum metrology. However, there are other fascinating phenomena that utilize atom interefernce and also photon interference(such as Quantum Imaging, Gravitational Wave Detection ,...) . These include quantum gradiometers for measuring gravitational field gradients, quantum lithography for high-precision patterning, and quantum optical coherence tomography for advanced imaging. You can find more details on these applications in Refs.[\onlinecite{Tsang,Shin,Peters2001,Boto,Dbrowska,Kok,Schnabel2010, Danilishin2012, abbott2016observation, Acernese_2015,Tse,punturo2010einstein, reitze2019,Saxena2022, Moodley:23,Degen, Genovese2016, Cameron2024,kolobov2014}] or other sources used in this section.\\
\section{Conclusion}
In conclusion, atomic interference serves as a cornerstone in the advancement of quantum optics, enabling sophisticated manipulation of light-matter interactions fundamental to modern quantum technologies. Through phenomena such as CPT, STIRAP, and EIT, we achieve remarkable control over quantum states in multi-level atomic systems. These interference-based techniques allow for intricate manipulation of atomic populations and transitions, enabling effects like transparency in otherwise opaque media and efficient quantum state transfer—capabilities essential for applications in quantum information processing and precision sensing.

The applications of atomic interference span several critical areas of quantum technology, including quantum computing, cryptography, and quantum metrology. In quantum computing, interference facilitates stable, high-fidelity qubit control, enhancing gate operations and supporting the development of resilient quantum algorithms. In cryptography, interference-based protocols like QKD leverage phase coherence to detect eavesdropping, securing communication channels. Meanwhile, quantum metrology harnesses atomic self-interference in highly sensitive interferometers capable of detecting minute variations in gravitational and magnetic fields. This sensitivity is pivotal for advanced applications in gravitational wave detection, magnetic field sensing, and fundamental physics investigations.

This study reinforces atomic interference as a bridge between theoretical quantum mechanics and practical technological applications, underscoring its role in advancing quantum control and coherence. By deepening our understanding of interference-based processes, we lay the groundwork for innovations in quantum metrology, secure communication, and quantum computation, establishing interference as a vital element in next-generation quantum technologies.
\section{References}
  \bibliography{MyReferences}
\end{document}